\documentclass[12pt]{article}

\AtBeginDocument{
  \addtocontents{toc}{\footnotesize}
}

\pdfoutput=1

\usepackage{amsmath,amssymb,amsfonts,amscd,mathrsfs}
\usepackage{xcolor}
\definecolor{darkblue}{rgb}{0.1,0.1,.7}
\usepackage[colorlinks, linkcolor=darkblue, citecolor=darkblue, urlcolor=darkblue, linktocpage]{hyperref} 
\usepackage[square, comma, compress,numbers]{natbib}
\usepackage[]{graphicx}
\usepackage{geometry}
\usepackage[margin=10pt,font=small,labelfont=bf]{caption}
\usepackage{ifthen}
\usepackage{tikz}

\usepackage{accents}
\newlength{\dhatheight}
\newcommand{\doublehat}[1]{%
    \settoheight{\dhatheight}{\ensuremath{\hat{#1}}}%
    \addtolength{\dhatheight}{-0.35ex}%
    \hat{\vphantom{\rule{1pt}{\dhatheight}}%
    \smash{\hat{#1}}}}
    
\usepackage{dsfont} 

\newcommand{\ket}[1]{|#1\rangle}
\newcommand{\bra}[1]{\langle #1|}
\newcommand{\expec}[1]{\langle #1 \rangle}
\newcommand{\no}[1]{:\! #1 \!:}

\newcommand{\mrm}[1]{{\mathrm #1}}

\def\pd{\partial}
\def\a{\alpha}

\def\DD{\Delta}
\def\Oo{\mathcal{O}}

\def\eps{\epsilon}

\def\bn{\mathbf{n}}
\def\unit{\mathds{1}} 

\newcommand{\reef}[1]{(\ref{#1})}

\def\eps{\epsilon}
\newcommand{\beq}{\begin{equation}} 
\newcommand{\eeq}{\end{equation}}
\def\del {\partial} 
\def\nn{\nonumber} 
\def\bZ {\mathbb{Z}} 
\def\bR {\mathbb{R}} 
 
\def\calO {{\cal O}}

\def\calM {{\cal M}} 
 
\def\calV {{\cal V}} 
\def\calA {{\cal A}} 
\def\calB {{\cal B}} 

\def\bn{{\mathbf{n}}}
\def\bZ {\mathbb{Z}} 
 
\def\half{{\textstyle\frac 12}}

\def\ge{\geqslant}
\def\le{\leqslant}

\newcommand{\diffop}[2]{\ifthenelse{\equal{#2}{1}}{\frac{\mrm{d}}{\mrm{d} #1}}{\frac{\mrm{d}^#2}{\mrm{d} #1^#2}}}
\def\LIR{\Lambda_{\rm IR}}
\def\LUV{{\Lambda_{\rm UV}}}
\def\HCFT{H_{\rm CFT}}
\def\Sd{{\rm S}_d}
\def\Nd{{\rm N}_d}
\def\Ric{{\rm Ric}}
\def\Dmax{\Delta_{\rm max}}

\newcommand{\NO}[1]{{:\!#1\!:}}

\numberwithin{equation}{section}
\begin{document}

\vspace*{-.6in} \thispagestyle{empty}
\begin{flushright}
CERN PH-TH/2014-155
\end{flushright}
\vspace{1cm} {\Large
\begin{center}
{\bf A Cheap Alternative to the Lattice?}\\
\end{center}}
\vspace{1cm}
\begin{center}
{\bf Matthijs Hogervorst$^{a,b}$, Slava Rychkov$^{b,a,c}$, Balt C.~van Rees$^{b}$}\\[2cm] 
{
$^{a}$ Laboratoire de Physique Th\'{e}orique de l'\'{E}cole normale sup\'{e}rieure, Paris, France\\
$^{b}$ CERN, Theory Division, Geneva, Switzerland\\
$^{c}$ Facult\'{e} de Physique, Universit\'{e} Pierre et Marie Curie, Paris, France
}
\end{center}

\vspace{4mm}

\begin{abstract}
  We show how to perform accurate, nonperturbative and controlled calculations in quantum field theory in $d$ dimensions. We use the Truncated Conformal Space Approach (TCSA), a Hamiltonian method which exploits the conformal structure of the UV fixed point. The theory is regulated in the IR by putting it on a sphere of a large finite radius. The QFT Hamiltonian is expressed as a matrix in the Hilbert space of CFT states. After restricting ourselves to energies below a certain UV cutoff, an approximation to the spectrum is obtained by numerical diagonalization of the resulting finite-dimensional matrix. The cutoff dependence of the results can be computed and efficiently reduced via a renormalization procedure. We work out the details of the method for the $\phi^4$ theory in $d$ dimensions with $d$ not necessarily integer. A numerical analysis is then performed for the specific case $d = 2.5$, a value chosen in the range where UV divergences are absent. By going from weak to intermediate to strong coupling, we are able to observe the symmetry-preserving, symmetry-breaking, and conformal phases of the theory, and perform rough measurements of masses and critical exponents. 
  As a byproduct of our investigations we find that both the free and the interacting theories in non integral $d$ are not unitary, which however does not seem to cause much effect at low energies. 
\end{abstract}
\vspace{.2in}
\vspace{.3in}
\hspace{0.7cm} September 2014

\newpage

{
\setlength{\parskip}{0.05in}
\renewcommand{\baselinestretch}{0.7}\normalsize
\tableofcontents
\renewcommand{\baselinestretch}{1.0}\normalsize
}
\newpage

\setlength{\parskip}{0.1in}

\section{Introduction}
\label{sec:intro}
\small
A renormalization group (RG) flow is fully specified by the UV fixed point where it starts, and by the perturbing relevant operator which gives the initial direction. Once this UV data is given, everything else about the flow must follow, including the IR properties, and should be computable. Is there a general algorithm suitable for performing such a computation? We have in mind in particular strongly coupled situations, when perturbation theory is not applicable. 

For strongly coupled flows starting at a free theory, IR physics is accessible by Monte Carlo simulations on the lattice. What about flows which start at a strongly interacting conformal field theory (CFT)? One \emph{could} try to put this CFT on the lattice first, as a critical point of some lattice model, or as a fixed point of another RG flow starting at a higher scale from a free theory. However, this is baroque, and may not always be possible. Moreover, this dodges the question. Nonperturbatively, the UV CFT is specified by the conformal data (the spectrum of its local operators and the structure constants of the operator algebra). The relevant perturbation is specified by the operators by which we are perturbing and their relative coefficients. As a matter of principle, all IR physics must be computable in terms of only this nonperturbative UV data.

The purpose of this paper is to thrust into the limelight one method designed to solve this problem---the Truncated Conformal Space Approach (TCSA). While the method is familiar in the statistical mechanics and condensed matter community, so far it has not had much attention from high energy physicists. The field of potential applications of TCSA is huge and poorly explored. In particular, all prior work has been done in $d=2$ spacetime dimensions, and one of the goals of this paper will be to provide a generalization to $d>2$.

Although the original purpose of the TCSA was to study flows starting at an interacting CFT, the method is perfectly applicable when the UV CFT is free and therefore represents an alternative to the lattice for studying such flows. In TCSA, statistical errors are absent and systematic errors are very different from the lattice Monte Carlo. Including fermions (both Dirac and chiral) is straightforward; including gauge fields is more complicated but seems doable.

Time will tell if TCSA is a viable alternative to the lattice for flows in more than two dimensions. To get an initial feeling, in this paper we study the Landau-Ginzburg theory---the free scalar theory in $d>2$ dimensions perturbed by two relevant operators $\phi^2$ and $\phi^4$. Depending on the relative value of the quartic coupling and the mass, this theory flows in the IR to a massive theory with the $\bZ_2$-symmetry preserved or spontaneously broken. For a critical value of the coupling the theory flows to an interacting IR CFT in the universality class of the $d$-dimensional Ising model. Using TCSA, we are able to observe this phase structure, and to obtain reasonably precise predictions about the finite-volume spectrum of the theory, at weak, intermediate, or strong coupling. The required  computational resources turn out to be quite minor---the total cost of all the computations in this paper is $O(10)$ single-core days on a desktop. A bonus is that the theory for any $d$ (including fractional $d$) can be studied within the same unified framework. All in all, our initial experience with the TCSA has been quite positive, and justifies further explorations of the method.

The paper is organized as follows. We start in section \ref{sec:tcsa} with a general discussion of TCSA and of its connection to the Rayleigh-Ritz method in quantum mechanics. Section \ref{sec:free} is devoted to the free massless scalar in $d$ dimensions quantized on the sphere. We focus on the CFT description of the spectrum of theory, via radial quantization, but we also discuss the relation to more conventional canonical quantization. As a byproduct of this discussion, we show that the free massless scalar in a fractional number of dimensions is actually a non-unitary theory---its Hilbert space contains negative norm states. This curious fact does not seem to have been noticed before.

We continue in section \ref{sec:tcsa-eigs} with more details about TCSA for perturbations of the free massless scalar. In particular, we describe an OPE-based method which allows to efficiently compute the matrix entering the TCSA eigenvalue problem. This method should be useful beyond the Landau-Ginzburg flows studied in this paper.

Section \ref{sec:RG} is central to the paper---it explains how TCSA results depend on the cutoff energy of the Hilbert space, and how this dependence can be reduced by applying a renormalization procedure. The results of this section are crucial for improving the accuracy of TCSA.

In sections \ref{sec:phi2} and \ref{sec:phi4} we apply TCSA to study the Landau-Ginzburg flow. We first discuss in section \ref{sec:phi2} the case when only the mass parameter is nonzero, i.e.~we study the free massive theory as a perturbation of the free massless scalar by the mass term. This trivial theory allows us to test TCSA and the renormalization procedure in a controlled situation where the exact answer is known. After this test, we proceed in section \ref{sec:phi4}
to study the general case when both the mass and the quartic are turned on. For reasons which will be explained below, we perform this study in $d=2.5$ dimensions. In spite of this exotic spacetime dimensionality, we find all the traits expected from the Landau-Ginzburg flow for $d=3$. In particular, depending on the values of the couplings, we observe both phases of the theory, separated by a continuous phase transition. We perform rough measurements of the mass spectrum of the theory in both phases, and of the critical exponents at the phase transition. We also comment about how the non-unitarity of the theory at fractional $d$ manifests itself via some high-energy eigenvalues acquiring imaginary parts.

We conclude in section \ref{sec:disc} with a discussion of future research directions. Appendices \ref{sec:2d} and \ref{sec:other} contain a brief review of prior work, on TCSA in $d=2$, and on other Hamiltonian truncation techniques in quantum field theory. Two more appendices contain derivations of auxiliary technical results.

\section{Truncated Conformal Space Approach}
\label{sec:tcsa}
We would like to study an RG flow obtained by perturbing a $d$-dimensional CFT by a relevant scalar operator $\calV$.\footnote{Generalization to several perturbing operators is straightforward.} To set up the calculation, we will first of all need an IR regulator. The most natural IR regulator for a CFT is to put the theory on the ``cylinder" $\bR\times S^{d-1}_R$, where the radius of the sphere $R$ serves as an IR scale. The map to the cylinder amounts to Weyl-transforming the metric of the flat Euclidean space. The CFT dilatation generator then maps to the 
Hamiltonian on the cylinder, and the CFT local operators $\calO_i$ of dimensions $\Delta_i$ map to
states $\ket{i}$ on the cylinder whose energies are given by
\beq
E_i= R^{-1}\Delta_i\,.
\eeq
In the theories we will be considering here, there will be a unique ground state corresponding to the unit operator, with energy zero.\footnote{\label{note:cas0}We will ignore the CFT Casimir energy density, nonzero in even dimensions. If needed, it's trivial to take it into account because it just shifts all eigenstates by $const/R$.}
The Hamiltonian of the perturbed theory on the cylinder is
\beq
H=H_{\rm CFT}+V,\quad V=g \int _{S^{d-1}_R} \calV(x)\,.
\label{eq:hfull}
\eeq
The key idea is to think about this Hamiltonian as an infinite matrix in the Hilbert space of unperturbed CFT states $\ket{i}$. The CFT Hamiltonian in this basis is diagonal and simply related to the CFT operator dimensions:\footnote{\label{note:notorth}We are assuming here that the states $\ket{i}$ form an orthonormal basis. In practical computations, a natural basis of operators gives rise to a basis of states which is not orthonormal. In this case $\delta_{ij}$ will have to be replaced by the Gram matrix, as discussed below.} 
\beq
\bra{i} H_{\rm CFT}\ket{j} = R^{-1} \Delta_j\,\delta_{ij}\,.
\eeq
The matrix elements of the perturbation on the other hand are related to the CFT three point functions and will provide an off-diagonal piece of the Hamiltonian. Schematically, we will have
\beq
\bra{i} V\ket{j} \propto R^{-1} (g R^{d-\Delta_\calV}) f_{\calO^\dagger_i\calV \calO_j},
\label{eq:iVj}
\eeq
where $f_{\calO^\dagger_i\calV \calO_j}$ is the coefficient of the CFT three point function of the unit-normalized local operators in flat space (it is thus $R$ independent). Precise prefactors and normalizations will be discussed below. Notice that the dependence on $R$ follows by dimensional analysis. The mass scale of the flow is given by
\beq
\Lambda_{\rm IR}\sim g^{\frac 1{d-\Delta_\calV}}\,.
\eeq
For $R\ll \Lambda_{\rm IR}^{-1}$ we are close to the UV regime, where $V$ is a small correction to $H_{\rm CFT}$ and perturbation theory is reliable. To probe the IR physics, we must take instead
\beq
R\gg \Lambda_{\rm IR}^{-1}\,.
\eeq
Here $V$ cannot be treated as a small perturbation, and the right thing to do would be to diagonalize the whole Hamiltonian $H_{\rm CFT} + V$. But how can we do this given that this matrix is infinite?

The trick is to introduce a UV cutoff $\Lambda_{\rm UV}$ and to \emph{truncate} the Hilbert space keeping only the states below this maximal energy: 
\beq
E_i \le \Lambda_{\rm UV}\,.
\eeq
If the cutoff is chosen so that
\beq
\LUV\gg \LIR\,,
\eeq
we can hope that the IR physics is not much affected. Let us furthermore assume that the UV CFT has a discrete spectrum, which will be true for most CFTs of interest. In this case, the truncated Hilbert space is finite-dimensional, $H$ is a finite matrix and can be numerically diagonalized. One then learns a wealth of information about the theory in the IR. For example:
\begin{itemize}
\item
the ground state dependence on $R$ gives the vacuum energy density;
\item
by looking at the number of exponentially degenerate ground states we can infer the symmetry breaking pattern;
\item
the excited states give the massive spectrum of the theory, including one-particle, many-particle, and bound states;
\item
studying the dependence of two-particle states energies on $R$ we can extract the $S$-matrix;
\item
for flows ending in conformal fixed points we can extract the spectrum of IR operator dimensions.
\end{itemize}
This, then, is the essence of the TCSA, first proposed in 1991 by Yurov and Al.~Zamolodchikov \cite{Yurov:1989yu}. 
In practice, one tries to take $\LUV$ as high as possible, but this will be limited by the rapid growth of the number of states with energy. The success of the method depends on whether we can get reasonable results with numerically tractable Hilbert space sizes. 

Both the original paper \cite{Yurov:1989yu} and all the subsequent TCSA literature known to us consider $d=2$, but here we presented directly the general $d$ case because the basic logic is very similar. Our focus in this paper will be on $d>2$. The current status of the TCSA research in $d=2$ is summarized in Appendix \ref{sec:2d}, to which the reader will be referred from time to time.

\subsection{A Digression: Truncation in Quantum Mechanics}
\label{sec:digr}
{\footnotesize
When one first hears about the TCSA, the usual reaction is incredulity. How can such a naive method solve such a hard problem? To demystify it a bit, recall that very similar techniques are routinely used in quantum mechanics under the name of the Rayleigh-Ritz method. For example, consider the problem of finding the spectrum of the anharmonic oscillator:
\beq
H=H_0+ \lambda\, q^4\,,\qquad H_0=\frac 12 p^2 + \frac {\omega^2}{2} q^2\,.
\eeq
Let us express this Hamiltonian as an infinite matrix in the Hilbert space spanned by the eigenstates $\ket{n}$ of the harmonic part $H_0$. Then truncate the basis by keeping only the states $\ket{n}$ below some cutoff, $n\le N$. The claim is that in the limit $N\to\infty$ the eigenvalues of $H$ diagonalized within the truncated Hilbert space tend to the exact anharmonic oscillator energy levels,
as obtained e.g.~by solving the Schr\"odinger equation. This exercise is extremely easy to carry out (we recommend it to the reader!), since the matrix elements of $q^4\propto (a+a^\dagger)^4$ are known in closed form. One also finds that in this example the convergence is exponentially fast.\footnote{This is not a generic feature of the method. It is related to the fact that $q^4$ raises/lowers energy by at most a finite amount (four units). In more complicated quantum mechanical examples, and in quantum field theory, perturbation matrix elements will have power-like tails at high energy. In this case the convergence rate will be power-like in the cutoff, as we discuss below.} The method works equally well for both small and large $\lambda$.\footnote{For large $\lambda$, exponentially fast convergence sets in once we include all harmonic energy levels below the anharmonic one we are trying to reproduce.} As is well known, perturbation theory would be divergent for the anharmonic oscillator problem for any value of the quartic, and requires Borel resummation. However, the Rayleigh-Ritz method is completely immune to this. In fact, Rayleigh-Ritz is probably \emph{the} most practical method to find energy levels of the anharmonic oscillator. Double well potentials can also be considered, by including a negative frequency term to the perturbation:
\beq
H_{\text{double well}}=H_0  - \frac {\omega'^2}{2} q^2 + \lambda\, q^4\,,\quad  \omega'^2>\omega^2.
\eeq

Rayleigh-Ritz is used everywhere in quantum mechanics where perturbation theory is insufficient. As a less trivial example, consider the helium atom Hamiltonian:
\beq
H=H_{1}+H_{2}+V_{12},
\eeq
where $H_{1}$ and $H_{2}$ describe the electrons in the Coulomb field of the nucleus, and $V_{12}$ is the electron-electron interaction. Here one could e.g.~work in the Hilbert space spanned by the tensor products of the $H_1$ and $H_2$ eigenstates, compute the matrix elements of $V_{12}$, truncate and diagonalize. In practice, it helps to choose a different basis, which does not diagonalize $H_1+H_2$, but takes into account that the two-electron wavefunction will be singular at the coincident points (this improves the convergence rate). This is how the helium energy levels were obtained with a precision adequate to test QED quantum corrections \cite{Pekeris}.  

In fact, Rayleigh-Ritz is nothing but a version of the time-honored variational method. The variational method is usually used for the ground state, and one is often content to get a reasonable estimate for its energy. With Rayleigh-Ritz, one usually aims at a \emph{precision} determination, and gets the excited states as well as the ground state (convergence being fastest for the low-lying states). The variational method approximates the ground state energy \emph{from above}. By the minimax principle, the same is true for all the truncated Rayleigh-Ritz eigenvalues:
\beq
E_i^{(N)}\searrow E^{(\infty)}_i\qquad(N\to\infty)\,.
\eeq

In quantum mechanics, truncation methods can often be put on solid ground by establishing rigorous results about its convergence. Simplest results of this type can be found e.g.~in \cite{reedIV}, section XIII.2, and many more are scattered through the literature. As in the helium example above, the convergence rate is influenced by how well singularities of the exact wavefunction in the coordinate representation are reproduced by the functions of the trial basis. For quantum field theory we are not aware of general mathematical results about the convergence of truncation methods, but there is a large body of evidence that TCSA does converge. The evidence in $d=2$ is summarized in Appendix \ref{sec:2d}, and examples in $d>2$ will be studied in this paper.
}

\subsection{A Case Study for TCSA in $d$ Dimensions}
\label{sec:case}
In this paper we will apply TCSA to study the Landau-Ginzburg theory, i.e.~the free massless scalar theory perturbed by a linear combination of $\NO{\phi^2}$ and $\NO{\phi^4}$ operators. This is perhaps the simplest $d$-dimensional flow.
A priori, we are interested in $2\le d< 4$. However, in this first work we will have to stay away from the extremes of this range, since the TCSA analysis becomes complicated near these extremes.

The reason why $d$ close to 2 is hard for TCSA is that the scalar dimension $\Delta_\phi=(d-2)/2$ approaches zero in this limit, and the free scalar spectrum becomes dense and eventually continuous in $d=2$.\footnote{\label{note:Mussardo}One way to work around this problem is to compactify the scalar on a circle of large radius, considering a periodic approximation to the Landau-Ginzburg potential \cite{Mussardo}. {\bf Note added:} The paper \cite{Coser:2014lla} discussing such an approach was submitted to the arXiv the same day as our work.} To have a sufficiently sparse spectrum, we will keep $d$ not too close to 2.

The reason why $d$ close to $4$ is hard is that $\phi^4$ becomes marginal in this limit. On the contrary, as we will see, TCSA works best for strongly relevant perturbing operators $\calV$. The more relevant the operator is, the better-behaved perturbation problem is in the UV. The best situation is realized when
\beq
\Delta_\calV<d/2\,,
\label{eq:uvfin}
\eeq
which for the perturbations considered here means
\beq
\label{eq:uvfin1}
d<8/3\quad(\calV=\NO{\phi^4}),\qquad d<4\quad(\calV=\NO{\phi^2})\,.
\eeq
When \reef{eq:uvfin} is satisfied, the perturbation is simply UV-finite. At $\Delta_\calV = d/2$ the vacuum energy becomes divergent, as can be seen at second order in perturbation theory. Other UV divergences appear if we further increase $\Delta_\calV$, and these also affect the couplings of nontrivial local operators (including $\calV$ itself). These short-distance divergences have to be handled in the usual QFT way -- by adding counterterms.
In this first work we would like to avoid dealing with UV divergences, so we will stay within the bounds \reef{eq:uvfin1}. This does not mean however that we will altogether ignore cutoff dependence. Even in the range \reef{eq:uvfin1} when there are no UV divergences, the accuracy of the method for a finite cutoff will be influenced by power-suppressed corrections. This important issue will be discussed below.

As the reader must have noticed, we are considering the case of fractional $d$ on equal footing with the physically interesting integer $d$.
We will see that the TCSA problem allows a natural continuation to general $d$.

\section{Free Scalar in $d$ Dimensions}
\label{sec:free}
In this section we will discuss the UV CFT at which our RG flows will be starting---the free massless scalar CFT in $d$ dimensions. These results presented here lay the groundwork for the numerical investigations and for the renormalization, studied in the subsequent sections.

\subsection{Scalar Operators}
\label{sec:free-scalar}
The local operators of the free boson theory are built by taking products of the fundamental field $\phi$ 
and of its derivatives, e.g.
\beq
\NO{\,\phi^2 \del \phi\, \del \del \phi\, \del\del\del\phi\,}\,,
\label{eq:optyp}
\eeq
where some or all of the vector indices on the derivatives may be contracted. The operators are all inserted at the same point, and the normal-ordering sign means as usual that we don't consider Wick contractions within the operator when computing its correlation functions with other operators. 

 We can classify the operators according to their spin, i.e.~their representation under $SO(d)$. When we put the theory on the cylinder, the spin of an operator becomes the spin of the state into which it maps under the state-operator correspondence. Eventually we will perturb the theory by adding to the Hamiltonian an integral of a scalar operator over the sphere, as in Eq.~\reef{eq:hfull}. Since this perturbation preserves rotation symmetry of the sphere, the Hamiltonian matrix will split into blocks corresponding to the states of the same spin. The scalar sector contains most of the states we are interested in: the ground state, one-particle states at rest, and two-particle states in the center-of-mass frame. In the large $R$ limit, many of the states of higher spin will correspond to spin 0 states slightly boosted along the sphere.\footnote{And it may be interesting to check that their energies relate to the scalar energies in agreement with the $d$-dimensional Lorentz invariance which should emerge in the large $R$ limit.} In principle, there could also exist additional states with intrinsic spin, which could be thought of as bound states of fundamental scalars at strong coupling. This would be analogous to vector mesons in gauge theories with matter. In this paper we however focus exclusively on the scalar sector.
 
\begin{figure}[htbp]
\begin{center}
\includegraphics[scale=2]{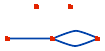}
\end{center}
\caption{The graph corresponding to the only scalar operator which can be obtained by contracting indices in \reef{eq:optyp}.}
\label{fig:exop1}
\end{figure}

Operators in the scalar sector will look like \reef{eq:optyp} with all the indices contracted. In \reef{eq:optyp} there is only one way to contract the indices to get a scalar operator, but in general there may be several inequivalent ways. It is useful to encode a scalar operator by a graph where $\phi$ with $n$ derivatives corresponds to a vertex with $n$ edges sticking out of it, and contracting two derivatives means joining two vertices with an edge. Notice that we can ignore operators containing contractions of derivatives acting on the same $\phi$, since $\del^2\phi=0$ by the equation of motion. On the other hand, two vertices may be connected by more than one edge, i.e.~our graphs can have parallel edges. In graph theory, graphs obeying these conditions are called \emph{multigraphs without loops}.\footnote{We can also replace $n$ parallel edges by a single edge with $n$ as a label. Then our graphs become simple edge-colored graphs.} Depending on how derivatives are contracted, the graphs may have one or several connected components. In particular, each $\phi$ without derivatives will give an isolated vertex, see figure \ref{fig:exop1}. It's also clear that isomorphic graphs correspond to identical operators, and should not be counted separately.

\subsection{Parity-Odd and Null States}
\label{sec:null}

The above discussion was incomplete in two ways. First, we could also consider contractions involving the $\eps$-tensor, which correspond to the parity-odd operators, as opposed to the parity-even operators discussed above. These two classes of operators won't mix under the parity-preserving $\phi^2$ and $\phi^4$ perturbations we will consider. In this paper we focus on the $P=+1$ sector. Notice that the $P=-1$ scalar operators have pretty high dimensions, e.g.~the lowest dimension one in $d=3$ is
\beq
\eps^{123} \delta^{45}\delta^{67} \phi_{,1} \phi_{,24} \phi_{,356} \phi_{,7}\,. 
\eeq
Consequently, the $P=-1$ states in the IR are also likely to be heavier than for $P=+1$.

Second, it's important to realize that for integer $d$ different graphs sometimes give rise to the same operator. The simplest example in $d=3$ is 
\beq
\raisebox{-12pt}{\includegraphics[scale=2]{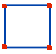}}\equiv
\frac 12 \raisebox{-12pt}{\includegraphics[scale=2]{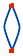}
\includegraphics[scale=2]{im/graph3.pdf}}\qquad (d=3)\,.
\eeq
This can also be also written as
\beq 
\label{eq:tracerel} 
\calO=\text{tr} \, M^4 - \half (\text{tr} \,M^2)^2 \equiv 0 \qquad (d=3)\,,
\eeq
where we consider $M_{\mu\nu}=\phi_{,\mu\nu}$ as a symmetric $3\times3$ matrix, traceless by the equations of motion. Then \reef{eq:tracerel} is just a statement about symmetric polynomials built out of its eigenvalues.

As one goes higher in energy, one encounters infinitely many relations of this type. For example, for $d=3$ all graphs $\text{tr} \, M^n$, $n\ge 4$, are expressible via linear combinations of the products of $\text{tr} \,M^2$ and $\text{tr} \,M^3$. There are also many other similar relations. One then has two alternative ways to proceed. Either one may decide to find all such relations below the UV cutoff one is working at, and explicitly eliminate all graphs which are not independent. Alternatively, one may decide not to eliminate anything, and work in an extended Hilbert space containing one state per graph. In this Hilbert space, Eq.~\reef{eq:tracerel} would be interpreted as a \emph{null state} condition, i.e.~it means that a certain linear combination of states has zero overlap with any other state (and in particular zero norm). 

In this paper we will follow the second approach, because we would like to treat integer and fractional $d$ within the same formalism. However, null state relations only exist for integer and finite $d$. At fractional $d$, all non-isomorphic graphs give rise to inequivalent states. In future work focussing on integer $d$, it will probably make sense to eliminate the null states, in order to reduce the dimension of the Hilbert space and speed up the subsequent matrix diagonalization.

\subsection{Non-Unitarity at Fractional $d$}
\label{sec:nonun}

The above discussion of null states raises an interesting question---what is the precise fate of the null states when one passes from integer $d$ to a nearby fractional $d$? As we mentioned, these states are then no longer null, but are they positive- or negative-norm? We claim that some of these states acquire a negative norm. 

To see a concrete example, let us take the operator $\calO$ in the LHS of Eq.~(\ref{eq:tracerel}). By an explicit computation, its two point function for a general $d$ is given by
\beq
\langle \calO(x) \calO(0) \rangle =C (d-3)(d-2)^5(d-1)^2 d^5(d+1)(d+2)(3d+8)|x|^{-2\Delta_\calO},\qquad C>0\,.
\eeq
Notice that the zero at $d=3$ is first order. For $2<d<3$ the two point function of $\calO$ is negative, and so $\calO$ must have an overlap with a negative-norm state. This example can be easily generalized to show that there are negative norm states for any fractional $d$, in fact infinitely many of them.\footnote{This argument is reminiscent of how Ref.~\cite{Maldacena:2011jn} showed that analytic continuations of $O(n)$ models to fractional $n$ contain negative-norm states.} 

The presence of negative-norm states means that the \emph{free scalar theory in fractional $d$ is not unitary}. To our knowledge, this observation has not been made before, although theories in fractional dimensions have been extensively studied, especially in relation to critical phenomena, where they form the basis of the $\eps$-expansion. As we will see below in section \ref{sec:phi4-complex}, the lack of unitarity will lead to the presence of complex energy eigenvalues, once the free theory is perturbed by the quartic coupling.
However, the mass term alone leaves all energy levels real---see section \ref{sec:can}.

It has to be said that the first negative norm state has a pretty high dimension:
\beq
\Delta_{\rm neg}=
\begin{cases}8+4\Delta_\phi&(2<d<3)\,,\\
 10+5\Delta_\phi&(3<d<4)\,.
 \end{cases}
\eeq
This must be the reason why they have not been noticed until now. A few negative norm states at high dimensions, hidden among lots of positive-norm states of comparable dimensions, probably do not have a strong effect on the low-energy physics. In a recent conformal bootstrap study of the Wilson-Fisher fixed point in fractional dimensions \cite{El-Showk:2013nia} it was assumed that these theories were unitary, and very reasonable results were obtained.\footnote{{\bf Note added:} Another evidence for the mildness of the unitarity violation is provided by very recent calculation~\cite{Giombi:2014xxa} of the free energy $F$ of the free scalar and the Wilson-Fischer fixed point on $S^d$ for non-integer $d$. It was found that $F_d$ changes monotonically along the flow, just as for unitary theories in integer dimensions.}

\subsection{Primaries and Descendants}
\label{sec:primaries}
 
CFT local operators can be divided into primaries and descendants. For $d=2$ descendants are obtained by acting on the primaries with the raising part of the Virasoro algebra. For general $d$ considered here, descendants are simply derivatives of the primaries.

The basis for the Hilbert space on the cylinder includes of course all states, those corresponding to primaries and to descendants. If the UV CFT is strongly coupled, the matrix elements~\reef{eq:iVj} between primary states are part of the non-perturbative conformal data, while matrix elements involving descendant states can be computed using the conformal algebra. This situation often occurs in 2d TCSA applications (see Appendix \ref{sec:2d}).

On the other hand, in our case the UV CFT is free and we will follow an alternative approach. Instead of relating the descendant matrix elements to those of the primaries, we will simply evaluate all the necessary matrix elements using the fact that our operators are built out of the fundamental scalar field. This procedure is much faster, because the additional work required to classify the states into primaries and descendants is significant. In particular, many scalar states will be descendants of primaries with spin, whereas our approach completely avoids introducing states with spin.

Let us finally remark that the current mathematical understanding of the underlying algebraic structure for CFTs in fractional $d$ appears to be rather incomplete. This is particularly relevant for the primary/descendant classification of local operators, which would require defining $\mathfrak{so}(d)$ algebras and their representations in fractional $d$.\footnote{An analytic continuations of $\mathfrak{sl}(d)$ algebras has been constructed in \cite{feigin1988}.} Fortunately we need not be concerned with this in practice. All computations involving scalar operators, which have all indices contracted, can be done directly for fractional $d$ by setting $\Delta_\phi = (d-2)/2$ and $\delta_\mu{}^\mu = d$.

\subsection{Relation to Canonical Quantization}
\label{sec:can}
So far, our discussion of the free massless scalar on the cylinder has been based entirely on the radial quantization and the state-operator correspondence. Here we would like to comment on a more low-brow approach---canonical quantization. The free scalar in curved space is described by the action
\beq
\label{eq:freemassive}
S=\frac 12\int d^d x \sqrt{g} [g^{\mu\nu}\del_\mu\phi\del_\nu\phi-(m^2+\xi\,\Ric)\phi^2]\,.
\eeq
As is well known, the theory is Weyl-invariant for $m^2=0$ and the coupling to the Ricci scalar $\xi=\frac{d-2}{4(d-1)}$. We now quantize canonically on the cylinder metric
\beq
ds^2 = d\tau^2+ R^2 d\bn^2\,,\qquad \bn\in S^{d-1}\,.
\eeq
The field $\phi$ is expanded in the eigenfunctions of the Laplacian on the sphere of radius $R$ (spherical functions):
\beq
\phi_{l,n},\quad l=0,1,2\ldots,
\eeq
where $l$ is the angular momentum quantum number, and $n$ numbers the states in the multiplet.
The Laplacian eigenvalue is $l(l+d-2)/R^2$. To each of these modes we will associate a harmonic oscillator of frequency
\beq
\label{eq:omegal}
\omega_l=\sqrt{m^2+{l(l+d-2)}/{R^2}+\xi\,\Ric}\equiv \sqrt{m^2+(l+\nu)^2/R^2}\,,\qquad \nu=\Delta_\phi=(d-2)/2\,,
\eeq
where we used the fact that $\Ric=(d-1)(d-2)/R^2$ for a round $d-1$ dimensional sphere of radius $R$. The Hilbert space is the Fock space of these oscillators.

The free massless scalar is recovered setting $m\to0$. In this limit, $\omega_l$ reduces to $(l+\Delta_\phi)/R$, which is the right energy for the state corresponding in the radial quantization to the operator obtained by acting on $\phi$ by $l$ derivatives.

What are the relative disadvantages and merits of the canonical vs radial quantization description of the free scalar Hilbert space?
One definite merit of canonical quantization is that it allows to take the mass into account nonperturbatively from the start. On the contrary, in the radial quantization approach we have to perturb around $m^2=0$. In section \ref{sec:phi2} below we will do the exercise of reproducing the free massive scalar spectrum within this framework.
 
On the other hand, an advantage of perturbing around the conformal point $m^2=0$ is that the matrix elements of the perturbation have a simple overall power-law dependence on the radius $R$, see Eq.~\reef{eq:iVj}. So we don't have to recompute the matrix when we change the radius. On the contrary, in canonical quantization with a general mass, matrix elements will depend on $R$ nontrivially via the $1/\sqrt{\omega_l}$ normalization factors accompanying the oscillators. Thus the matrix will have to be computed separately for every $R$. 

In this paper, we will perturb around the conformal point and use exclusively TCSA (i.e.~radial quantization). In the future, it would be interesting to see if canonical quantization with $m^2\ne 0$ gives better results.
In upcoming papers \cite{Lorenzo1,Lorenzo2}, we will use canonical quantization to study the Landau-Ginzburg flows in $d=2$. As explained in section \ref{sec:case}, TCSA is not directly applicable to these flows (see however footnote \ref{note:Mussardo}).

\subsection{The Gram Matrix}
\label{sec:gram}

As made clear in the above discussion, we will be working in the Hilbert space of scalar states on the cylinder, in the basis which in the radial quantization can be identified with scalar operators acting on the vacuum:
\beq
\ket{i}\equiv \ket{\calO_i}=\calO_i(0)\ket{0}\,.
\eeq
As already mentioned in footnote \ref{note:notorth}, this basis in general will not be orthonormal and not even orthogonal. Rather, we will have a nontrivial Gram matrix. This Gram matrix is an essential ingredient in the existing implementations of the TCSA in $d=2$ (see Appendix \ref{sec:2d}). 

In our case, the Gram matrix will not play a crucial role. In fact, as we will see below, the perturbed spectrum computation can be organized without using the Gram matrix at all. Nevertheless, the Gram matrix is a conceptually important object, so we would like to discuss in some detail its definition and evaluation.

First we need a map from the states to their conjugates. As usual in radially quantized CFT, this map is defined with the help of the inversion transformation $R:x_\mu\to x_\mu/x^2$. The Gram matrix is then defined as
\beq
G_{ij}\equiv \langle\calO_i \ket{\calO_j}=\lim_{x\to0}\langle [\calO_i(x)]^\dagger \calO_j(x) \rangle\,, 
\eeq
where the conjugate operator $[\calO_i(x)]^\dagger$ is inserted at the point $Rx$. The rules for construction of the conjugate operators are as follows ($\phi$ is the fundamental scalar; $\calA,\calB$ any two fields in the theory):\begin{enumerate}
\item $[\phi(x)]^\dagger = |x|^{-2\Delta_\phi} \phi(Rx)$\,\quad(since $\phi$ is a primary),
\item $[\calA(x)_{,\mu}]^\dagger = \frac\del{\del x_{\mu}}[\calA(x)]^\dagger$\, \quad(since conjugation is antilinear),
\item $[\NO{\,\calA(x)\calB(x)\,}]^\dagger =\ \NO{[\calA(x)]^\dagger[\calB(x)]^\dagger}$\,.\footnote{To show this, start with
$
[\calA(x)\calB(y)]^\dagger =[\calA(x)]^\dagger[\calB(y)]^\dagger
$, where the operators are inserted at the same radial quantization ``time", so that no ordering issue arises. From here by induction in the number of fundamental fields we get
$
[\NO{\,\calA(x)\calB(y)\,}]^\dagger =\ \NO{[\calA(x)]^\dagger[\calB(y)]^\dagger}
$, and Rule 3 follows by taking the coincident point limit.}

\end{enumerate}
Starting from Rule 1 and using Rule 2 repeatedly we can conjugate all derivatives of $\phi$. Then by applying Rule 3 we can conjugate all normal-ordered products of derivatives, and in particular all scalar operators forming our basis.

Computation of the Gram matrix is thus reduced to evaluating two-point functions of operators made of several $\phi$'s acted upon by various derivatives. In principle, this is straightforward to do using Wick's theorem. The number of Wick contractions to perform can be dramatically reduced by using selection rules. To begin with, the only nonzero entries are those for which
(a) $\calO_i$ and $\calO_j$ contain equal number of $\phi$'s\,, and (b) $\Delta_i=\Delta_j$\,. These two rules are subsumed by the following much more powerful rule. Let $N_l(\calO)$ be the number of times the~$l^{\rm th}$ derivative $\del^l\phi$ occurs in the operator $\calO$ (irrespectively of how its indices are contracted). Then the Gram matrix entry $\langle\calO_i \ket{\calO_j}$ can be nonzero only if
\beq
\label{eq:selrule}
N_l(\calO_i)=N_l(\calO_j)\text{ for all }l=0,1,2\ldots
\eeq
This rule can be easily understood using the relation with canonical quantization described in the previous section. In this description, $N_l$ maps to the total occupation number of oscillators with angular momentum $l$. To get nonzero overlap, all angular momentum modes should have the same occupation number. 

Putting it all together, the Gram matrix is thus evaluated as follows. First one computes the overlaps
\beq
\langle\del^l_{\{\mu\}}\phi \ket{\del^l_{\{\nu\}}\phi}\,,
\label{eq:basic}
\eeq
by using the above prescription, or by using the conformal algebra, as explained e.g.~in \cite{Pappadopulo:2012jk}. These are particular invariant tensors, symmetric and traceless in both groups of indices $\{\mu\},\{\nu\}$. 
Overlaps between general scalar states are then computed by contracting the basic overlaps \reef{eq:basic} between their constituents.

Using this direct algorithm, we could compute the Gram matrix up to a rather high cutoff in operator dimension. However, we found it expensive to compute in this way the Gram matrix all the way up to the cutoffs we will be using in the numerical analyses below. An alternative, indirect, method for computing the Gram matrix will be described in section \ref{sec:ope}. That method is much faster and easily yields the Gram matrix up to the required values of the cutoff. In any case, as we will see below, the spectrum computations can be organized avoiding the use of the Gram matrix. In this paper, the Gram matrix has been used only in one instance---to count the number of negative norm states plotted in figure \ref{fig:states25d}. 

\section{TCSA Eigenvalue Problem}
\label{sec:tcsa-eigs}
\subsection{Simple versus Generalized Eigenvalue Problem}
\label{sec:tcsa-vs}

Let us formalize a bit more our problem. Energy levels on the cylinder are solutions of the eigenvalue problem
\beq
H\ket{\psi}=E\ket{\psi}\,.
\label{eq:eigabs}
\eeq
We will be looking for scalar eigenstates, expanding them in a basis of states $\ket{j}$:
\beq
\ket{\psi}=c^j\ket{j} 
\eeq
The states $\ket{j}$ will be in one-to one correspondence with the scalar local operators of the UV CFT (in this paper, the free massless scalar theory).
The Hamiltonian in this basis will be represented by a matrix: 
\beq
\label{eq:Hi}
H\ket{j}=H^i{}_j\ket{i}\,.
\eeq
In terms of this matrix, Eq.~\reef{eq:eigabs} becomes a simple eigenvalue problem
\beq
H^i{}_j\, c^j=E c^i\,.
\label{eq:eigsim}
\eeq
Notice that the matrix $H^i{}_j$ is not hermitean. To transform the problem to a hermitean form, we consider the matrix elements
\beq
H_{ij}=\bra{i}H\ket{j}\,.
\label{eq:Hij}
\eeq
We of course have
\beq
\label{eq:Hij1}
H_{ij}=G_{ik} H^{k}{}_{j},
\eeq
where $G_{ik}=\bra{i}k\rangle$ is the Gram matrix discussed above. The matrix $G_{ij}$ is hermitean, and $H_{ij}$ is hermitean if the Hamiltonian is hermitean, which is the case for the Landau-Ginzburg flows with real couplings considered here. Actually, for the operator bases considered in this work, these matrices will be real symmetric. We then have an equivalent symmetric generalized eigenvalue problem:
\beq
H_{ij}\, c^j= E\, G_{ij}\, c^j\,.
\label{eq:eiggen}
\eeq

In the existing $d=2$ TCSA implementations (see Appendix \ref{sec:2d}), one starts by computing the matrices $G_{ij}$ and $H_{ij}$, which naturally leads to the generalized eigenvalue problem \reef{eq:eiggen}. One then usually multiplies both sides by $G^{-1}$ and transform to \reef{eq:eigsim}.\footnote{Strictly speaking, this is not necessary, since numerical methods for solving generalized eigenvalue problems are readily available.}
Here, we will choose an alternative path. Namely, we will directly compute the matrix $H^i{}_j$ and find eigenvalues from \reef{eq:eigsim}. The method for computing $H^i{}_j$ is described below in section \ref{sec:ope}.

\subsection{Working in Presence of Null States}
\label{sec:tcsa-null}
As mentioned in section \ref{sec:null}, we will be working in a basis which, for integer $d$, will contain null states. In presence of null states the above discussion needs to be reconsidered. In particular, the Hamiltonian matrix
is then ambiguous since we can add an arbitrary null state to the RHS in \reef{eq:Hi}:
\beq
H\ket{i}\to H\ket{i} +\ket{\rm null}\,.
\eeq 
Also, the eigenvalue problem \reef{eq:eigabs} has to be considered modulo addition of an arbitrary null state in the RHS.
In practice, however, we won't have to deal with these subtleties. We will compute the Hamiltonian matrix as if there were no null states,\footnote{We perform this computation in {\tt Mathematica} keeping $d$ as a free parameter, and set $d$ to the desired value before the diagonalization.} and solve the original eigenvalue problem \reef{eq:eigabs}. Our final spectrum for integer $d$ will thus contain both physical and null state eigenvalues. It's easy to see that the physical state eigenvalues are the same as in the more rigorous treatment.\footnote{A key to this argument is that null states can only be mapped into null states by the Hamiltonian. In principle, the fact that we don't solve \reef{eq:Hi} modulo the appearance of a null state could lead to some physical eigenvectors disappearing, due to the Jordan block phenomenon. However, this is very non-generic and would be easily detectable as we are varying parameters such as couplings and the radius of the cylinder. We have never observed it happen.} The null eigenvalues are unphysical---they have to be separated and thrown out. There are many ways to do this in practice: one can follow a null eigenvalue from the UV where its value is known; one can detect it by the presence of crossings with physical states (physical eigenvalues don't cross in RG flows which are not integrable); one can check the nullness of the corresponding eigenvector. For the low-lying spectrum this issue does not even arise, since the first null state has a pretty high dimension.

\subsection{Matrix Element Evaluation: OPE Method}
\label{sec:ope}
The CFT piece of the Hamiltonian matrix is diagonal:
\beq
\label{eq:hcft}
(\HCFT)^{i}{}_j = R^{-1}\Delta_j \delta^i{}_j\,.
\eeq
The nontrivial part is to compute the matrix of the perturbation. We compute this by using the following \emph{OPE method}. Namely, in the radial quantization we are supposed to compute
\beq
\left(\int_{|x|=1} \calV(x)\right) \calO_j(0)\,,
\label{eq:OOj}
\eeq
where $\calV$ is the perturbing operator (in our examples it will be $\NO{\phi^2}$ or $\NO{\phi^4}$), and $\calO_j$ is the operator corresponding to the state $\ket{j}$. Consider the OPE
\beq
\calV(x)\calO_j(0) = \sum_k C_{k,{\{\mu\}}}(x) \calA^{\{\mu\}}_k(0)\,
\eeq
where $A_k(0)$ are local operators inserted at the origin, and $C_k(x)$ are $c$-number coefficient functions. Since we are in a free theory, this OPE can be worked out explicitly. Notice that while $\calV$ and $\calO_j$ will be scalars, many of the operators $\calA_k$ will be tensors, and $\{\mu\}$ stands collectively for their indices, contracted with those of $C_k$. Now to evaluate \reef{eq:OOj} we just integrate the OPE term by term, which amounts to integrating the coefficients:
\beq
\sum_k \left(\int_{|x|=1} C_{k,{\{\mu\}}}(x)\right) \calA^{\{\mu\}}_k(0)\,.
\label{eq:CAk}
\eeq
By rotation invariance, the integrals will produce invariant tensors, i.e.~a number of Kronecker deltas connecting the indices in $\{\mu\}$. Contracting these with the indices of $\calA^{\{\mu\}}_k$ will give scalar operators. Expressing the RHS of \reef{eq:CAk} in the original basis, we read off the matrix $V^{i}{}_{j}$. In the above discussion we were effectively setting $R$ and $g$ to unity. To restore the dependence in these parameters, we need to multiply the resulting matrix by $R^{-1}(g R^{d-\Delta_{\calV}})$.

This, then, is how we compute the matrix entering the eigenvalue problem. Notice that this approach is more economical (involves fewer Wick contractions) than the direct computation of the three-point functions $\bra{i}H\ket{j}$.

The matrices $V^i{}_j$ computed by the OPE method can be subjected to a check. We know that if we multiply them by the Gram matrix as in \reef{eq:Hij1}, the resulting matrix $V_{ij}$ must be symmetric. As mentioned at the end of section \ref{sec:gram}, we can compute the Gram matrix directly up to a rather high cutoff. Up to this cutoff, we can then check the symmetry of $V_{ij}$ for the $\phi^2$ and $\phi^4$ perturbations---this check works. 

For still higher cutoff, we found it expensive to compute the Gram matrix directly. However, we can ask the following question: given our matrices $V^i{}_j$, is there a symmetric matrix $G_{ij}$, subject to the selection rule \reef{eq:selrule}, and with the property that $V_{ij} = G_{ik} V^k{}_j$ is symmetric, for both the $\phi^2$ and the $\phi^4$ perturbations? It turns out that up to the highest cutoffs explored in this work, such a matrix always exists and, moreover, is unique! This provides an alternative, indirect, method to compute the Gram matrix. It is by this method that we computed the Gram matrix used to count the negative norm states in figure \ref{fig:states25d}.

\section[Cutoff Dependence and Renormalization]{Cutoff Dependence and Renormalization\footnote{The reader may wish to skip this section and come back to it while studying sections \ref{sec:RG-phi2} and \ref{sec:RG-phi4} below.}}
\label{sec:RG}

\subsection{General Remarks}
\label{sec:RG-gen}
In the following sections we will see in concrete examples that TCSA converges as $\LUV$ is increased. We will also see that the rate of convergence is power-like. We would like to examine here why the method converges, and how its convergence rate can be improved. We thus have to understand the effect of removing the high energy states from the Hilbert space on the low energy spectrum. Effective field theory intuition tells us that this effect should be small, and can be corrected for. 

The basic equations are as follows. We work in the Hilbert space of the unperturbed CFT on the cylinder, which is divided into the low ($l$) and high ($h$) energy parts:
\beq
\mathcal{H}=
\mathcal{H}_l \oplus \mathcal{H}_h\,,
\eeq
where $\mathcal{H}_l$ includes all states of energy up to $\LUV$.
The full Hamiltonian is a block matrix:
\beq
H=\left(\begin{array}{cc} H_{ll}& H_{hl}\\
H_{lh} & H_{hh}
\end{array}
\right)\,.
\eeq
where $H_{ab}$ maps $\mathcal{H}_b$ into $\mathcal{H}_a$. The TCSA truncated Hamiltonian is the upper left corner: $H_{ll}=H_{\rm TCSA}$. 

The full eigenvalue problem is
\beq
H.c=E c, \quad c=(c_l,c_h)^t\,,
\label{eq:fullSchr}\eeq
or, in components,
\beq
H_{ll}.c_l+H_{lh}.c_h=E c_l\,,\qquad H_{hl}.c_l+H_{hh}.c_h=E c_h\,.
\eeq
Let us now eliminate $c_h$ by using the second equation. We get:
\beq
\label{eq:ex}
\left(H_{ll}- H_{lh}.(H_{hh}-E)^{-1}.H_{hl}\right).c_l = E c_l\,,
\eeq 
This exact equation should be compared to the truncated equation used in TCSA:
\beq
\label{eq:h00}
H_{ll}.{\bar c}_{l}= {\bar E} {\bar c}_{l}\qquad(\rm TCSA)\,.
\eeq
Here, we write ${\bar E}, {\bar c}$ rather than $E,c$ to indicate that these are solutions to the \emph{truncated} equation rather than Eq.~\reef{eq:fullSchr}.

Conclusion: TCSA will converge if the matrix correction in \reef{eq:ex} can be neglected in the limit $\LUV\to\infty$. Naively, this seems likely since it is suppressed by $H_{hh}-E$, and we are assuming that $E$ belongs to the low-energy spectrum, while the eigenvalues of $H_{hh}$ will be presumably large. However, the precise statement will depend also on the size of the matrix elements mixing $\mathcal{H}_h$ into $\mathcal{H}_l$. This mixing being due to the perturbation, we can expect that the importance of corrections will depend on $\Delta_{\calV}$.

Let us view the problem from a practical angle. Suppose we know an eigenvalue $\bar E$ and the corresponding eigenvector $\bar c$ of the truncated problem \reef{eq:h00}. How can we correct $\bar E$ to get closer to the solution of the exact eigenvalue equation? Let us write the full Hamiltonian as
\beq
H=H_0+H_1,\quad 
H_0=
\left(\begin{array}{cc} 
H_{\rm TCSA}& 0\\
0 & H_{\text{CFT},h}
\end{array}
\right)\,,
\quad
H_1=\left(\begin{array}{cc} 0& V_{lh}\\
V_{hl} & V_{hh}
\end{array}
\right)\,.
\eeq
We took into account that the off-diagonal elements $H_{hl}$ and $H_{lh}$ are associated only with the perturbation $V$. The eigenvalues of $H_0$ are known---these are the TCSA eigenvalues and the unperturbed eigenvalues of the diagonal $H_{\text{CFT},h}$. We will now view $H_1$ as a perturbation and compute corrections to the TCSA eigenvalues. By the usual Rayleigh-Schr\"odinger perturbation theory we get:
\begin{align}
E&=\bar E+\bra{\bar c} \Delta H \ket{\bar c}\label{eq:ourcorr0}\,,\\
\Delta H  &= -V_{lh}.(H_{\text{CFT}}-\bar E)^{-1}.V_{hl}+\ldots\,\label{eq:ourcorr}
\end{align}
Further corrections terms are simple to write down. For example, the next one is given by 
\beq
\label{note:om}
V_{lh}.(H_{\text{CFT}}-\bar E)^{-1}.V_{hh}.(H_{\text{CFT}}-\bar E)^{-1}.V_{hl}\,.
\eeq 
We will only use the term shown in \reef{eq:ourcorr} in this paper, but in the future increasing the accuracy of the renormalization procedure will likely require mastering \reef{note:om} and perhaps even further terms. 

Our job is not yet finished, since evaluating the correction term \reef{eq:ourcorr} requires an infinite summation over the states in $\mathcal{H}_h$. It would be desirable to find a simplified approximate form for this correction:
\beq
\Delta H \approx \sum_c V_c\,,
\eeq
where $V_c$ act simply on $\mathcal{H}_l$. For example, $V_c$ might be of the same form as $V$ itself, i.e.~an integral of a local operator $\calV_c$ over the sphere. Then adding $\Delta H$ to the TCSA Hamiltonian can be thought of as renormalizing the couplings. In the language of usual perturbative quantum field theory, the $V_c$ might be called \emph{counterterms}. The difference is that in perturbation theory, we usually worry only about the counterterms which diverge when the UV cutoff is taken to infinity. Here we care also about the correction terms which are power-suppressed---we will want to add them in order to improve the accuracy of the method. The Hamiltonian with added correction terms can be called an \emph{improved TCSA Hamiltonian}. This is analogous to ``improved actions" in Lattice QCD.

\subsection{Computation of $\Delta H$}
\label{sec:RG-count}

To find the correction terms, we examine the matrix element of $\Delta H$ between two states $i,j\in \mathcal{H}_l$:
\beq
\label{eq:cc1}
(\Delta H)^i{}_j=- \sum_{E_n >\LUV}
\frac
{(M_n)^i{}_j}
{E_n-\bar E} \,,\qquad (M_n)^i{}_{j}
\equiv\sum_{k:\Delta_k=\Delta_n} V^{i}{}_k V^k{}_j \,,
\eeq
where $E_n=\Delta_n/R$ stands for the unperturbed CFT energy. We will estimate the large energy asymptotics of $M_n$. The key idea is to consider the correlation function:
\beq
C(\tau)=\bra{i}V(\half \tau)V(-\half \tau)\ket{j} =g_a g_b \bra{i}\int_{S^{d-1}}d\bn\, \calV_a (\bn, \half\tau) \int _{S^{d-1}} d\bn' \calV_b (\bn', -\half \tau)\ket{j}\,.
\label{eq:ctau}
\eeq
Inserting the resolution of unity, this correlation function can be represented through the same $M_n$ as:
\beq
C(\tau)= \sum_{n}(M_n)_{ij} e^{-[\Delta_n-(\Delta_i+\Delta_j)/2] \tau}\,.
\eeq
The large energy behavior of $M_n$ can then be extracted from the part of $C(\tau)$ which is non-analytic as $\tau\to0$, since the low energy states give rise to an analytic contribution. 

A moment's thought shows that nonanalyticity for $\tau\to0$ can appear only from the region where the nonintegrated correlator has a singularity, i.e.~from $\bn$ close to $\bn'$. In this region we can use the OPE
\beq
\calV_a(x) \calV_b(y)\approx \sum_c f_{abc} \frac{\calV_c(\half(x+y))}{|x-y|^{h}},\qquad h=h_{abc}=\Delta_a+\Delta_b-\Delta_c\,.
\label{eq:leadOPE}
\eeq
To the accuracy needed below, it will be sufficient to use only the shown leading term in the OPE. Moreover, we will be considering only scalars in the RHS of the OPE. With a Poincar\'{e}-invariant cutoff, non-scalar operators are not induced in the renormalization group flow. However, the TCSA regulator is more subtle. We break the Poincar\'{e} group to $SO(d)$ times dilatations. Furthermore, since we are working in a Hamiltonian formalism, we may find integrals of tensorial operators induced by the RG flow. As an example, the appearance of the stress tensor $T_{\mu \nu}$ on the RHS gives (after integrating over the sphere) a contribution of the form
\beq
\int_{S^{d-1}} n^\mu n^\nu T_{\mu \nu} \propto \HCFT
\eeq
so it leads to a renormalization of the coefficient of $\HCFT$ in the TCSA Hamiltonian.\footnote{This term is the analogue of wave function renormalization in ordinary perturbation theory.} However, since the stress tensor and other operators with spin have high dimension, their effects will be suppressed compared to the effects of the scalars by a higher power of $\LUV$.

Each term in the OPE will give rise to a term in the $\tau\to0$ asymptotics of the correlator. The prefactor will be given by the matrix element of $\calV_c$ integrated over the sphere, while the dependence on $\tau$ will come from
the integral of the OPE kernel. Up to $O(\tau^2)$ accuracy we have (see Appendix \ref{sec:tausq}):
\begin{gather}
C(\tau)\supset  B(h) \Gamma(h-d+1) \tau^{d-h-1}[1+O(\tau^2)]\times g_a g_b f_{abc} \bra{i}\int _{S^{d-1}}\calV_c(x)\ket{j}\,,\nn\\
B(h) = 
\frac{ 2^{d-h}\pi^{d/2}}
{\Gamma(h/2)\Gamma(h/2-\nu)}\,.
\label{eq:Bh}
\end{gather}
This non-analytic behavior can be reproduced provided that the large-dimension distribution of the coefficients $M_n$ contains a component with a power law:
\beq
\label{eq:pl}
[M(\Delta)]_{ij}\supset  \frac{B(h) }{[\Delta-\half(\Delta_i+\Delta_j)]^{d-h}} g_a g_b f_{abc} \bra{i}\int _{S^{d-1}}\calV_c(x)\ket{j} \,.
\eeq
It should be kept in mind that $M_n$ is a discrete sequence, and so the given continuous distribution is supposed to approximate it only on average. Below we will discuss the accuracy of this approximation in more detail. Also, for the renormalization of the $\phi^2$ flow we will work out the asymptotics of the sequence $M_n$ via an alternative method.

For the moment, to get an expression for $\Delta H$, we introduce the shown asymptotics into \reef{eq:cc1} and perform the sum approximating it by an integral. Gathering all the prefactors, reinstating the dependence on the coupling constant and on $R$, we obtain the following formula for the correction term:
\begin{gather}
V=\sum_a g_a \int_{S_R^{d-1}} \calV_a(x)\quad\Rightarrow\quad \Delta H \approx - \sum_{ab} g_a g_b K_{abc} \int_{S_R^{d-1}} \calV_c(x)\\
K_{abc}=f_{abc}\, B(h) \int_\LUV^\infty 
\frac {dt}{[t-\half (\Delta_i+\Delta_j)/R]^{d-h}(t-\bar E)},\quad h=h_{abc}\,.
\label{eq:master}
\end{gather}
For very large $\LUV$ we are allowed to drop the corrections due to $\bar E$ and $\Delta_{i}+\Delta_{j}$ in the denominator. However, below we will find it useful to keep track of these subleading corrections, at least approximately. 

\subsection{Renormalization Group Improvement}
\label{sec:RG-improv}

In the above discussion we were assuming that $\Delta H$ is very small, and correcting eigenvalues by the leading-order perturbation formula \reef{eq:ourcorr0} is adequate. For this, $\LUV$ has to be taken sufficiently large so that the renormalizations of all the couplings implied by \reef{eq:master} are small compared to their values in the bare TCSA Hamiltonian.
This condition is rather restrictive and in fact in our main example below---the Landau-Ginzburg flow---we will not be able
to satisfy it, as the mass renormalization due to the quartic will sometimes be comparable to the bare mass.

The way out in such a situation is to perform an RG improvement of the correction procedure. This is inspired by the usual RG in perturbative quantum field theory, which resums large logarithms. Here we don't have logarithms but power-suppressed terms, but the logic is very similar.

As usual in RG, we imagine performing a sequence of Hilbert space reductions with cutoffs $\Lambda_1>\Lambda_2>\ldots$  To first order in $\Delta H$, Eq.~\reef{eq:ourcorr0} is equivalent to diagonalizing $H+\Delta H$. So instead of using \reef{eq:ourcorr0} we will just keep adding $\Delta H$ to the original Hamiltonian, and iterate. We can imagine changing the cutoff in each step from $\Lambda$ to $\Lambda-\delta\Lambda$ for $\delta\Lambda\ll \Lambda$. Concretely, using the form of $\Delta H$ given in \eqref{eq:master} and assuming that the corrections due to $\Delta_i + \Delta_j$ and $\bar E$ can be ignored, this procedure results in RG equations of the form:
\beq
\label{eq:rgmaster}
\frac{\delta g_c (\Lambda)}{\delta \Lambda} = \sum_{ab} g_a (\Lambda) g_b (\Lambda) f_{abc}\, B(h) \frac {1}{\Lambda^{d-h+1}} \qquad \qquad (\Lambda \gg \bar E, (\Delta_i + \Delta_j)/R )\,.
\eeq
In this way we obtain a flow in the space of Hamiltonians, which we can integrate all the way down to the desired cutoff $\LUV$. It may be expected that, under certain circumstances, the final `resummed' Hamiltonian obtained by such a procedure will have a larger range of applicability (i.e.~work for smaller $\LUV$) than the first-order correction formula. This will be the case if the subleading on the right-hand side of \reef{eq:ourcorr}, such as \reef{note:om}, are less important than the terms we are proposing to resum. Again, we can draw analogy from the usual perturbative RG, when the beta function is the sum of one-loop, two-loop etc terms. Our RG improvement is like integrating the one-loop beta function while dropping all higher-loop terms. 

As already mentioned, in the examples considered below we will want to keep track of the corrections due to $(\Delta_i+\Delta_j)/R$ and $\bar E$ in \reef{eq:master}. These corrections are state-dependent, and taking them into account completely would require a separate RG flow for every value of these parameters---a complication that we wish to avoid. Instead, we would like to find a practical way to represent them by operators. The easiest way to do so is to expand in powers of the inverse cutoff and keep only the first-order terms. In that case we can replace $\Delta/R$ by $\HCFT$ and $\bar E$ by $H$. For example, in the case where $\calV_c=\unit$, the first of these two subleading corrections can be thought of as `wave function' renormalization of the coefficient of $\HCFT$ in the TCSA Hamitonian,\footnote{This shows once again that corrections due to integrals of non-scalar operators, $T_{\tau\tau}$ in this case, can be induced by the flow with the TCSA cutoff. In the previous subsection, we pointed out that the correction due to the direct appearance of $T_{\mu\nu}$ in the OPE would be suppressed, since the corresponding coefficient $h$ is quite large. However, here we are discovering another way for the appearance of this correction---as a subleading term accompanying the unit operator in the OPE.} while the second correction becomes a uniform overall rescaling of all couplings. Both of these can be taken into account easily by a slight modifications of \eqref{eq:rgmaster}. The situation is more complicated if $\calV_c\ne \unit$. In this case, the expansion generates terms of the form\footnote{Notice that these corrections, as written, preserve the hermiticity of the Hamiltonian.}
\beq
\HCFT.V_c+V_c.\HCFT,\qquad H.V_c+V_c.H,\qquad V_c\equiv  \int_{S^{d-1}_R}\calV_c\,.
\label{eq:nonlocal}
\eeq
Not only are these terms not present in the original Hamiltonian, they are also of a qualitatively different type---they are not given as an integral of a local operator over the sphere. In other words, these terms are nonlocal. While this may seem confusing, a moment's thought shows that this was to be expected. The reason is that the TCSA regulator---throwing out all states above a certain energy---is not a fully local UV regulator.\footnote{The TCSA regulator reproduces exact correlators as long as the insertion points are separated in the \emph{time} direction by $\gg\LUV^{-1}$. In particular, correlation functions on a constant time slice are not faithfully reproduced no matter how far the points are separated in the space direction. By a fully local UV regulator we mean a regulator which reproduces exact correlation functions as long as points are separated in \emph{some} direction, time or space, by $\gg\LUV^{-1}$. E.g.~the point splitting procedure, used in conformal perturbation theory, is a fully local regulator.}
So we have to learn to live with nonlocal correction terms. Fortunately, from the practical point of view the terms 
\reef{eq:nonlocal} pose no problem. First of all, they are easily computable, since they are given by products of matrices which we anyway have to compute in the earlier stages of the TCSA procedure. Secondly, although in principle the non-local terms would appear also on the right-hand side of the RG flow equations, which would substantially complicate the flow, in practice we found that they remain rather small compared to the local terms. This happens because their running is suppressed by one extra power of the cutoff. Therefore, in this work we will ignore backreaction of the non-local terms on the other running couplings.

Although the above procedure correctly takes into account the leading $\bar E/ \Lambda$ dependence, we realized that expanding in $\bar E / \Lambda$ is actually not a reasonable thing to do at large $R$. The point is that the ground state energy $E_0$ grows at large $R$ like $R^{d-1}$, and even for moderately large $R$ becomes non-negligible compared to $\LUV$. Whether this is a problem depends on the sign of $E_0$. If $E_0$ were to become large and positive, there would be no magic way out---the correction procedure would break down as soon as $E_0\sim \LUV$, as seen e.g.~by the blow up of the integral in \reef{eq:master}. Fortunately, the ground state energy density at large $R$ is usually negative.\footnote{The second-order correction to the ground state energy is negative. Assuming that higher-order corrections don't change the situation, we may expect negative energy density at large $R$. Studying many examples of RG flows known in $d=2$, this seems to be invariably true. The only exceptions happen when the dimension of the perturbing operator exceeds $d/2$. In this case the renormalized ground state energy density may be positive, although the non-renormalized, divergent energy density is still negative. In both concrete examples of $d>2$ flows studied in this work, the ground state energy density is negative at large $R$.} In this case, although $E_0$ becomes large in absolute value, nothing bad occurs with the correction in \reef{eq:master}; it even decreases with respect to the $E_0=0$ case. However, were one to expand in $\bar E/ \Lambda$, one would unnecessarily introduce large corrections even in this benign case.

We will therefore adopt the following prescription. We will replace the estimate $\bar E$ in \reef{eq:master} by $E_r+(\bar E-E_r)$ where $E_r$ is a convenient reference energy that we estimate to be close to the expected value of $\bar E$. For example, we may choose $E_r$ to be around the ground state energy as obtained by extrapolation from lower values of the radius, or around the energy of the first excited state. In fact, the end results for the spectrum should not depend much on the chosen value of $E_r$, which provides a consistency check for the method. We then expand not in $\bar E / \Lambda$ but instead in the difference $(\bar E-E_r)/\Lambda$, which is not expected to become large in the large volume limit. The RG evolution is then performed keeping track of the exact dependence on $E_r$ (no expansion) through the simple substitution $\Lambda^{d-h+1} \to \Lambda^{d-h} (\Lambda - E_r)$ in the denominator of \eqref{eq:rgmaster}. Since we will expand in $(\bar E-E_r)/\Lambda$, the leading correction in \eqref{eq:nonlocal} should be modified by replacing $H \to (H-E_r)$. Below we will see a concrete example of how this works, when discussing the Landau-Ginzburg flow. 

\subsection{Comments on Earlier Treatments of Renormalization}
\label{sec:comments}
Cutoff dependence and renormalization have been discussed in the context of the $d=2$ TCSA studies, most importantly in \cite{Giokas:2011ix} (following \cite{Feverati:2006ni,Watts:2011cr}) but see also Appendix \ref{sec:2d} for other references. In particular, Section 3 of \cite{Giokas:2011ix} discusses in detail how the cutoff dependence can be analyzed using the OPE, and gives renormalization group equations similar to our \eqref{eq:rgmaster} for the couplings of the local operators. At leading order, then, their results are basically equivalent to ours.\footnote{A factor $\half$ seems to be missing in their Eq.~(3.7). Even having corrected this misprint, we did not manage to reproduce their figure 1(c).} 

Ref.~\cite{Giokas:2011ix} also initiated a discussion of subleading terms. For example, the first of the two subleading terms in \eqref{eq:nonlocal} may be discerned in their equations, for the special case where $\calV_c$ is the identity operator. However, significant differences do exist between us and them at how these subleading effects are implemented.

According to the prescription in Section 4.2 of \cite{Giokas:2011ix}, on top of leading RG improvement, each IR state should get a subleading correction factor computed from the conformal perturbation theory applied to a UV state from which the IR state in question originates. This prescription, as well as a more recent detailed discussion in Section 3 of \cite{Lencses:2014tba}, are designed to fix up, order by order in the coupling, the discrepancies between TCSA and conformal perturbation theory. On the other hand, our discussion uses from the very beginning the fact that the true expansion parameter is the inverse cutoff rather than the couplings, which become large in the IR.

Let's illustrate the differences by looking at the correction in Eq.~\reef{eq:ourcorr0}. Our derivation demonstrates clearly that one should compute $\Delta H$ with the \emph{nonperturbative} energy $\bar E$ and take the matrix element between the \emph{nonperturbative} states $\bar c$. A similar correction in Eq.~(3.11) of \cite{Lencses:2014tba} uses the UV energy and the UV state in place of $\bar E$ and $\bar c$. At small $R$ the two methods would give very similar results, but at large $R$ the difference will be significant. Indeed, the IR states at large $R$ will have a complicated composition, in which the original UV state carries little weight (see e.g.~figure \ref{fig:phi2comp}). Also the energy in the IR will get a very large correction, implying a large change in the denominator in \eqref{eq:ourcorr}.
As a result the whole correction may be modified at $O(1)$. Out of curiosity, we compared our method to that of \cite{Lencses:2014tba} for the $\phi^2$ flow discussed in the next section. We found that at large $R$ our method is more effective in reducing the discrepancy from the exact results.

For completeness it should be noted that \cite{Lencses:2014tba}, using their renormalization prescription, achieved an excellent agreement of TCSA data with the results obtained by exact integrability methods applicable for the model they studied. This success is puzzling to us, since as we explained we believe that their prescription is problematic at large $R$. This question deserves further analysis.

\section{The $\phi^2$ Flow}
\label{sec:phi2}

\subsection{Theoretical Expectations} 
\label{sec:phi2-exp}

We are now ready to do our first TCSA calculation in $d$ dimensions---the flow starting at the free massless scalar and perturbing by the pure mass term $\half m^2\NO{\phi^2}$. Needless to say, this RG flow is considered for illustrative purposes only. We expect to find the free massive scalar theory, whose spectrum in canonical quantization was discussed in section \ref{sec:can}. We will restrict ourselves to the spin 0 states of that spectrum, corresponding to one or several particles at rest or in relative motion along the sphere in such a way that the total angular momentum adds up to zero.

In addition to the spectrum, another observable is the ground state energy. For the canonically quantized massive scalar it's given by the zero point energy of all oscillators
\beq
E_{0,{\rm can}}=\frac 12\sum_{l=0}^\infty D_d(l) \omega_l\,,
\label{eq:E0can}
\eeq
where $D_d(l)$ is the size of the spin $l$ symmetric traceless representation of $SO(d)$ (see e.g.~\cite{Cardy:1991kr}):
\beq
D_d(l)=f_d(l)+f_d(l-1),\quad f_d(l)=\frac{(d+l-2)!}{l!(d-2)!}\,.
\eeq
Here we will be computing the ground state energy by perturbing a CFT, and so our expected answer is not $E_{0,{\rm can}}$ but a simple modification thereof. First of all, in our treatment the unperturbed CFT ground state energy on the sphere is set to zero (see footnote \ref{note:cas0}). Furthermore, the $O(m^2)$ term in the ground state energy must vanish, being proportional to a CFT one-point function which is zero. We expect however that all terms higher-order in $m^2$ will agree between CFT and canonical quantization. Thus, the perturbed CFT ground state energy should be given by \reef{eq:E0can} with terms of the zeroth and first order in $m^2$ dropped.\footnote{Notice that this vacuum energy is \emph{not} the same as the one considered in the studies of the Casimir effect, where one renormalizes by subtracting the vacuum energy density of the same theory in flat space. The latter procedure is relevant if one is interested in the dependence of the vacuum energy on the geometry keeping the mass fixed. Here we are interested in the dependence on the mass itself.} This can be written as follows:
\begin{align}
E_{0}=\frac12\sum_l D_d(l) \Bigl[\omega_l-\omega_l|_{m^2=0}-m^2\frac {\del \omega_l}{\del m^2}\Bigl|_{m^2=0}\Bigr]=-\frac{\mu^2}{2\Gamma(2\nu+1) R}\sum_l H(l+\nu), \label{eq:E0}
\end{align}
where $\mu=mR$ and
\beq
H(z) = \frac{\Gamma(z + \nu)}{\Gamma(z+1-\nu)} \frac{\sqrt{z^2 + \mu^2} - z}{\sqrt{z^2 + \mu^2} + z}\,.
\eeq
Because of subtractions, the general term in the series behaves at large $l$ as $l^{d-5}$. So the ground state energy is finite for $d<4$, in agreement with the criterion \reef{eq:uvfin}, \reef{eq:uvfin1}.\footnote{The two subtractions in \eqref{eq:E0} remove the divergences that originate from the normal ordering of the operators in the bare CFT Lagrangian and the bare $\phi^2$ operator, respectively. These divergences are intrinsic to the CFT and not associated with the RG flow.}
 
For general $R$, we can compute $E_0$ by numerically summing the series \reef{eq:E0}. In the large volume limit $\mu\gg 1$, the sum will be dominated by large $l$ terms and can be approximated by an integral. The leading behavior in this limit scales as the volume of the sphere, with a constant density set by the mass:
\begin{gather}
E_0\approx - C_d\, m^d R^{d-1} \qquad(R\gg m^{-1})\,,\nn\\
C_d= \frac{1}{2\Gamma(2\nu+1)} \int_0^\infty x^{2\nu-1} \frac{\sqrt{1+x^2}-x}{\sqrt{1+x^2}+x} \mrm{d}x 
= \frac{\Gamma (1-\nu ) \Gamma \left(\nu +1/2\right)}{4 \sqrt{\pi } \nu  (\nu +1) \Gamma(2\nu + 1)}\qquad(2<d<4)\,. 
\label{eq:E0lim}
\end{gather}
One can show that the first correction in this formula arises at order $1/R^2$. The same is true for the rate of approach of masses of particle states to their infinite volume limit, as can be seen from the explicit formulas in section \ref{sec:can}. The presence of these power-like in $R$ corrections is due to the curvature of the general $d$-dimensional sphere. Here we are observing them in a free theory, and we expect them to be present in an interacting situation as well. This can be contrasted with what happens when a QFT is put on a torus $T^{d-1}$ (which for $d=2$ is of course the same as $S^{d-1}$). In this case it's been observed long ago \cite{Luscher:1985dn} that masses in an interacting theory are affected by terms which are exponentially small in the size of the torus.

\subsection{TCSA Setup}
\label{sec:phi2-setup}
Which value of $d$ shall we choose in our numerical study of the $\phi^2$ flow? As already mentioned in section \ref{sec:case}, the case $d\to2$ is expected to be difficult, as the CFT spectrum is becoming dense in the limit. For $d>4$ the vacuum energy will be divergent. Here we will show results for the physical value $d=3$. We have also performed checks for other nearby values of $d$ and they work equally well.

We will construct the truncated Hilbert space by including all scalar operators below a certain dimension
\beq
\Delta_i\le \Dmax\,.
\eeq
For practical reasons, this maximal dimension will be held fixed when varying $R$. This means that we will be working with a sliding UV cutoff
\beq
\label{eq:sliding}
\LUV=\Dmax/R\,.
\eeq
TCSA can be expected to reproduce the IR spectrum roughly below this cutoff. As we increase $R$, the sliding cutoff decreases and eventually becomes comparable with $m$. At this point TCSA results can no longer be trusted.\footnote{The range of validity of TCSA will be somewhat extended due to the fact that the induced ground state energy is negative, which reduces the size of correction terms, as we discussed in section \ref{sec:RG-improv}.}

\begin{figure}[htbp]
\begin{center}
\includegraphics[scale=1.0]{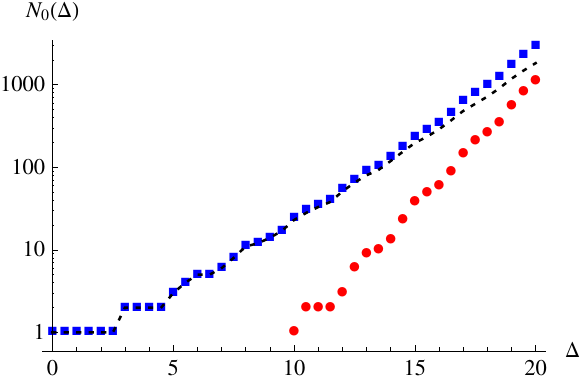}
\end{center}
\caption{The number of scalar $P$-even states in the extended Hilbert space of free massless scalar theory in $d=3$ on the cylinder.
Blue squares: all states (physical + null). Red dots: null states. Black dotted curve: just physical states. The proportion of null states grows quickly: at $\Delta=18$, which is the maximal cutoff we will be working with, about a quarter of all states are null. In future studies one should perhaps separate the null states to speed up the numerics.}
\label{fig:states3d}
\end{figure}

The number $N_0(\Delta)$ of scalar, parity even states in the Hilbert space as a function of $\Delta$ is shown in figure \ref{fig:states3d}.
The dotteed line gives the number of physical states, counted using group theory. The \emph{total} number of states in a $d$-dimensional CFT grows with $\Delta$ exponentially \cite{Cardy:1991kr}\footnote{See \cite{Pappadopulo:2012jk} for a review.}:
\beq
N(\Delta)\sim \exp(C\, \Delta^{1-1/d})\,,
\eeq
where $C$ is a theory dependent constant related to the prefactor in the free energy density dependence on the temperature.\footnote{$C=d[\zeta(d)]^{1/d}/(d-1)^{1+1/d} $ for a free massless scalar \cite{Cardy:1991kr}.}
It's not hard to see that the number of scalar states will also grow exponentially with the same exponent, although with a smaller prefactor. In figure \ref{fig:states3d} we can clearly see this exponential growth. The fast growth of the number of states implies that it will be hard to increase $\Delta_{\max}$. The success of any TCSA calculation will depend on whether reasonable results can be obtained with a manageable $\Delta_{\max}$. As we will see, achieving numerical accuracy for such $\Delta_{\max}$ will require the use of renormalization corrections discussed in section \ref{sec:RG}.

 As discussed in section \ref{sec:null}, in this paper we will be working in an extended Hilbert space which for integer $d$ is somewhat larger than the physical Hilbert space, since it includes some null states. The extended Hilbert space states are in one-to-one correspondence with non-isomorphic multigraphs; their number is shown by blue squares, while red dots show null states. We see that the first null state occurs at $\Delta=10$; this is the state \reef{eq:tracerel}. 

\subsection{Numerical Results}
\label{sec:num-phi2}
We will now show our numerical TCSA results and compare them with theoretical expectations. 
The TCSA computation starts by constructing the truncated Hilbert space. We consider cutoffs up to $\Dmax=18$, which corresponds to 4573 scalar $P$-even states. We then construct the Hamiltonian matrix $H^{i}{}_{j}$. The CFT part is given by the diagonal matrix \reef{eq:hcft}. The perturbing part is computed using the OPE method from section \ref{sec:ope}, for the operator $\calV=\NO{\phi^2}$. We then diagonalize the Hamiltonian matrix to find the spectrum of the perturbed theory.

Since the perturbation preserves the $\bZ_2$ symmetry which maps $\phi\to-\phi$, the Hamiltonian matrix does not mix $\bZ_2$-even and $\bZ_2$-odd states. The two sectors have roughly equal number of states, and it makes sense to do the computation separately in each of them, reducing the size of the matrices to be diagonalized by factor $\sim 2$. The ground state belongs to the $\bZ_2$-even sector. 

\begin{figure}[htbp]
\begin{center}
\includegraphics[scale=0.4]{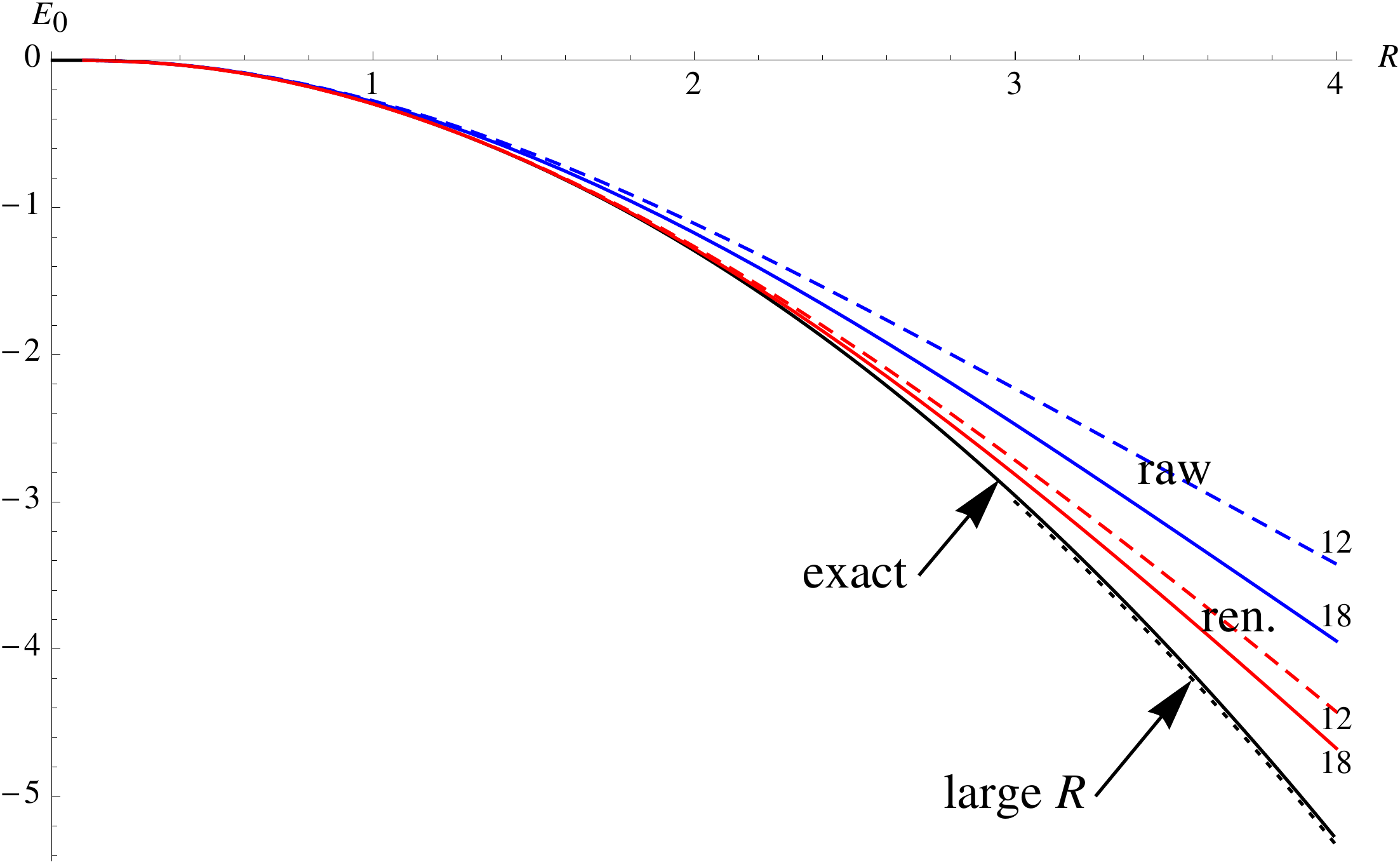}
\caption{The ground state energy of the $\phi^2$ flow in $d=3$ as a function of $R$ (we set $m=1$). Solid black curve: theory prediction \reef{eq:E0}. Dotted black: theory limit at large $R$, Eq.~\reef{eq:E0lim}.
Blue curves marked `raw': raw TCSA results, i.e.~before applying any correction. Red curves marked `ren.': renormalized TCSA results, see section \ref{sec:RG-phi2}. Dashed and solid TCSA curves correspond to cutoff $\Dmax = 12(18)$.}
\label{fig:E0phi2}
\end{center}
\end{figure}

In figure \ref{fig:E0phi2}, we plotted $E_0$ as a function  of $R$. In this and other plots in this section, we set $m=1$, which means that we measure $R$ in units of $m^{-1}$ and energies in units of $m$. The black solid curve shows the theoretical prediction for $E_0(R)$ obtained by summing the series in Eq.~\reef{eq:E0}. Let us focus first on the `raw' TCSA results, i.e.~the results obtained without any renormalization corrections (blue curves marked `raw'). We see that the agreement is good up to $R\sim 1$, while for larger $R$ there are noticeable deviations. As the cutoff is increased (we show $\Delta_{\max}=12(18)$ in dashed(solid) blue), the numerical results are moving towards the theoretical prediction, but the convergence is not very fast.

Cutoff dependence of TCSA predictions was discussed in section~\ref{sec:RG}. As we have seen, the errors induced by omitting the states with $E>\LUV$ are expected to go down as a series of power-laws with known exponents. These errors can be understood analytically and then subtracted away, greatly improving the accuracy of the TCSA calculations. The red curves marked `ren.' in figure \ref{fig:E0phi2} have been produced using such a renormalization procedure (see section \ref{sec:RG-phi2} for details). The agreement with the exact results is greatly improved; it now extends up to $R\sim 2.5$. Notice that the corrected results also exhibit a smaller dependence on the cutoff. This is because we are subtracting the leading correction, and the remaining ones are suppressed by extra powers of $\LUV$. 

Is $R\sim 2.5$ a large or a small radius? For two reasons, it should be considered as large. First of all, the corresponding sphere circumference, $L=2\pi R$, is much larger than the inverse mass, so that there is plenty of room for a massive particle wavefunction to fit into the sphere. Second, it takes us much beyond the radius of convergence of conformal perturbation theory $R_c=(d-2)/2$.\footnote{As determined by the leading singularity in the exact expressions for the vacuum energy density and the massive spectrum, located at $m^2 R^2 =-\nu^2$.}

\begin{figure}[htbp]
\begin{center}
\includegraphics[scale=0.7]{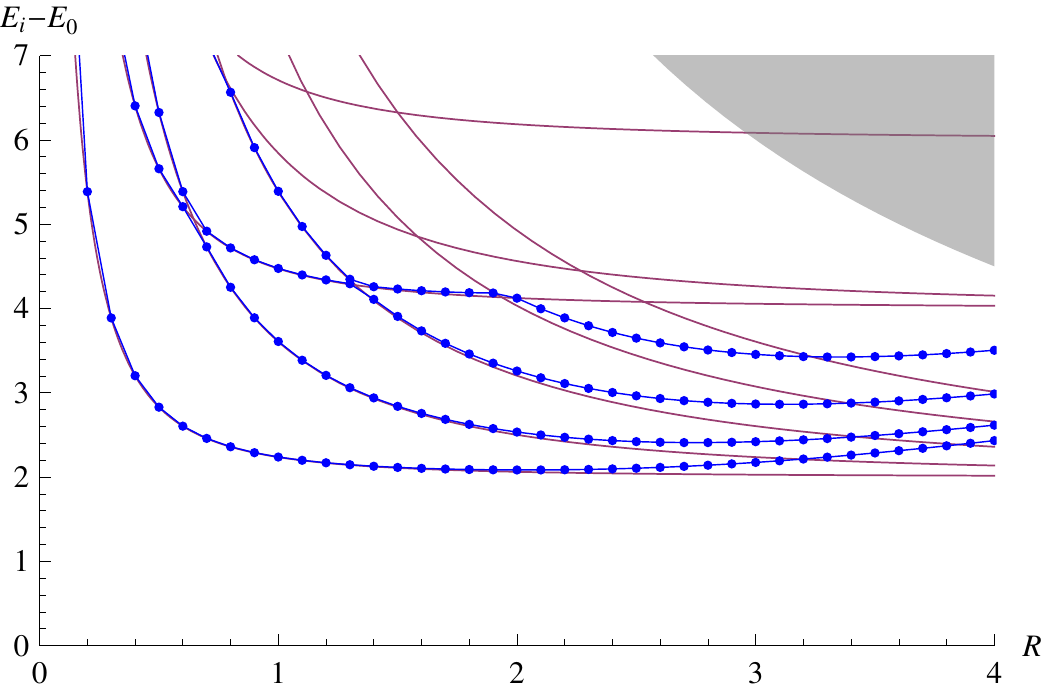}\quad
\includegraphics[scale=0.7]{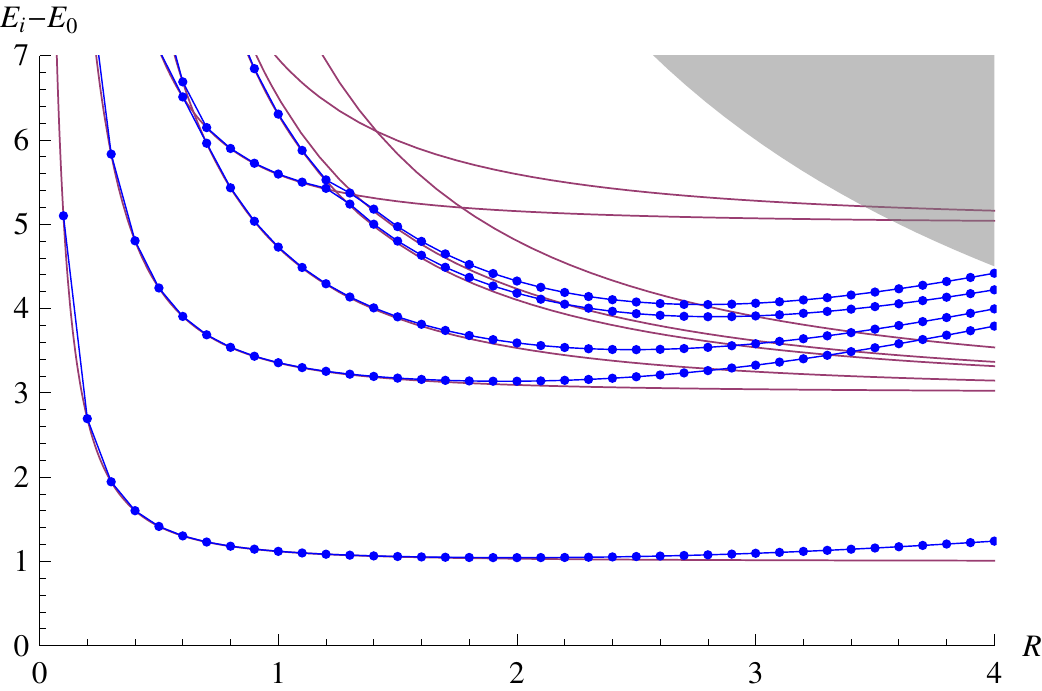}
\caption{A few lowest massive excitations from the raw TCSA spectra at $\Delta_{\max}=18$ (blue dots connected with a line to guide the eye) vs exact spectrum (magenta lines). Left(right): $\bZ_2$-even (odd) sector. The gray region indicates the sliding UV cutoff \reef{eq:sliding}.}
\label{fig:phi2raw}
\end{center}
\end{figure}

We now turn to the excitations above the vacuum. In figure \ref{fig:phi2raw} we plot the energies of these excitations, subtracting the vacuum energy. We have two plots, one for the $\bZ_2$-even and one for the $\bZ_2$-odd sectors. To keep the plots from cluttering, we show the lowest five eigenvalues in each sector. Notice that in both cases we subtract the same quantity $E_0$, which is the lowest energy in the $\bZ_2$-even sector.
Blue dots are computed using TCSA for $\Delta_{\max}=18$, while lines joining them are added to guide the eye. 

In the same plot thin magenta lines show the exact free massive scalar spectrum, computed by combining oscillator energy levels from section \ref{sec:can}. In the $\bZ_2$-even sector the lowest state corresponds to two particles at rest, while the states above it 
correspond to two particles with some angular momentum on the sphere combined in a state of total spin zero. Then there comes the state with four particles at rest etc. In the $\bZ_2$-odd sector we recognize one particle at rest, three particles at rest, then three-particle states in relative motion, etc. 

\begin{figure}[htbp]
\begin{center}
\includegraphics[scale=0.7]{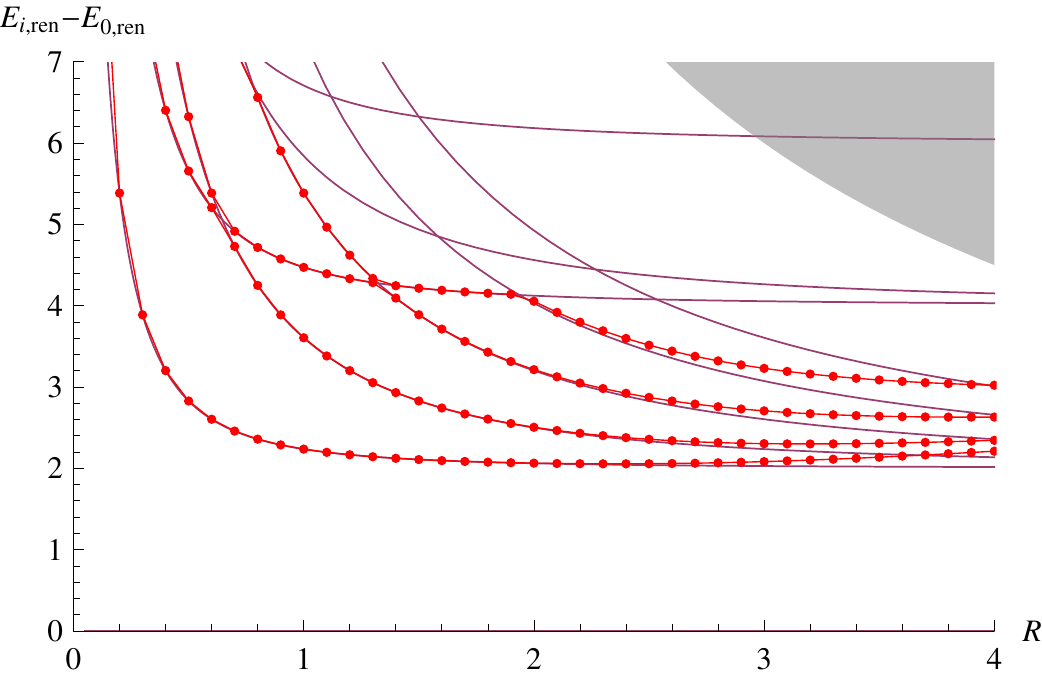}\quad
\includegraphics[scale=0.7]{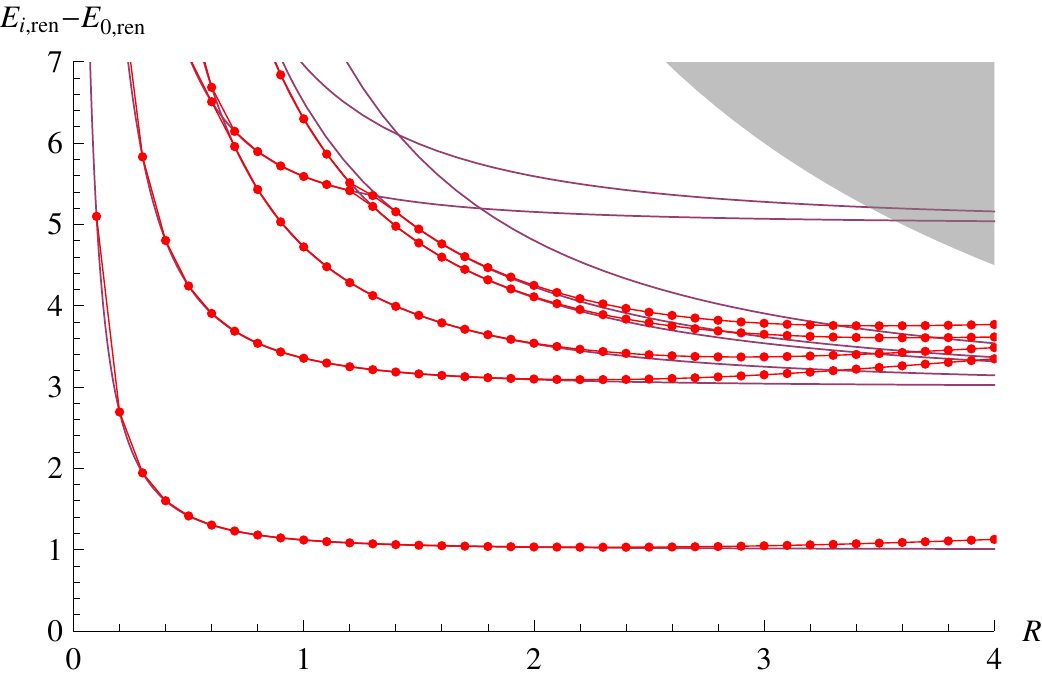}
\caption{Same as the previous figure, but for the renormalized TCSA spectra.}
\label{fig:phi2ren}
\end{center}
\end{figure}

In figure \ref{fig:phi2ren} we show the same but for the spectra computed using the renormalized TCSA eigenvalues.
The details of the renormalization procedure will be discussed in the next section. We see from these plots that renormalization extends the range of $R$ where TCSA is in agreement with the exact results from $R\lesssim 2$ to $R\lesssim 3$.

\subsection{Renormalization Details}
\label{sec:RG-phi2}

The general method of renormalization was presented in section \ref{sec:RG}. Here we will describe particular issues which arise when the procedure is applied to the $\phi^2$ flow. The leading contributions to the correction term $\Delta H$ are expected to come from the low-dimension operators in the $\phi^2 \times \phi^2$ OPE:
\beq
\NO{\phi^2(x)}\times \NO{\phi^2(0)} =  \frac{2\Nd^2}{|x|^{2(d-2)}}\unit +\frac{2\Nd}{|x|^{d-2}}\NO{\phi^2}+\NO{\phi^4}+\ldots\,,
\label{eq:phi2phi2}
\eeq
Here ${\rm N}_d$ is the normalization factor in the two point function of the canonically normalized massless scalar:
\beq
\langle \phi(x)\phi(0)\rangle=\Nd/{|x|^{d-2}},\quad \Nd=1/{[(d-2)\Sd]}\,,
\eeq
where $\Sd= 2\pi^{d/2}/\Gamma(d/2)$ is the area of the unit sphere in $d$ dimensions. 

Now, curiously, although the operators $\phi^2$ and $\phi^4$ appear in the OPE \reef{eq:phi2phi2}, their contributions to the renormalization corrections vanish. Indeed, the coefficient $B(h)$ given by \reef{eq:Bh} is zero for the corresponding $h$'s. The reasons this happens are not difficult to understand; they are ultimately related to the fact that the UV CFT we are perturbing is free. Starting with $\phi^4$, notice that since the dimensions factorize, $\Delta(\phi^4)=2\Delta(\phi^2)$, and the OPE kernel is just a constant. Clearly, the $\tau\to0$ limit discussed in section \ref{sec:RG-count} is perfectly analytic in this case, and so $B(h)$ must vanish. For $\phi^2$, although the OPE kernel is singular, it is a harmonic function of $x-y$. By the mean value property, the integral of a harmonic function over a sphere is equal to its value at the center of the sphere. This implies that also in this case the $\tau\to0$ limit is analytic, and $B(h)=0$.\footnote{\label{note:nm}This argument shows that, more generally, corrections will vanish for the $\phi^{n+m}$ and $\phi^{n+m-2}$ operators in the $\phi^n\times \phi^m$ OPE. This observation will be useful for the general Landau-Ginzburg flow in section \ref{sec:RG-phi4}.}

Thus the only leading non-vanishing correction is for $\calV_c=\unit$.\footnote{This also implies that the RG improvement discussed in section \ref{sec:RG-improv} is not of much use for this particular example: the mass parameter never appears on the right-hand side of the renormalization group equations \eqref{eq:rgmaster}, so their solution is straightforward and essentially given by \eqref{eq:master}, i.e. the unimproved equation.} We will have to include this correction taking into account the subleading dependence on $\Delta_i+\Delta_j$ and $\bar E$. Indeed, were we to drop these subleading parts, we would get a constant counterterm which would shift all eigenvalues in the same way. This would have a chance to improve the agreement for the ground state energy, but would have no effect on the spectrum of massive excitations. However, the raw TCSA massive spectra in figure \ref{fig:phi2raw} do show noticeable deviations, which we would also like to improve.

In fact, we will be able to do even better. Not only will we include the above-mentioned subleading effects, but we will also take into account the discreteness of the sequence $M_n$. Recall that the general formula \reef{eq:pl} gives this sequence only on average. However, it turns out that for the $\phi^2$ flow the tail of the $M_n$ sequence can be worked out explicitly, independently of the argument in section \ref{sec:RG-count}. As we show in  appendix \ref{sec:phi2tail}, $M_n$ at $\Delta_n\gg \Delta_j$ is nonzero only if $\Delta_n-\Delta_j-\Delta(\phi^2)=2p$ is an even integer, in which case it's given by 
\beq
(M_n)^i{}_j
=
\frac{2(2\nu)_p(\nu)_p}{p!(\nu+1)_p} ( {\rm N}_d {\rm S}_d )^2 \delta^i{}_j \stackrel{d=3}{\to} \frac{2}{2p+1}( {\rm N}_d {\rm S}_d )^2 \delta^i{}_j \,.
\label{eq:Mnij}
\eeq
It's not difficult to see that on average this sequence does agree with the continuous distribution \reef{eq:pl}, which also provides a check for the general argument.

We next evaluate $\Delta H$ via Eq.~\reef{eq:cc1}. When doing the sum, we use the expression \reef{eq:Mnij} for all terms. This is not quite true, since \reef{eq:Mnij} was derived under the assumption $\Delta_n\gg \Delta_j$, which does not hold for the external states $i,j$ just below and $n$ just above the cutoff. However, the caused error cannot be large, since the states $i,j$ close to the cutoff will anyway contribute little to the renormalization, having small weight in the eigenvector $\bar c$; see figure \ref{fig:phi2comp} below. So we will tolerate this little imprecision. The infinite sum over $p$ becomes a ${}_4 F_3$ hypergeometric sum, and specializing to $d=3$ we obtain the digamma function $\psi(z)$. Reinstating the coupling and radius dependence, we get:
\beq
(\Delta H)^i{}_j\approx -(\half m^2)^2 \frac{R^3}{(d-2)^2} 
\frac{\psi( (K_j+\Delta_j- R\bar E)/2)-\psi(K_j/2)}{\Delta_j- R\bar E} \delta^i{}_j\,,
\label{eq:corrphi2}
\eeq
where $K_j$ is defined as the smallest odd integer such that $\Delta_j+K_j>\Delta_{\max}$. 

The leading term in $\Delta H$ for large $\Delta_{\max}$ is a state-independent correction $\propto \Delta_{\max}^{-1}$. As mentioned above, keeping only this correction would not be adequate. Instead, we will use the full expression \reef{eq:corrphi2} to compute corrected (`renormalized') eigenvalues $E_{\rm ren}$ from the raw TCSA eigenvalues $\bar E$ via the formula \reef{eq:ourcorr0}:
\beq
\label{eq:ErenIR}
E_{\rm ren}=\bar E+ \bar c_i (\Delta H)^i{}_j \bar c^j\,.
\eeq
It is these `renormalized' results which were used to produce figures \ref{fig:E0phi2},\ref{fig:phi2ren}. Here $\bar c$ is the eigenvector corresponding to the raw TCSA eigenvalue $\bar E$. In this approach, each energy level is corrected separately. 

Note that to apply formula \reef{eq:ErenIR} we need to compute both \emph{right} eigenvector $\bar c^j$, as in \reef{eq:eigsim}, and the \emph{left} eigenvector $\bar c_i$: 
\beq
H^i{}_j \bar c^j =\bar E \bar c_j\,,\quad \bar c_i H^i{}_j=\bar E \bar c_i\,.
\label{eq:twoeig}
\eeq
The eigenvectors are assumed normalized via $\bar c_i \bar c^i=1$. Of course these two eigenvectors are related, up to normalization, via the Gram matrix:
\beq
\label{eq:cleft}
 \bar c_i \propto G_{ij}  \bar c^j\,.
 \eeq
As mentioned in section \ref{sec:gram}, computing the full Gram matrix may be expensive, although we did find an indirect way to do it, described in section \ref{sec:ope}. If one has access to the Gram matrix, one can use it to compute the left eigenvectors via \reef{eq:cleft}. Without the Gram matrix, one simply finds $\bar c_i$ from the second eigenvalue problem in \reef{eq:twoeig}.\footnote{It should be noted that the nonsymmetric eigenvalue problems are somewhat more difficult to solve numerically than the symmetric ones, and more prone to numerical instabilities. In our work we overcome the instabilities by applying the transformation $H\to (H-\sigma)^{-1}$ to the matrix $H$ before diagonalization. This transformation focuses on the eigenvalues nearest to $\sigma$. We also checked some of our results by working at a higher number of digits. In future work, it would be interesting to keep looking for other, more numerically efficient diagonalization procedures.}
 
\begin{figure}[htbp]
\begin{center}
\includegraphics[scale=0.7]{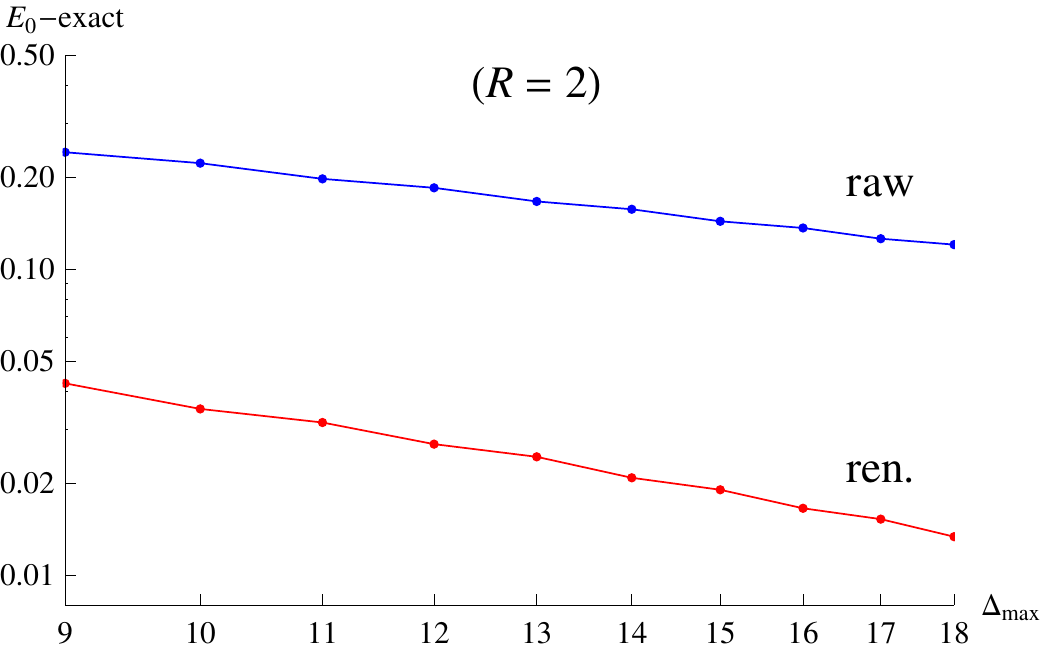}\ 
\includegraphics[scale=0.7]{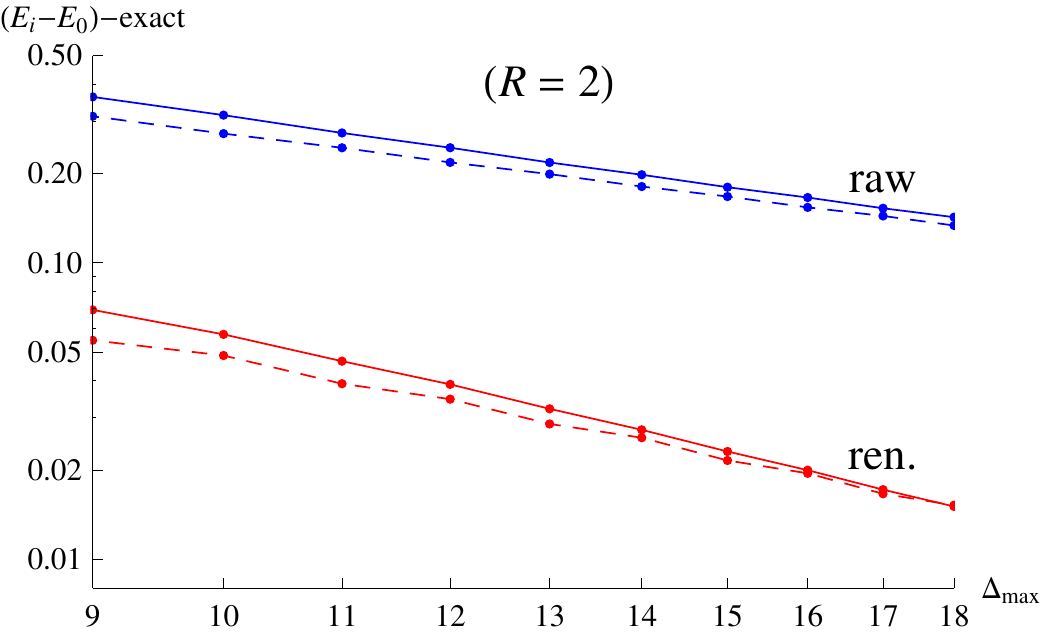}
\caption{Convergence rate before and after renormalization. Left: ground state energy. Right: lightest states in the massive spectrum, $\bZ_2$-even (solid) and $\bZ_2$-odd (dashed).}
\label{fig:phi2conv}
\end{center}
\end{figure}

Figures \ref{fig:E0phi2} and \ref{fig:phi2ren} do demonstrate that our renormalization procedure works---upon applying the renormalization corrections, discrepancy from the exact results is reduced compared to the raw TCSA data. Figure \ref{fig:phi2conv} demonstrates the same as a function of the UV cutoff: we show how the TCSA ground state energy and the massive spectrum converge to their exact values with the gradual increase of $\Delta_{\max}$. We do this plot for one value $R=2$, but the picture is qualitatively the same for all $R$. This figure shows that not only the accuracy is greatly improved after the renormalization, but the convergence rate is also improved. This is because the error terms remaining after the leading renormalization subtractions are suppressed by higher powers of $1/\LUV$.

One last aspect we would like to discuss here is an assumption implicit in the entire procedure of renormalization, namely that the contribution of high energy states to low energy observables is suppressed. It is possible to get a feel about this assumption by studying the distribution of eigenstate components in energy, defined as:
\beq
w(\Delta)=\sum_{i:\Delta_i=\Delta} \bar c_i\, \bar c^i\,.
\label{eq:w}
\eeq
In figure \ref{fig:phi2comp}, we plot this distribution for the lowest $\bZ_2$-even massive excitation (the one which is interpreted as a state of two particles at rest) and for several values of $R$. As expected, for small $R$ the distribution is strongly peaked at $\Delta=\Delta(\phi^2)=1$. As $R$ is increased, the distribution becomes wider and wider, but its high-energy tail does remain suppressed. The same qualitative behavior is true for the other states. One can wonder what it would mean if for very large $R$ the distribution becomes flat or even peaked at high $\Delta$. Does this ever happen for CFTs perturbed by a relevant operator? Presumably the method would completely break down for such $R$, but for the values of $R$ explored in this work this does not happen.
\begin{figure}[htbp]
\begin{center}
\includegraphics[scale=0.7]{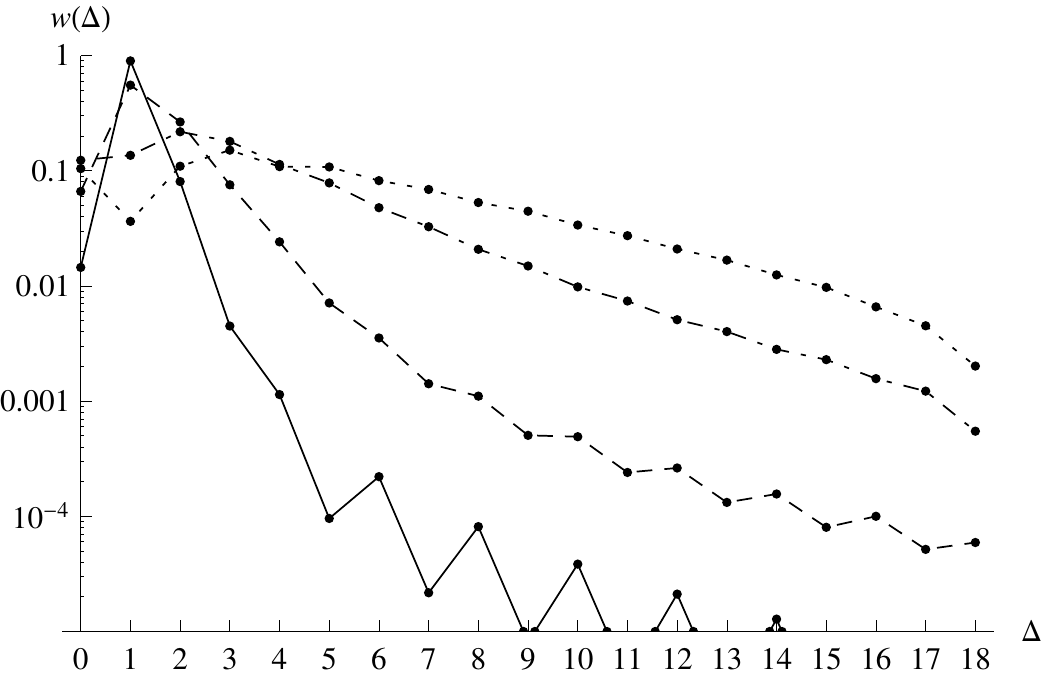}\caption{The distribution of eigenstate components in energy, Eq.~\reef{eq:w}, plotted for the lowest $\bZ_2$-even massive excitation, for $R=0.5$ (solid), 1 (dashed), 2 (dot-dashed), 3 (dotted).}
\label{fig:phi2comp}
\end{center}
\end{figure}

\section{The Landau-Ginzburg Flow}
\label{sec:phi4}
\subsection{Theoretical Expectations}
\label{sec:phi4-th}
  
In the free massive flow considered in the previous section, we could compare TCSA results with the exact theoretical answers for all observables. 
In a significant fraction of the $d=2$ TCSA literature (see Appendix \ref{sec:2d}), the method is applied to integrable flows, where the exact results are also available (of course, through much harder work than for the free massive scalar). 
We would however like to encourage the use of TCSA in situations when no other technique is readily available. This will be the case for most physically interesting strongly coupled theories, since exact integrability is possible only in $d=2$, and even then it's an exception, not a rule. 

In this spirit, we will now use TCSA to study the Landau-Ginzburg flow---the flow obtained by perturbing the free massless scalar CFT by
\beq
\label{eq:lgpotential}
\int d^dx (\half m^2 \NO{\phi^2} +\lambda \NO{\phi^4})\,.
\eeq  
The quartic $\lambda$ should be positive to have a stable vacuum, while $m^2$ can be positive or negative. The IR physics of this theory depends on the dimensionless ratio 
\beq
t \equiv m^2/\lambda^{2/(4-d)}.
\eeq
The theory is not integrable even in $d=2$, and here we will study it in $d>2$.

The case of small quartic coupling corresponds to $|t|\gg 1$. In this regime the theory in the IR describes weakly interacting massive particles, and predictions can be obtained from perturbation theory. For positive (and still large) $t$, the perturbative vacuum is at $\phi=0$, and the $\bZ_2$ symmetry $\phi\to-\phi$ is preserved. On the other hand, for negative $t$, perturbation theory is developed around one of two degenerate vacua of the double-well potential. Thus, the $\bZ_2$ symmetry is spontaneously broken.

\begin{figure}[htbp]
\begin{center}
\includegraphics[scale=1.3]{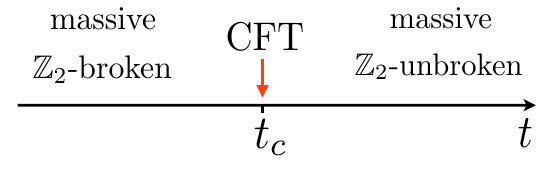}\\[3pt]
\caption{The commonly accepted phase diagram for the Landau-Ginzburg flow. Our calculations will indicate that $t_c>0$ in $d=2.5$ dimensions.}
\label{fig:phi4-phase}
\end{center}
\end{figure}

It is then interesting to know what happens for $t=O(1)$, when the IR theory is strongly coupled, and perturbation theory is not useful.\footnote{It has been rigorously shown in the constructive field theory literature that for positive $m^2$, perturbation theory is Borel-summable for all couplings in $d=2$ \cite{eckmann1974} and $d=3$ \cite{magnen1977}.
} One generally expects that the $\bZ_2$ broken and preserving phases extend into the strongly coupled region, where they are separated by a second-order phase transition at $t=t_c$, see figure \ref{fig:phi4-phase}. At $t=t_c$ the theory is expected to flow in the IR to a CFT, belonging to the Wilson-Fisher family of fixed points in the Ising model universality class.

In this paper we will only study the $m^2>0$ (i.e.~$t>0$) part of the phase diagram.
Instead of varying $t$ as in figure~\ref{fig:phi4-phase}, we will find it convenient to work in the units $m=1$, and vary $\lambda$. Using TCSA, we will compute how the finite volume spectrum of the theory depends on $\lambda$. As we will see, for small $\lambda$ the spectrum will be consistent with preserved $\bZ_2$ symmetry, while for $\lambda>\lambda_c$ our calculations will indicate spontaneous $\bZ_2$ symmetry breaking.
Thus we will obtain qualitative confirmation of the phase diagram in figure~\ref{fig:phi4-phase}, and quantitative information about the massive spectrum in the strongly coupled region.  We will determine the critical value $\lambda_c$ with some precision. For $\lambda=\lambda_c$ we will observe the mass gap going to zero, indicating that the IR theory is conformal. We will be able to get a rough estimate of the leading critical exponents at the phase transition point and compare them with the known values in the Ising universality class.

Notice that since our calculations indicate a positive value of $t_c=1/\lambda_c^{(4-d)/2}$, the whole region $t<0$ is expected to be in the $\bZ_2$-broken phase. However, we have not explored this region numerically. 

\subsection{Numerical Results}
\label{sec:phi4-num}
\begin{figure}[htbp]
\begin{center}
\includegraphics[scale=0.7]{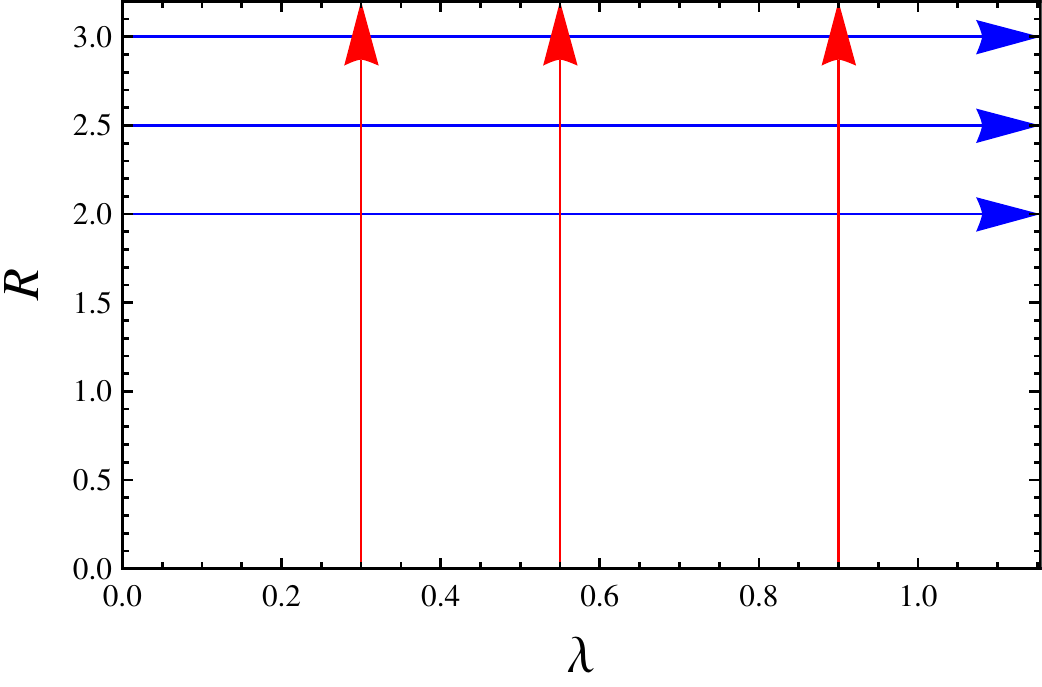}\\[3pt]
\caption{The range of $R$ and $\lambda$ explored in our study. Subsequent figures will show the spectrum dependence along the vertical and horizontal sections of this region, shown by the arrows.}
\label{fig:phi4range}
\end{center}
\end{figure}

We will now perform TCSA analysis of the Landau-Ginzburg flow in $d=2.5$ dimensions, working with cutoff up to $\Delta_{\max}=17$, which corresponds to 5494 (4907) $\bZ_2$-even (odd) states. As already mentioned above, we will set $m^2=1$. The spectrum will depend on $\lambda$ and the TCSA radius $R$. We will explore the region $R\lesssim 3$ and $\lambda\lesssim 1.15$ (see figure~\ref{fig:phi4range}). Raw TCSA without renormalization corrections would give converged predictions only in the lower left corner of this region, corresponding to weak coupling and small physical volume. To extend the range of applicability of the method, we will apply a renormalization procedure as described in section \ref{sec:RG}, with the theory specific details described in section \ref{sec:RG-phi4} below. To reduce the number of plots, we will only show results with all renormalization corrections taken into account. 

\subsubsection{Spectrum for a fixed $R$ and varying $\lambda$}
\label{sec:fixedR}

\begin{figure}[htbp]
\begin{center}
\includegraphics[scale=0.7]{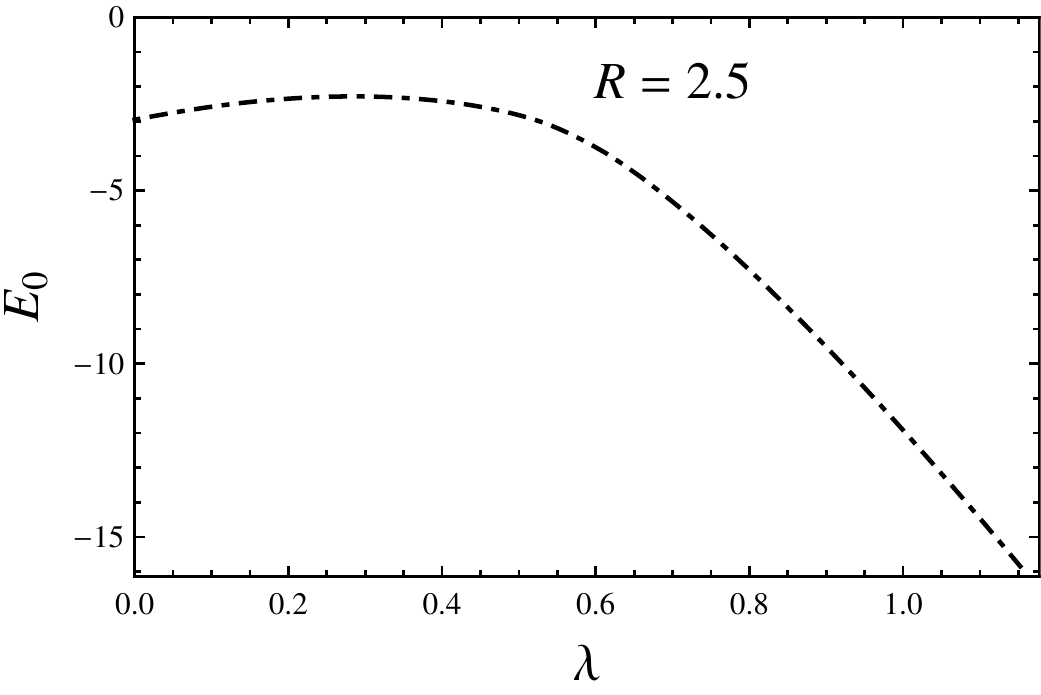}\quad
\includegraphics[scale=0.7]{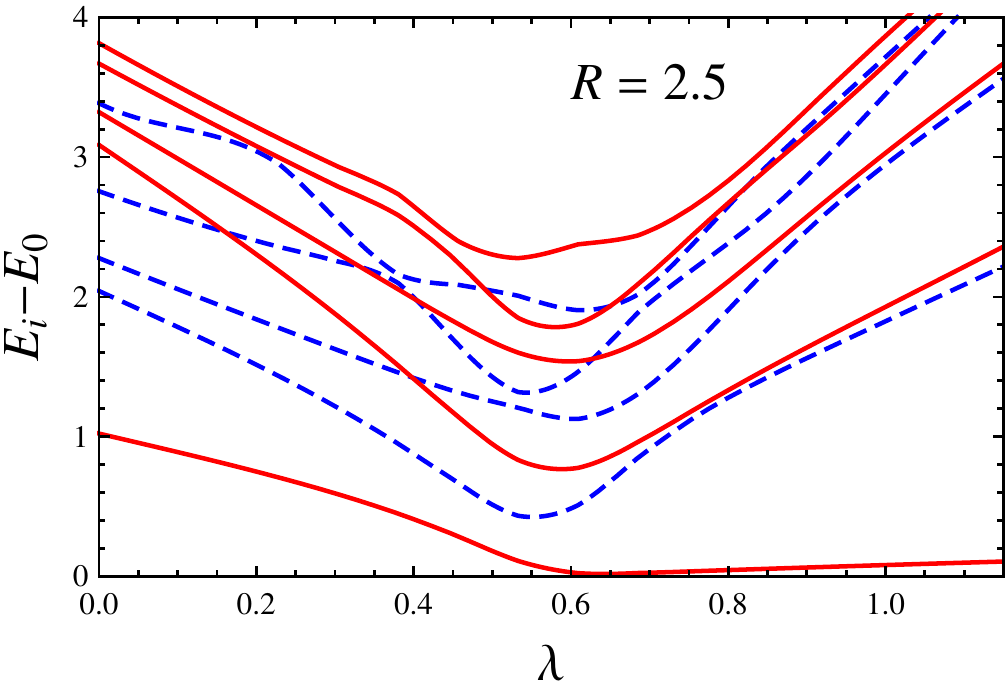}\\[3pt]
\caption{The ground state energy (left panel) and the spectrum of low-lying massive excitations (right panel) as a function of $\lambda$ for $R=2.5$. We plot 4 lowest $\bZ_2$-even (solid red) and 5 lowest $\bZ_2$-odd (dashed blue) states. }
\label{fig:phi4R2.5}
\end{center}
\end{figure}

To visualize the spectrum dependence, we will plot it along a number of vertical and horizontal sections in the two-dimensional range in figure~\ref{fig:phi4range}. Let us start with plots at a fixed $R$ and varying $\lambda$. In figure \ref{fig:phi4R2.5} we show the results for $R=2.5$. The ground state energy $E_0$ is defined as the lowest energy in the $\bZ_2$ even sector. The excitation spectra are given by $E_i-E_0$, in the $\bZ_2$-odd and the $\bZ_2$-even sectors, respectively. 

We see that as $\lambda$ is increased, the excitation energies first decrease, and then, for $\lambda>\lambda_c\approx 0.5\text{ - }0.6$, start increasing again. An interesting feature of the spectrum at $\lambda>\lambda_c$ is an approximate double degeneracy of states in the $\bZ_2$-even and odd sectors, well visible for the vacuum and the first couple of excited levels. This behavior is the telltale sign that the theory for $\lambda>\lambda_c$ is in the phase of spontaneously broken $\bZ_2$ symmetry. We then expect that the theory at $\lambda=\lambda_c$ is conformal. This expectation will be further tested below.

It may be somewhat counterintuitive that the $\bZ_2$ symmetry breaks for a \emph{positive} value of $m^2$ (remember that we fixed $m=1$).
In fact, there is no paradox. The $m^2$ is a UV parameter defining the initial direction of the flow, while the breaking is an IR phenomenon.
As we flow from UV to IR, $m^2$ is renormalized and the effective squared mass may become negative.\footnote{See section \ref{sec:RG-phi4} for the RG equations for the Landau-Ginzburg flow.} In other words, we may imagine that a double-well potential is generated by quantum effects.  In this case there will be two degenerate vacua and all excitations above the vacua should be degenerate as well. The degeneracy would be exact in infinite volume. In finite volume we expect some mixing due to the potential barrier tunneling,\footnote{Such tunneling effects were for example studied in TCSA in Ref.~\cite{Bajnok:2000wm}.} so that the exact eigenstates are $\bZ_2$-even or $\bZ_2$-odd and split by a small amount (exponentially small for large volume). The mixing and the splitting are expected to become more important for higher energy states, for which the tunneling is not suppressed. All these intuitive expectations are confirmed by figure \ref{fig:phi4R2.5}. 

Another interesting feature of the spectra in figure~\ref{fig:phi4R2.5} is the absence of level crossing---eigenstates belonging to the same $\bZ_2$ sector don't cross. 
There are several values of $\lambda$ when a pair of same $\bZ_2$-parity eigenstates come close to each other, but then repel. This should be contrasted with the free massive flow spectra, which do show level crossings, reproduced by TCSA calculations. The difference stems from the fact that the $\phi^2$ flow is integrable, while the Landau-Ginzburg flow is not. This way of distinguishing integrable and non-integrable flows has long been noticed in the $d=2$ TCSA literature (see e.g.~\cite{Lassig:1990xy,Brandino:2010sv}), and here we are observing it in $d>2$.

\subsubsection{Mass gap as a function of $\lambda$ and determination of $\lambda_c$}
\label{eq:lambdac}

\begin{figure}[htbp]
\begin{center}
\includegraphics[scale=0.7]{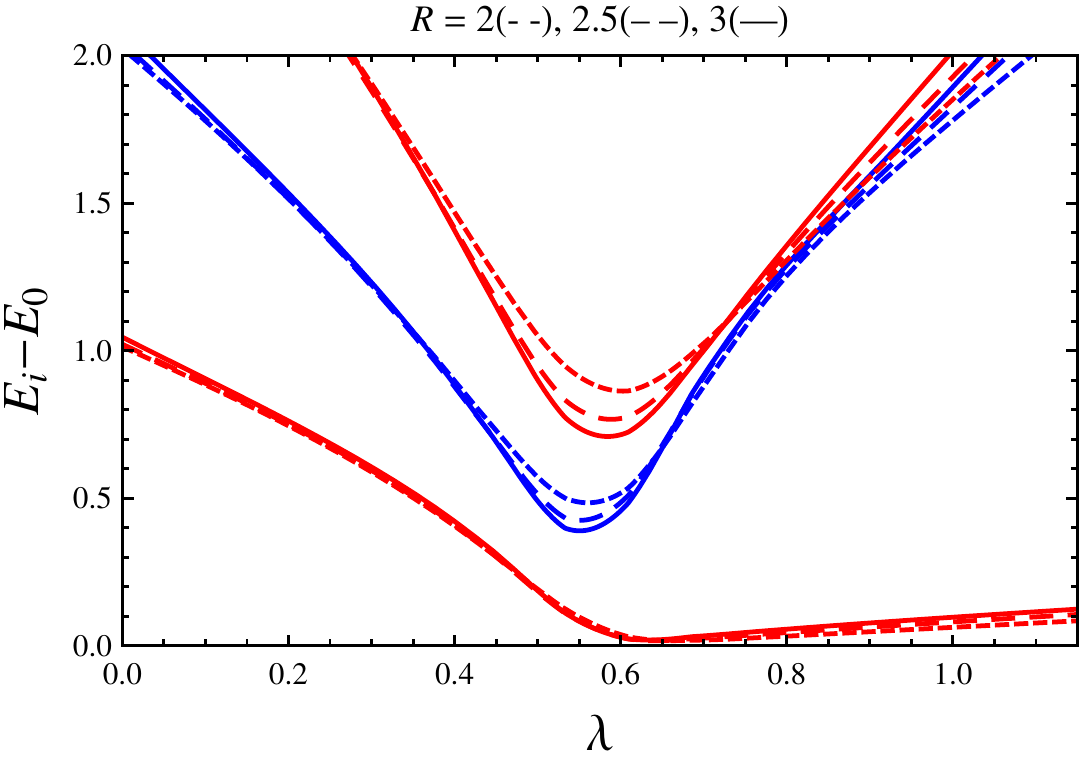}\quad
\caption{The spectrum of three lowest excitations as a function of $\lambda$ for three values of $R$: 2 (short dashed), 2.5 (longer dashed), 3 (solid). }
\label{fig:phi4low}
\end{center}
\end{figure}

We will now further test the expectation that the theory at $\lambda=\lambda_c$ is conformal. In figure \ref{fig:phi4low} we plot the low-lying spectrum of excitations (just the first three states) for $\lambda$ varying from 0 to 1.15 and for three values of $R=2,2.5,3$.
We see that the excitation energies are decreasing with $R$ for $\lambda$ near $\lambda_c$. This is especially noticeable for the second and third excited level. Away from $\lambda_c$ the spectrum is relatively stable with $R$.\footnote{Or even slightly increasing. 
We observed that this slight increase of the spectrum with $R$ is reduced when raising the cutoff, so it must be attributed to truncation effects.}

The decrease of the spectrum with $R$ at $\lambda=\lambda_c$ is what one should expect if the critical theory is conformal. Indeed, for a flow ending in a conformal fixed point the excitation energies should behave at large $R$ as $\Delta_i^{\rm IR}/R$ where $\Delta_i^{\rm IR}$ are the IR CFT operator dimensions. We will test this expectation in the next section.

\begin{figure}[htbp]
\begin{center}
\includegraphics[scale=0.7]{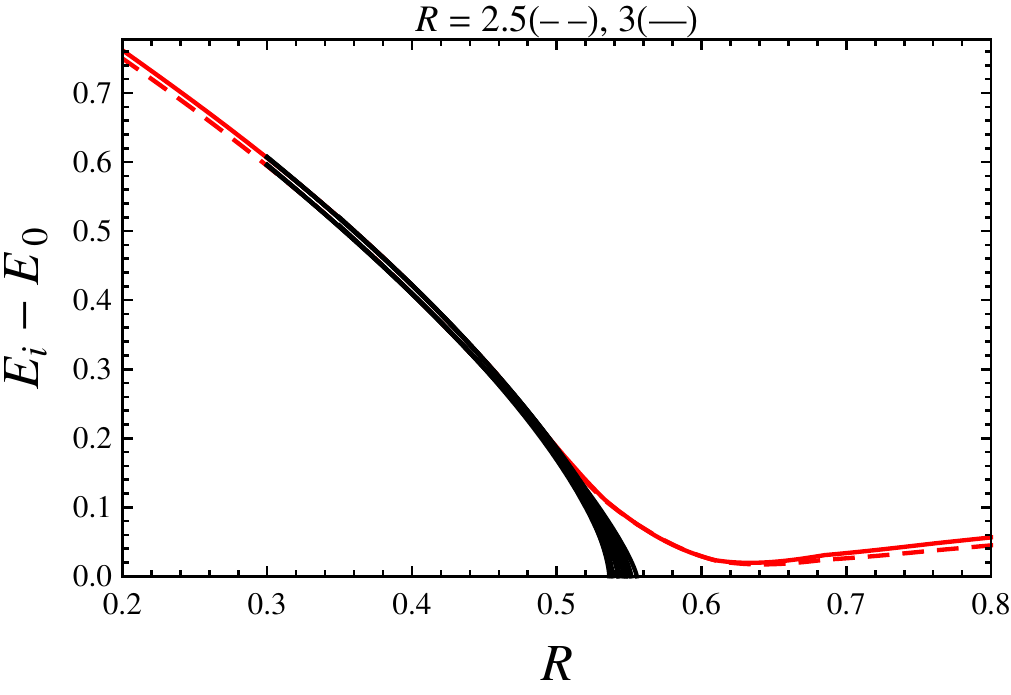}\quad\includegraphics[scale=0.58]{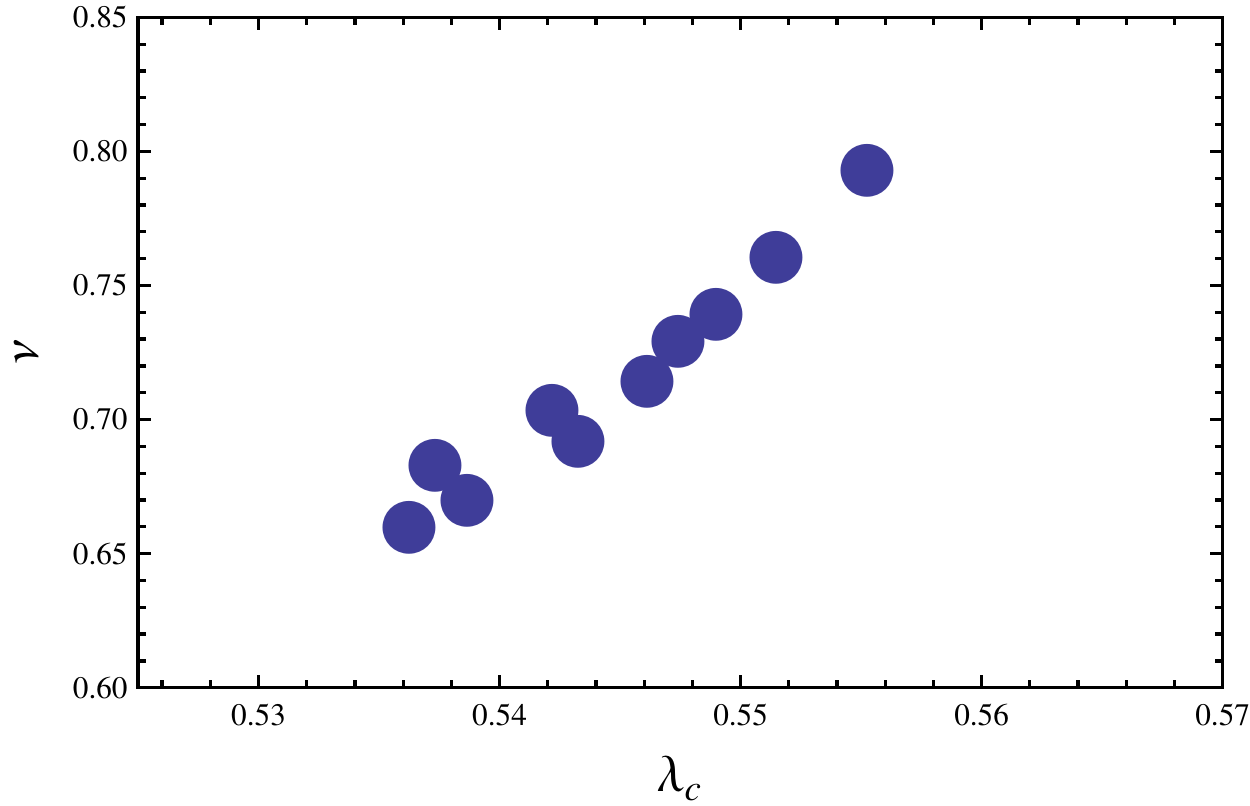}
\caption{Left panel: We fit the $R=2.5$ and $R=3$ mass gap in the region $[\lambda_1,\lambda_2]$, for several $\lambda_1$ and $\lambda_2$ values chosen within the ranges $0.3\ldots0.4$ and $0.45\ldots 0.5$, respectively. Right panel: a scatter plot for the $\lambda_c$ and $\nu$ parameters resulting from these fits.
 }
\label{fig:phi4lowfit}
\end{center}
\end{figure}

Let us now determine the critical value of the coupling $\lambda_c$ with some precision. According to the standard renormalization theory, the mass gap for $\lambda$ near $\lambda_c$ should depend on $\lambda$ as
\beq
M_{\rm gap}\approx C|\lambda-\lambda_c|^\nu\,,
\label{eq:mgap}
\eeq
where $\nu$ is a critical exponent,\footnote{This critical exponent $\nu$ should not be confused with the shorthand $\nu \equiv (d-2)/2$ that was introduced in Eq.~\reef{eq:omegal}.} which in the case at hand is related to the dimension of $\eps$---the first $\bZ_2$-even scalar operator at the Wilson-Fisher fixed point:
\beq
\nu=1/(d-\Delta_\eps)\,.
\eeq
In our case the mass gap is $E_1-E_0$ for $\lambda<\lambda_c$ and  $E_2-E_0$ for $\lambda>\lambda_c$. We will fit $E_1-E_0$ in the region $\lambda<\lambda_c$ to determine $\lambda_c$ and $\nu$. We exclude the region $\lambda>0.5$ from the fit since it is clearly affected by finite $R$ effects which smear out the expected power-law behavior. We also exclude the region $\lambda<0.3$ since Eq.~\reef{eq:mgap} is expected to be valid only in the $\lambda\to\lambda_c$ limit. We thus perform the fit in an interval $[\lambda_1,\lambda_2]$, and to estimate the systematic uncertainty we vary $\lambda_1$ between $0.3$ and $0.4$ and $\lambda_2$ between $0.45$ and $0.5$. This gives the following rough estimates for the critical coupling and the exponent $\nu$ (see figure~\ref{fig:phi4lowfit}):
\beq
\lambda_c=0.535\text{ - }0.555\,,\quad \nu=0.65\text{ - }0.8\,,
\eeq
with a positive correlation between $\lambda_c$ and $\nu$.

We will now compare our determination of $\nu$ with the results by other approaches. The dimension $\Delta_\eps$ for $d=2.5$ dimensions can be extracted from the Borel-resummed epsilon-expansion series 
\cite{LeGuillou:1987ph}. 
It can also be determined from the conformal bootstrap under the assumption that the Wilson-Fisher fixed point lives at a kink in the region of the $(\Delta_\sigma,\Delta_\eps)$ plane \cite{El-Showk:2013nia}. The latter analysis was done under the assumption that the Wilson-Fisher fixed point in fractional dimensions is unitary, which as we now know is not quite true. However, as noticed in section \ref{sec:nonun}, 
a small fraction of high-dimension negative-norm states should not have strong influence on the conformal bootstrap predictions. This probably explains why \cite{El-Showk:2013nia} found no disagreement with the results of \cite{LeGuillou:1987ph}. Both analyses predict:
\beq
\Delta_\eps \approx 1.175\qquad (d=2.5)\,,
\label{eq:deltaeps}
\eeq
which gives a value $\nu \approx 0.755$, close to the upper end of the confidence interval for $\nu$ determined by our fitting procedure above. Assuming this precise value of $\nu$ and repeating the fits leads to a somewhat more accurate determination of the critical coupling:
\beq
\lambda_c\approx 0.55\text{ - }0.56\,.
\eeq

\subsubsection{Spectrum for a fixed $\lambda$ and varying $R$}
\label{sec:phi4-fixedlambda}
\begin{figure}[htbp]
\begin{center}
\includegraphics[scale=0.7]{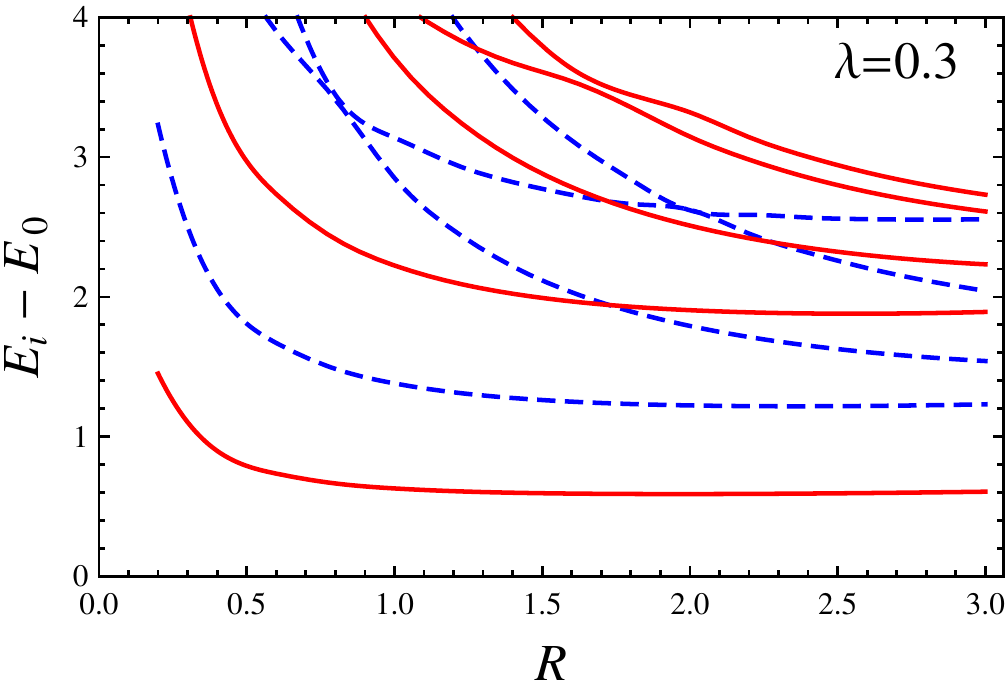}\quad \includegraphics[scale=0.7]{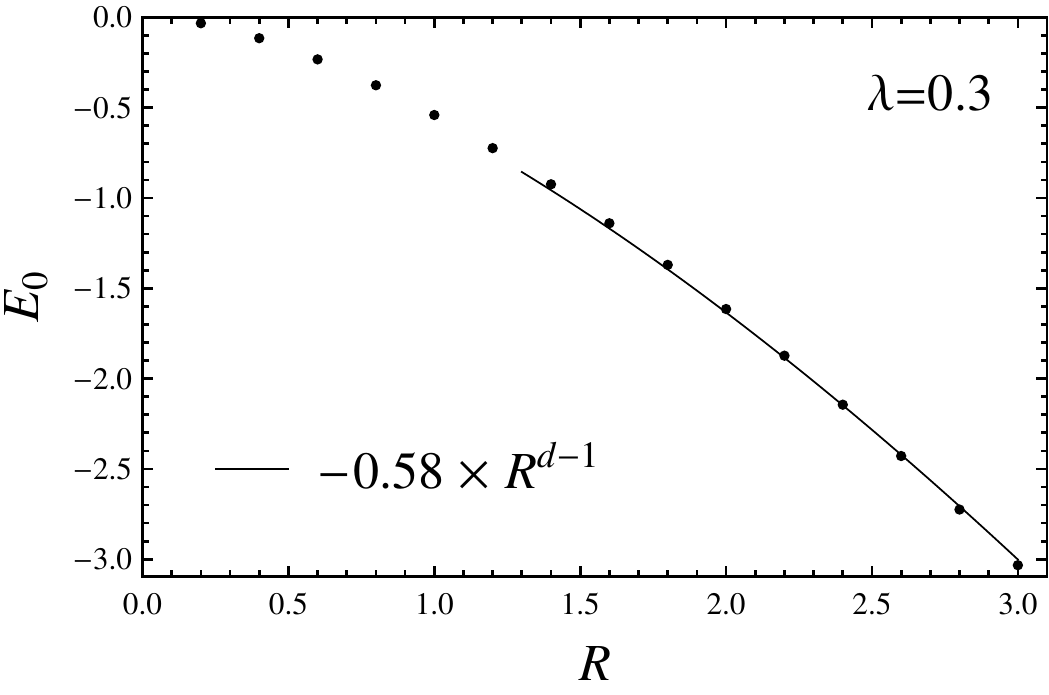}
\\[3pt]
\includegraphics[scale=0.7]{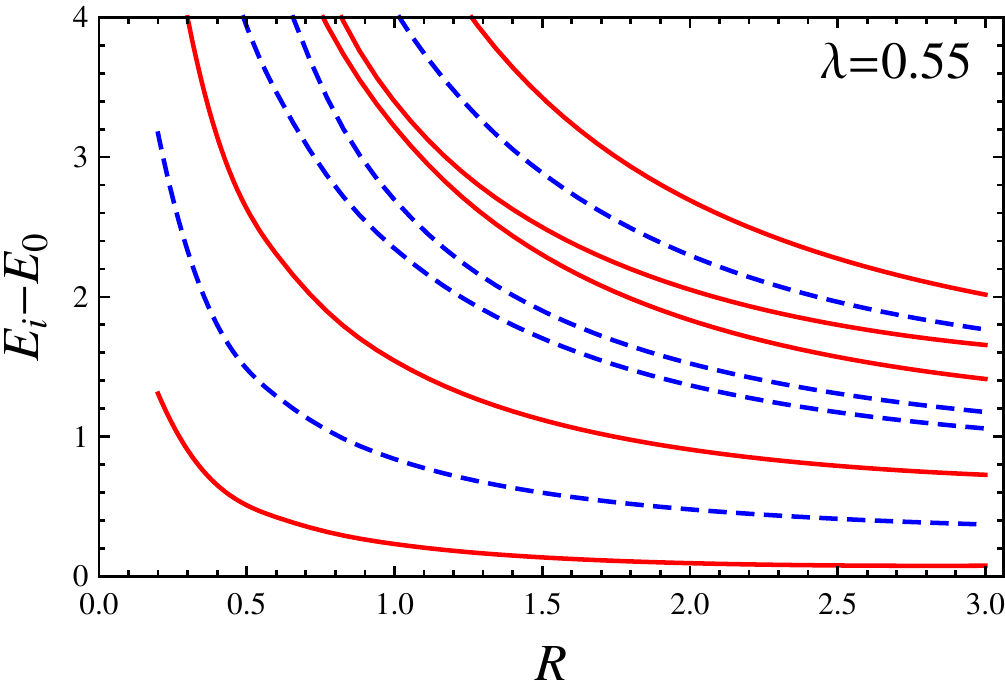}\quad \includegraphics[scale=0.7]{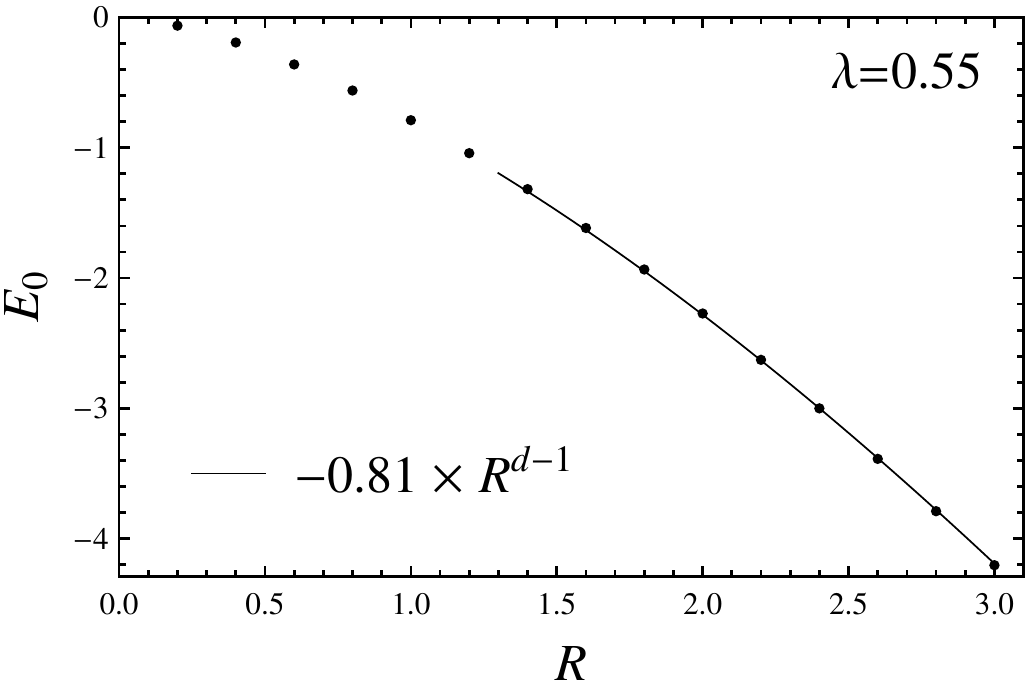}
\\[3pt]
\includegraphics[scale=0.7]{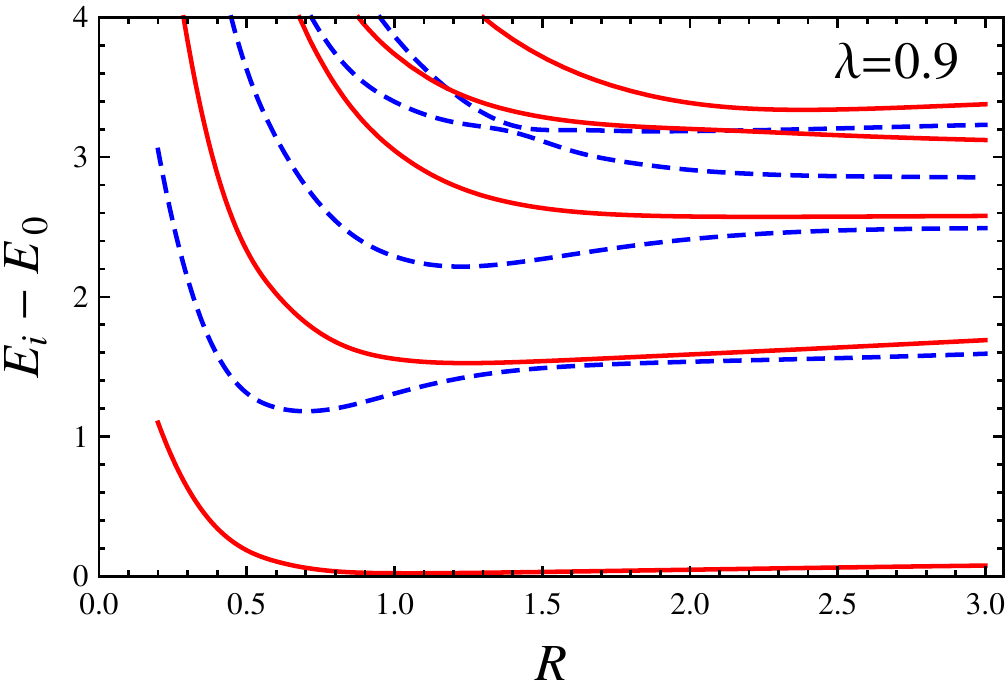}\quad \includegraphics[scale=0.7]{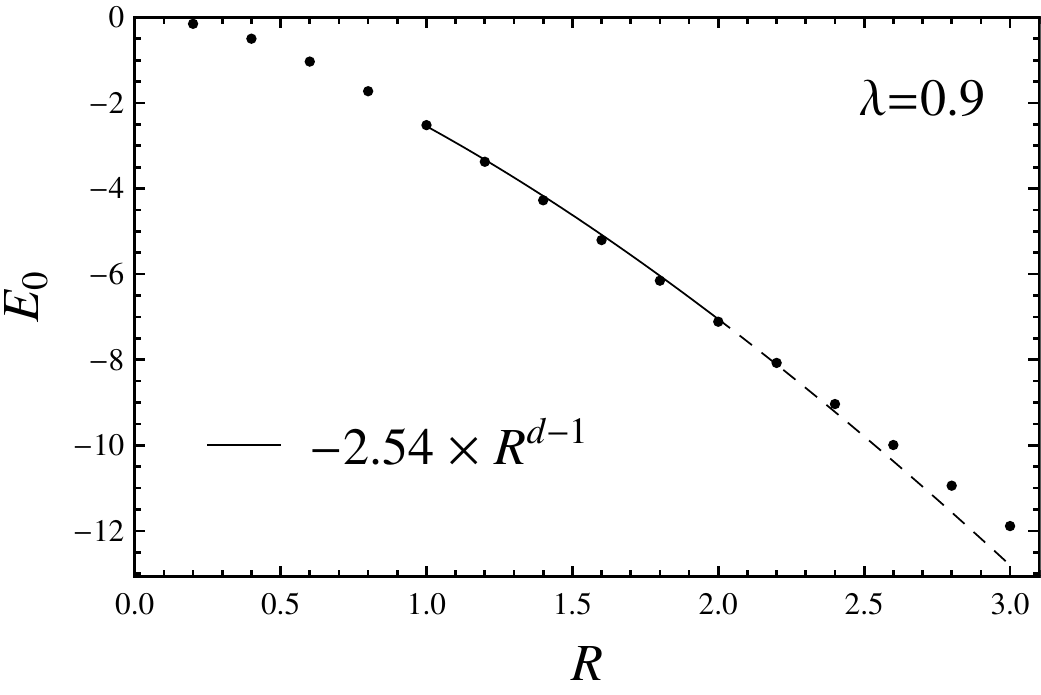}
\caption{Left panels: The spectrum of excitations as a function of $R$ for three values of the coupling: $\lambda=0.3,0.55,0.9.$ Solid red (dashed blue): $\bZ_2$-odd (-even) states. Right panels: The ground state energy for the same couplings. Dots: numerical data. Black curves: fits of the data by $const.R^{d-1}$ in the range $R=1.4\text{ - } 3$ ($R=1\text{ - } 2$ for $\lambda=0.9$).}
\label{fig:phi4lambda}
\end{center}
\end{figure}

We next present how the spectrum depends on $R$ for a fixed value of $\lambda$ (figure~\ref{fig:phi4lambda}). We pick three representative values of the quartic: 
$\lambda=0.3$ for the $\bZ_2$-preserving phase, $\lambda=0.55$ near the presumed critical point, and $\lambda=0.9$ in the $\bZ_2$-breaking phase. We will now comment upon what we see in those plots, first for the ground state energy, and then for the massive spectra.

\noindent{\bf Ground state energy}
\nopagebreak

\noindent The ground state energy is expected to grow for large $R$ as a constant times $R^{d-1}$, corresponding to a finite energy density (cosmological constant) induced by the RG flow. This behavior is clearly visible in the data for $\lambda=0.3,0.55$, while for $\lambda=0.9$ the fit is not so good and there are significant deviations for $R\gtrsim 2$. These deviations decrease with $\Delta_{\max}$ and are thus due to truncation effects. Jumping a bit ahead, notice that there are no comparably flagrant deviations in the massive excitation spectrum for $\lambda=0.9$ and large $R$. This is because the largest truncation effects are expected in the coefficient of the unit operator, which has the smallest possible dimension (0), and the unit operator affects the ground state energy but not the spectrum.

\noindent{\bf Excitations for $\lambda=0.3$} 
\nopagebreak

\noindent\nopagebreak Since the energies are observed to tend to finite nonzero limits for $R\to\infty$, we conclude that the IR theory is massive.
The lightest $\bZ_2$-odd state $E_1$ is a scalar particle of mass 
\beq
M=\lim_{R\to\infty} (E_1-E_0)\sim 0.6\,,
\eeq 
The next two excitations, belonging to the $\bZ_2$-even sector, are readily interpreted as two-particle states. The former, of mass $\approx 2M$, must have both particles at rest, while in the latter the particles must be in relative motion with respect to each other, with total angular momentum zero. Higher up, we observe a state of three particles at rest, of mass $\approx 3M$, and orbital excitations thereof. 

The appearance of this hierarchy of states, quantized in units of the lowest excitation, is a nontrivial consistency test on the results. It is also a prediction for the absence of bound states. \emph{At weak coupling}, the two-particle interaction is repulsive in the $\bZ_2$-symmetric phase of the $\phi^4$ theory,
so we wouldn't expect bound states. Our results show that this conclusion remains valid at strong(er) coupling.
Notice that the physical mass $M$ is significantly less than the bare mass $m$, so that the theory we are examining is presumably moderately to strongly coupled. 

It is interesting to study the rate with which the excitation energies approach their infinite volume limits. Focussing first on the lowest excitation, the leading correction is expected to arise from coupling to curvature and scale as $1/R^2$:
\beq
\label{eq:E1E0}
E_1-E_0=M+\Delta M_{\rm curv}+\ldots,\quad \Delta M_{\rm curv}=\frac{A (d-2)^2}{ 8 M R^2}\,,
\eeq  
where $A$ is a (theory-dependent) constant. Indeed, when the theory is put in a weakly curved background, it should be possible to describe corrections to the lightest state energy by an effective Lagrangian of the same form as the free massive scalar Lagrangian \reef{eq:freemassive} with $m\to M$ and an effective $\xi$ which will, in general, be different from $\xi_{\rm Weyl}$ in the UV. Then we get \reef{eq:E1E0} with $A=\xi/\xi_{\rm Weyl}$.\footnote{Notice that for $d=2$ the curvature vanishes, and the modification of the mass spectrum is entirely due to boundary conditions. The leading correction in this case is exponentially small \cite{Luscher:1985dn}:
$E_1-E_0=M+O( e^{-\sqrt{3}/2 M L})$, $L=2\pi R$\,.
}

\begin{figure}[htbp]
\begin{center}
\includegraphics[scale=0.7]{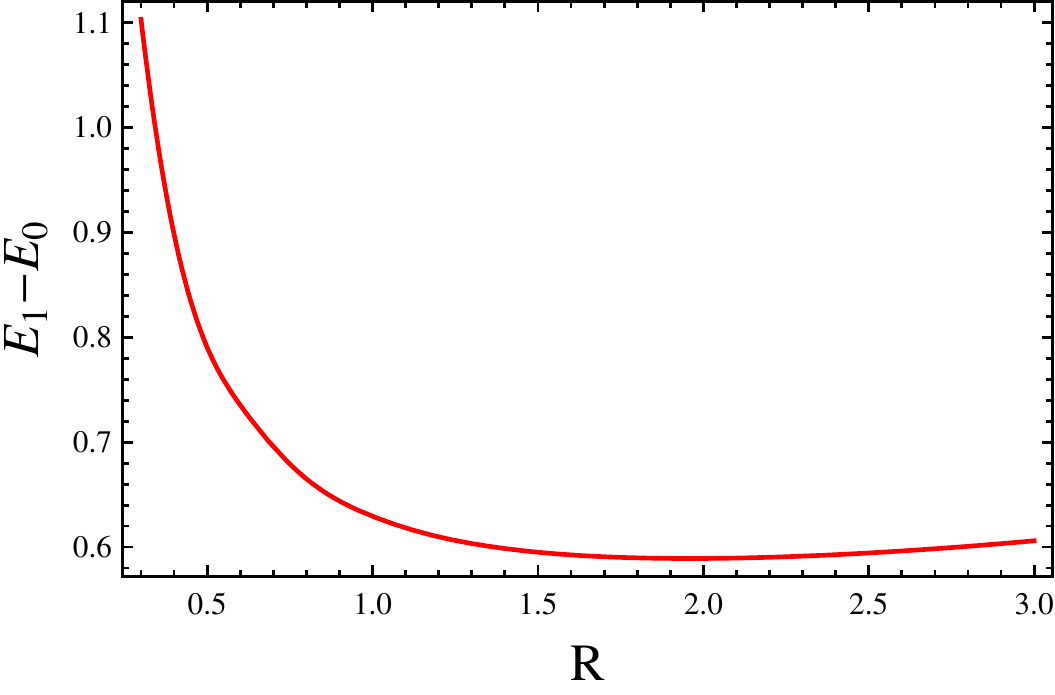}
\includegraphics[scale=0.7]{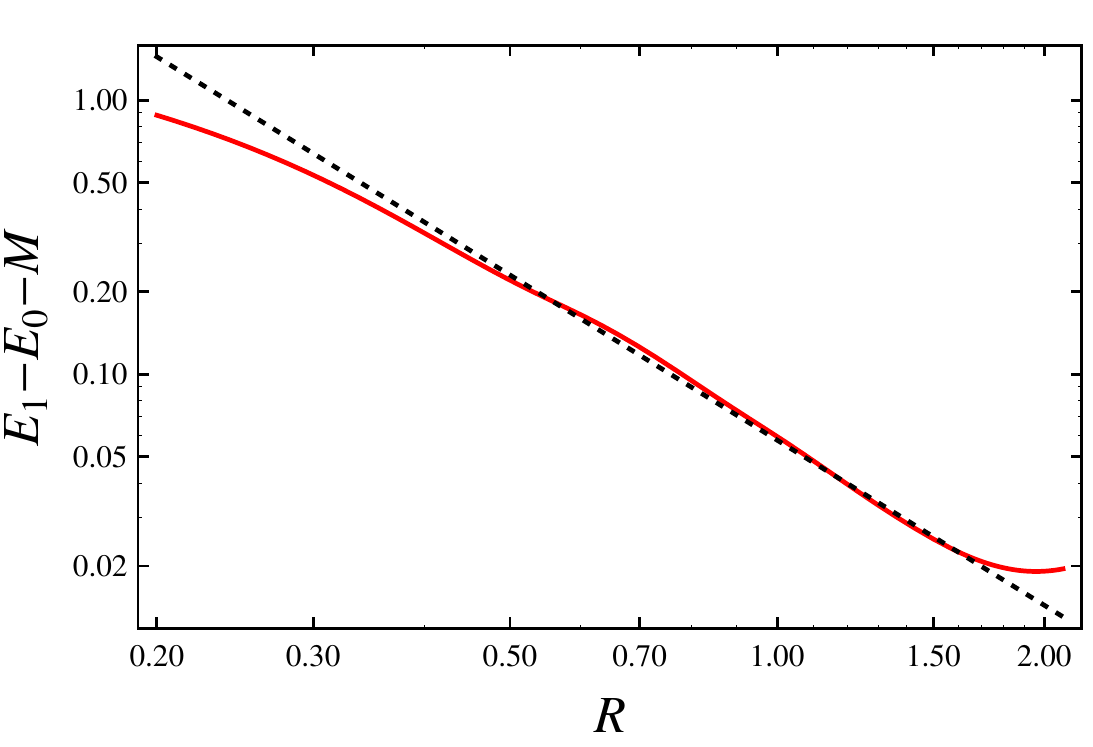}
\caption{Left panel: the lowest $\bZ_2$-odd excitation of the $\lambda=0.3$ spectrum. We see that the excitation energy decreases for $R\lesssim 2$ and then starts somewhat increasing, likely due to truncation effects. In the right panel we test Eq.~\reef{eq:E1E0} in the range $R\lesssim 2$. Red curve: $E_1-E_0-M$ (log-log scale). Dotted line: $\Delta M_{\rm curv}$. The parameters $M=0.57$, $A=1.05$ have been determined by performing a fit in the range $R=0.4\text{ - }1.5$. The agreement is good. }
\label{fig:phi4lambda03E1E0}
\end{center}
\end{figure}

In figure \ref{fig:phi4lambda03E1E0} we test Eq.~\reef{eq:E1E0} for the lowest excitation of the $\lambda=0.3$ spectrum. We see that it describes the large $R$ approach reasonably well up to $R\sim 1.5$, where the truncation effects apparently kick in and make the error to increase rather than decrease with $R$. Fitting the correction in the range $R=0.4\text{ - }1.5$ we determine $M\approx 0.57$, $A\approx 1.05$.

One could wonder why $A$ is so close to one in the case at hand. As already mentioned, we don't expect that $A$ should be universal. We will encounter $A<0$ below in the $\bZ_2$-broken phase. Moreover, the coupling to curvature will be suppressed if the state in question is a pseudo-Goldstone boson, as may happen for more complicated flows with a continuous global symmetry. So, for the pion in QCD we expect $A\sim (m_\pi/\Lambda_{\rm QCD})^2\ll 1$.

\begin{figure}[htbp]
\begin{center}
\includegraphics[scale=0.7]{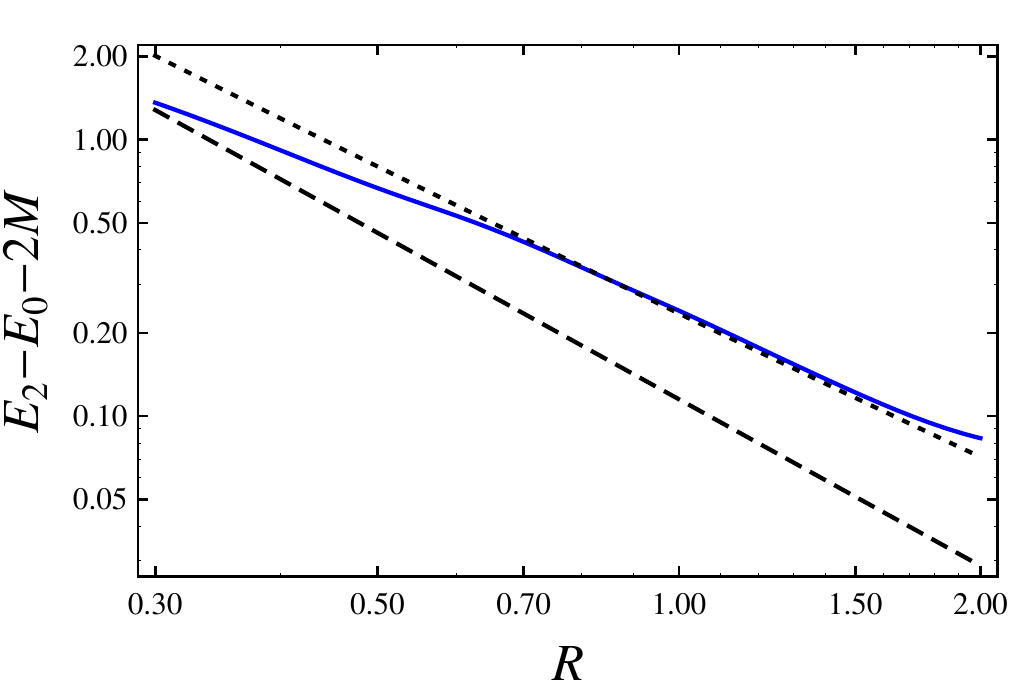}
\caption{Blue curve: $E_2-E_0-2M$ for the `two-particles at rest' state in the  $\lambda=0.3$ spectrum (log-log scale). Dashed black line: $2\times \Delta M_{\rm curv}$. We use the best fit value $M=0.57$, $A=1.05$. Dotted black: $2\times \Delta M_{\rm curv}$ plus $\Delta M_{\rm scat}=0.12/R^{d-1}$.}
\label{fig:phi4lambda03E2E02M}
\end{center}
\end{figure}

We next discuss the rate of approach for the two-particle states, starting with the two-particle state at rest. The energy of this state can be approximated as
\beq
E_2-E_0= 2(M+\Delta M_{\rm curv})+ \Delta M_{\rm scat}\,,
\label{eq:E2E0M}
\eeq
where the last correction is due to the interaction (scattering) between two particles put into a finite volume. Since the interaction is short-range, we expect the leading correction of this type to scale as the inverse volume of the box \cite{Luscher:1986pf}. In figure 
\ref{fig:phi4lambda03E2E02M} we plot $E_2-E_0-2M$ for the $\lambda=0.3$ spectrum. We see that the difference is not well described by the finite-volume correction $2\,\Delta M_{\rm curv}$ alone. A much better agreement can be obtained including a correction with the scaling $\propto 1/R^{d-1}$, as would be expected from a scattering correction. Notice that the sign of the so determined scattering correction is positive, corresponding to a repulsion between the constituent particles. Indeed, as we already mentioned above, at weak coupling in the unbroken phase, $\lambda \phi^4$ interaction is repulsive; here we see the same effect persisting at strong coupling. In principle, it should be possible to relate the size of the scattering correction to the scattering length, as was done for a flat torus by L\"uscher \cite{Luscher:1986pf}. In would be interesting to work out the corresponding theory for the sphere.

\begin{figure}[htbp]
\begin{center}
\includegraphics[scale=0.7]{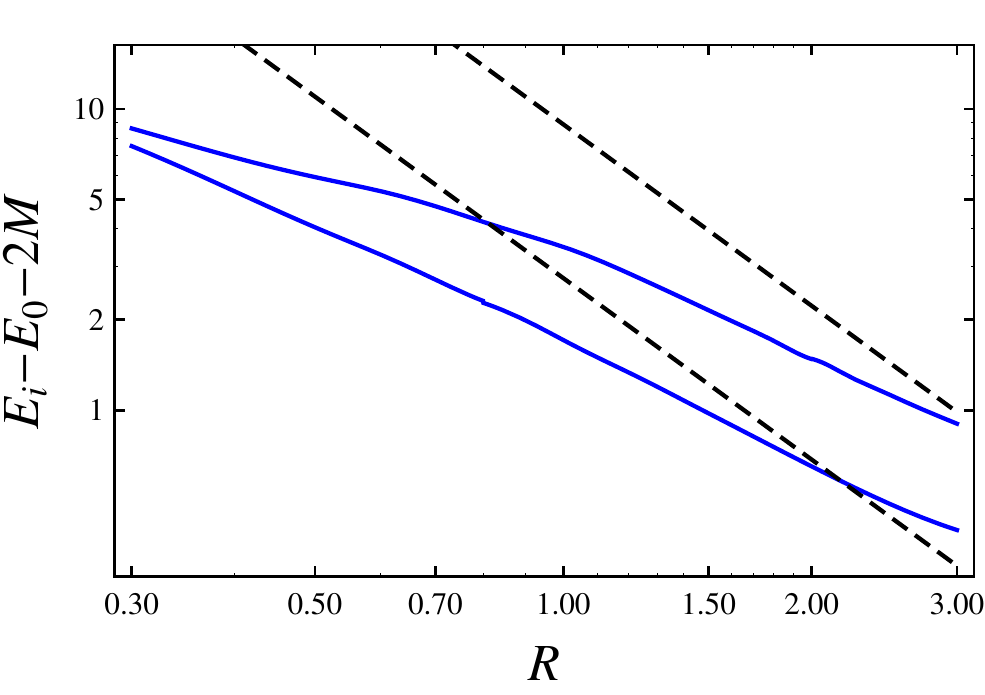}
\caption{Blue curves: $E_i-E_0-2M$ for the lowest two orbital excitations in the two-particle sector of the  $\lambda=0.3$ spectrum corresponding (log-log scale). We excised by hand a state which asymptotes to $4M$ and thus looks like a four-particle state at rest.  Dashed black lines: $2( \Delta M_{\rm curv}+\Delta M_l)$ for $l=1,2$. }
\label{fig:phi4lambda03EiE02M}
\end{center}
\end{figure}

Finally, in figure \ref{fig:phi4lambda03EiE02M} we plot the difference $E_i-E_0-2M$ for the lowest two orbital excitations in the two-particle sector, which should consist of two particles moving in the $l=1$ and $l=2$ angular momentum modes, combined to have the total angular momentum zero. Thus their finite volume mass correction should have an extra orbital term (see Eq.~\reef{eq:omegal})
\beq
2\times \Delta M_{l},\quad  \Delta M_{l}= l(l+d-2)/(2 M R^2)\,.
\eeq
As is clear from figure \ref{fig:phi4lambda03EiE02M}, $E_i-E_0-2M$ decrease way too slowly with $R$ to be described in the asymptotic region by just the sum of the curvature and orbital corrections. It must be that the difference is due to the scattering correction, although it looks hard to make a quantitative conclusion using the data in the $R<3$ region. 

\noindent{\bf Excitations for $\lambda=0.55$} 

\noindent A very different behavior presents itself in the spectrum dependence on $R$ for $\lambda=0.55$, which is close to the critical coupling. Instead of energies tending to finite limits, we see them all gradually decrease with $R$. 

\begin{figure}[htbp]
\begin{center}
\includegraphics[scale=0.2]{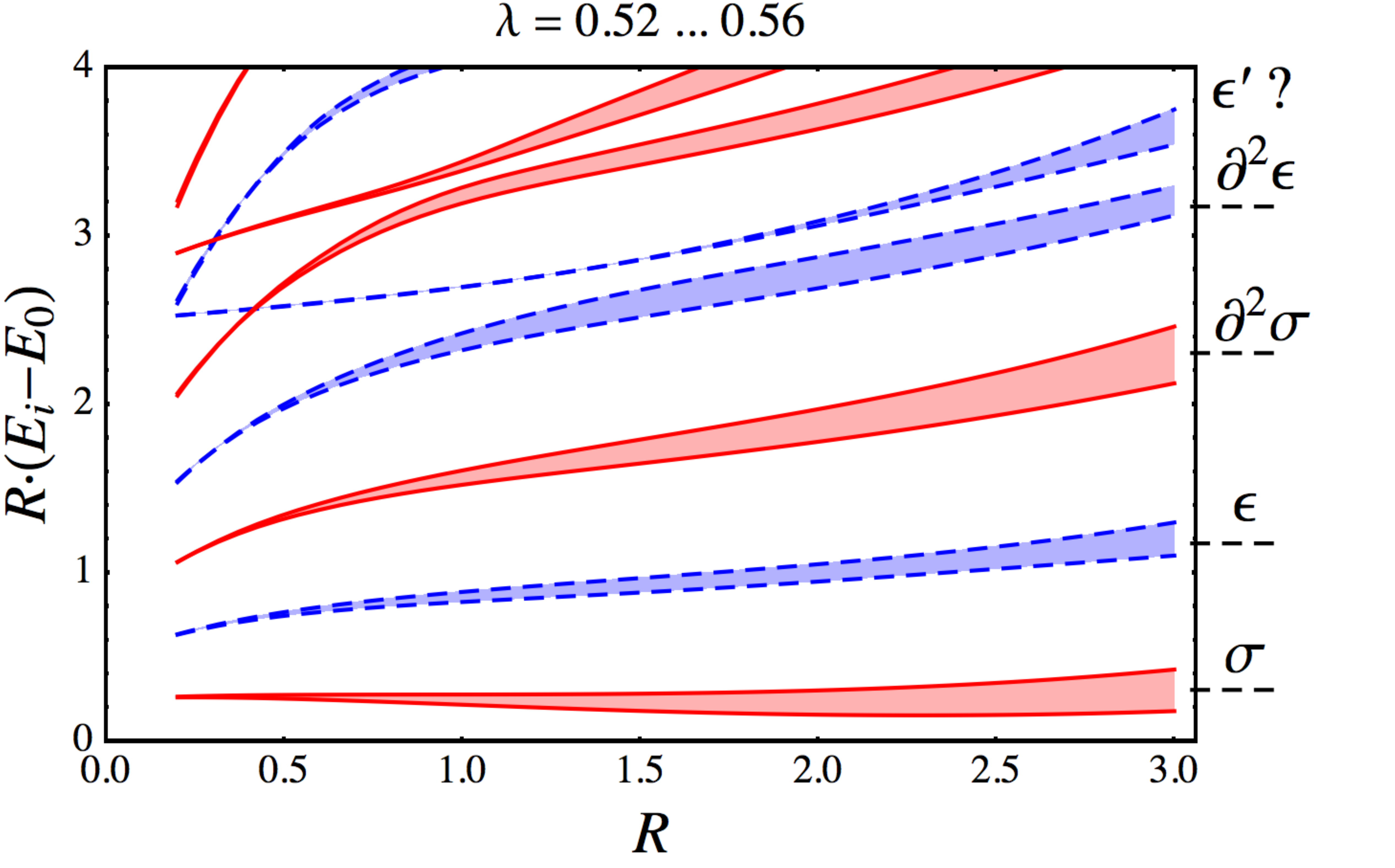}
\caption{$\bZ_2$-odd (red solid) and $\bZ_2$-even (blue dashed) excitation energies multiplied by $R$, as a function of $R$. The shaded regions show the variation when $\lambda$ varies in the range $0.53\ldots 0.56$ (when the coupling is increased all the excitation energies go down). On the right border of the plot we indicate the expected dimensions of the lowest-lying states of the Wilson-Fisher fixed point in $d=2.5$ dimensions (see the text). The curves must tend to the indicated finite limits. They do reach these limits for $R\approx3$, but would overshoot them (except for $\sigma$) for larger $R$. }
\label{fig:phi4delta}
\end{center}
\end{figure}

As already mentioned in section~\ref{eq:lambdac}, at $\lambda=\lambda_c$ we expect the excitation energies to scale at large $R$ as $\Delta_i^{\rm IR}/R$, where $\Delta_i^{\rm IR}$ are the IR CFT operator dimensions. To test this expectation, we plot in figure \ref{fig:phi4delta} the excitation energies times $R$. We vary $\lambda$ in the range $0.53\ldots 0.56$, roughly the range determined in section \ref{eq:lambdac} to contain the critical coupling.
Apart from the state $\eps$ in the $\bZ_2$-even sector, of dimension $\approx 1.175$ (see Eq.~\reef{eq:deltaeps}), we expect to see a $\bZ_2$-odd operator $\sigma$ of dimension $\approx 0.305$ (as extracted again from \cite{LeGuillou:1987ph,El-Showk:2013nia}). We also expect to see the states corresponding to operators $\del^2\eps$ and $\del^2\sigma$, of dimension two units higher. Finally, we may hope to see the next primary $\bZ_2$-even operator $\eps'$, whose dimension for $d=2.5$ is not precisely known but may be expected to lie between 3.5 and 4.\footnote{It's 4 in $d=2$ and in $d=4-\eps$, and $\approx 3.83$ in $d=3$ \cite{Guida:1998bx,Campostrini,Hasenbusch2010,El-Showk:2014dwa}.}

Interestingly, for $R\approx 3$ we can observe all of the above mentioned states in the spectrum in figure \ref{fig:phi4delta}, at the dimensions where they are supposed to be and with the right $\bZ_2$ quantum number. The agreement of theory and our numerical results remains imperfect in that the curves don't really approach finite limits very well. Perhaps one could claim that for the lowest two states $\sigma$ and $\eps$, whose variation with $R$ is not huge. However, the higher states definitely exhibit growth with $R$ and would overshoot the theoretical prediction for their dimension, were we to extend this plot to higher values of $R$. 
We hope that this issue will get resolved in the future by improving the accuracy of the method (see section \ref{sec:RG-phi4}).

\noindent{\bf Excitations for $\lambda=0.9$} 

\noindent Finally, we consider the spectrum for $\lambda=0.9$.\footnote{{\bf Note added:} It's interesting to compare the discussion below with section VI of the contemporaneous work \cite{Coser:2014lla} devoted to Landau-Ginzburg flows in $d=2$; see also footnote \ref{note:Mussardo}.} The first eye-catching feature of the spectrum is the approximate degeneracy of $\bZ_2$-even and odd states in the region of large $R$. This degeneracy is clearly visible for the first excitation, which becomes degenerate with the vacuum, and for two more pairs of states. The interpretation of this phenomenon was already discussed in section \ref{sec:fixedR}---it means that the $\bZ_2$-symmetry is spontaneously broken. In the infinite $R$ limit we expect a pair of degenerate vacua $\ket{0}_{L,R}$, each with its own tower of excitations $\ket{i}_{L,R}$. The spectrum is thus exactly doubly degenerate. For a finite large $R$, the height of the potential separating the two vacua is no longer infinite---it is expected to scale as the volume of space $R^{d-1}$. This allows tunneling and the eigenstates become the $\bZ_2$-even and odd linear combinations:
\beq
\ket{i,\pm}=\frac{1}{\sqrt{2}} (\ket{i}_{L}\pm \ket{i}_{R})\,,
\eeq
of energies $E_i\pm \delta_i$ where $\delta_i$ is the tunneling matrix element. As usual in quantum mechanics, we expect that the $\bZ_2$-even combination will have a smaller energy than the $\bZ_2$-odd one. This is confirmed by the spectrum in figure \ref{fig:phi4lambda}---in each of the three approximately degenerate pairs, it's the $\bZ_2$-even state which is the lower one.

The above tunneling argument seems to predict that the even/odd state pairs should be split roughly symmetrically with respect to the infinite volume limiting value. In fact, since the excitation energies are defined as $E_i-E_0$ and $E_0$ belongs to the $\bZ_2$-even sector, we expect the $\bZ_2$-even excitations to shift down by $(\delta_i-\delta_0)$, while the $\bZ_2$-odd ones to move up by $(\delta_i+\delta_0)$. Since the tunneling probability strongly depends on the energy, we expect $\delta_0\ll \delta_i$, and the shifts should be roughly symmetric. However, that's not what we see in the $\lambda=0.9$ plots in figure \ref{fig:phi4lambda}---it rather looks that the negative shift of the $\bZ_2$-even excitations is much larger than the positive shift of the $\bZ_2$-odd ones. For the first pair of massive excitations, it even looks like both the $\bZ_2$-odd and the $\bZ_2$-even state have a negative shift. 

The most natural explanation of this phenomenon is that we are forgetting the modification of the mass spectrum via coupling to curvature, see Eq.~\reef{eq:E1E0}. This effect goes as $1/R^2$ and for low-lying states should be larger than the splitting, which is exponentially small in the volume of sphere. The fact that both states in the first $\bZ_2$-odd/even pair have a negative shift can then be explained by taking $A$ negative in Eq.~\reef{eq:E1E0}.

In fact, it is not totally unexpected that $A$ should be negative for the lowest massive excitation at $\lambda=0.9$. The same occurs for the Landau-Ginzburg flow in the \emph{weakly coupled} part of the $\bZ_2$-broken phase, i.e.~for negative $m^2$ and a small quartic coupling. We did not study this part of the phase diagram numerically, but it's easy to understand what happens analytically. The full mass parameter of the UV theory, including the coupling to curvature, is $m^2+\xi_{\rm Weyl}{\rm Ric}$. Since $m^2<0$, we have to reexpand the Lagrangian around the true vacuum, and when we do this, the mass parameter picks up the usual $-2$ factor. We thus conclude that $\xi=-2\xi_{\rm Weyl}$, giving $A=-2$ at weak coupling in the $\bZ_2$ broken phase.

This finishes the discussion of splittings at finite $R$. Next, we would like to say a few words about the overall structure of the massive spectrum at large $R$. We identify the two near-degenerate pairs of even/odd states with two massive excitations, of mass
\beq
M_1\approx 1.6\,,\qquad M_2\approx 2.5\,.
\eeq
A very interesting feature of this spectrum is that $M_2<2 M_1$. This is unlike the spectrum at $\lambda=0.3$, which was neatly quantized in the units of the lightest excitation. In the situation at hand, the state of mass $M_2$ should probably be interpreted as a bound state of two $M_1$ particles. 

The appearance of such bound states was found long ago, and their masses measured, in the lattice simulations of the broken phase of the Ising model and of the $\phi^4$ theory in $d=3$ dimensions \cite{Caselle:1999tm}. In the weakly coupled regime, the existence of these states follows from the fact that the $\lambda \phi^4$ interaction becomes attractive in the broken phase, the cubic exchange diagrams overwhelming the repulsive effect of the contact term interaction \cite{Caselle:2000yx}. Their binding energy, exponentially small at weak coupling, is known in the leading and first subleading exponential approximation \cite{Caselle:2000yx,Caselle:2001im}. Apparently, here we are observing the same effect in $d=2.5$ dimensions and at strong coupling.

\subsection{Renormalization Details}
\label{sec:RG-phi4}
The renormalization in the Landau-Ginzburg flow is determined through the leading OPEs of the deforming operators in \eqref{eq:lgpotential}, given by
\beq
\label{eq:phi2phi4opes}
\begin{split}
\phi^2(x)\times \phi^2(0) &=  \frac{2\, \Nd^2}{|x|^{h_{220}}}\unit +\ldots\,,\\
\phi^4(x)\times \phi^4(0) &=  \frac{24\, \Nd^4}{|x|^{h_{440}}}\unit +\frac{96\, \Nd^3}{|x|^{h_{442}}}\phi^2+\frac{72\, \Nd^2}{|x|^{h_{444}}} \phi^4+\ldots\,,\\
\phi^2(x)\times \phi^4(0) &=  \frac{12\, \Nd^2}{|x|^{h_{422}}}\phi^2+\ldots\,,
\end{split}
\eeq
where $h_{ijk}=\Delta(\phi^i)+\Delta(\phi^j)-\Delta(\phi^k)=(i+j-k)\nu$. We have $\nu=1/4$ in $d=2.5$.

In the RHS of \eqref{eq:phi2phi4opes} we omitted the operators whose associated function $B(h)$ vanishes because of the remark in footnote \ref{note:nm}. All the retained operators have nonzero $B(h)$ and will be relevant for the renormalization. The effect of $\phi^2$ and $\phi^4$ in the RHS will be to make the couplings $\lambda$ and $m^2$ non-trivial functions of the cutoff. We are in a position to use the RG-improved formalism of section \ref{sec:RG-improv}.

A crucial ingredient in the renormalization is the relation between the above OPEs and the asymptotics of the matrix $(M_n)^i{}_{j}$ in \eqref{eq:cc1}. For the $\phi^2$ flow we were able to determine the asymptotics of this matrix exactly including the discrete structure, see equation \eqref{eq:Mnij}. In the future, such exact asymptotics may be also worked out with the $\phi^4$ coupling switched on, see the end of Appendix \ref{sec:phi2tail} for a discussion, although the task looks more challenging. In this work, we will use the continuum approximation \eqref{eq:pl}. We checked the accuracy of this approximation for many choices of external states $i,j$ against the exact expression for $M_n$ within the range $\Delta_n\le \Delta_{\max}$ where we know $V$ and can compute $M_n$ numerically. These checks convinced us that the approximation is adequate. 

\begin{figure}[htbp]
\begin{center}
\includegraphics[scale=0.9]{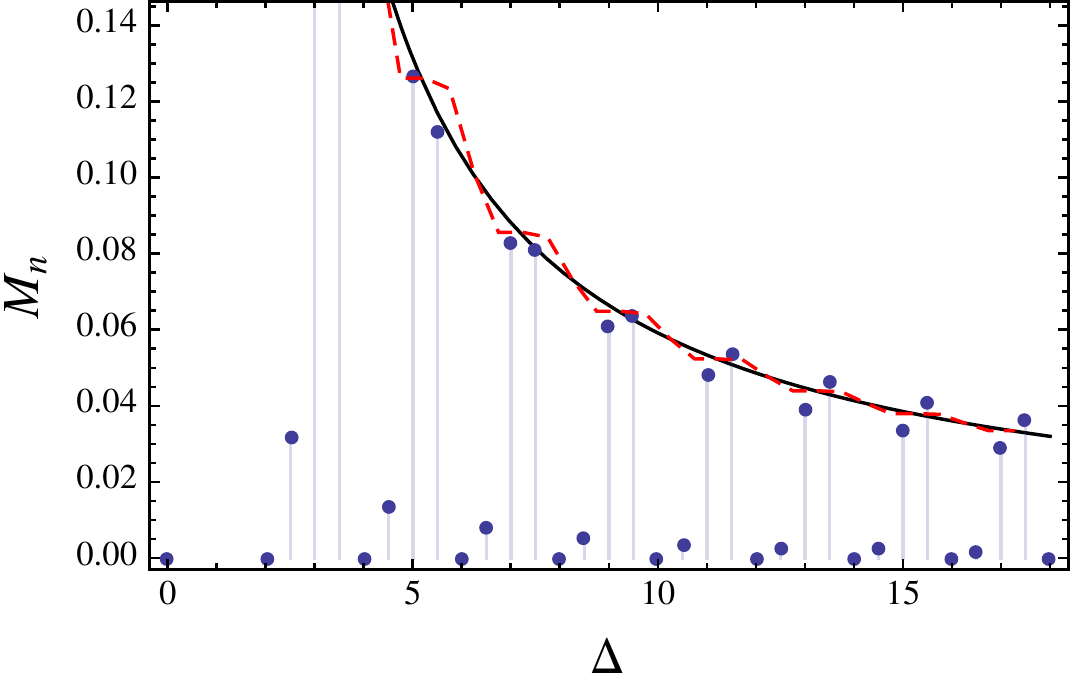}\\[3pt]
\caption{The behavior of $(M_n)^i{}_j$ for the $\phi^4$ deformation and one particular choice of $i,j$ (see text). Exact values are given in blue (isolated dots), a moving average in red (dashed line), and our continuum approximation in black (smooth curve).}
\label{fig:phi4-asympt}
\end{center}
\end{figure}

One such check is shown in figure~\ref{fig:phi4-asympt}, where we plot both the exact and the approximate behavior of $(M_n)^i{}_{j}$ for $\calV = \NO{\phi^4}$ and for $\calO_i = \calO_j = \NO{\phi^2}$. The blue dots represent the individual values (notice that $M_{n}$ is nonzero only for half-integer $\Delta_n$). The red dashed line shows moving average of these values within the interval $[\Delta-1,\Delta+1]$. The solid black line is our approximation as given in \eqref{eq:pl}, including the contributions of all three leading operators in the $\phi^4\times\phi^4$ OPE shown in \reef{eq:phi2phi4opes}. 
We see that the agreement between the moving average and the approximation becomes very good at $\Delta\sim 17$, which is also the cutoff we used in this study. The agreement for other choices of $\calO_i$ and $\calO_j$ is similarly good.

From the OPEs \eqref{eq:phi2phi4opes} we can directly generate the RG equations discussed in section \ref{sec:RG-improv} for the couplings of the local operators. They are given by:
\beq
\label{eq:rgflowlocalphi4}
\begin{split}
\frac{\delta g_0 (\Lambda)}{\delta \Lambda} &= \frac {\tilde f_{220}\, g_2^2 (\Lambda)}{\Lambda^{d-h_{220}}(\Lambda - E_r)} + 
\frac {\tilde f_{440}\, g_4^2 (\Lambda) }{\Lambda^{d-h_{440}}(\Lambda - E_r)}\,,\\
\frac{\delta g_2 (\Lambda)}{\delta \Lambda} &= \frac {\tilde f_{442}\, g_4^2 (\Lambda)}{\Lambda^{d-h_{442}}(\Lambda - E_r)} + 
\frac {2 \tilde f_{422}\, g_4 (\Lambda) g_2(\Lambda)}{\Lambda^{d-h_{422}}(\Lambda - E_r)}\,,\\
\frac{\delta g_4 (\Lambda)}{\delta \Lambda} &= \frac {\tilde f_{444}\, g_4^2 (\Lambda)}{\Lambda^{d-h_{444}}(\Lambda - E_r)}\,,
\end{split}
\eeq
where we denoted by $g_0$, $g_2$ and $g_4$ the coupling associated to $\unit$, $\NO{\phi^2}$, and $\NO{\phi^4}$. We also introduced
\beq
\tilde f_{abc} = f_{abc} B(h_{abc})
\eeq
with the OPE coefficients $f_{abc}$ given in \eqref{eq:phi2phi4opes} and $B(h)$ was defined in \eqref{eq:Bh}. The renormalized couplings are then found by integrating these equations numerically from $\Lambda=\infty$ to the desired value of the cutoff $\Lambda = \LUV=\Delta_{\max}/R$. We impose boundary conditions at infinity such that $g_0 = 0$ and $g_2$ and $g_4$ are given by their bare UV values:
\beq
g_4(\infty) = \lambda\,,\qquad g_2(\infty) = \half m^2\,.
\eeq 

As we explained in section \ref{sec:RG-improv}, the above RG equations depend on a reference energy $E_r$. In our study we found it convenient to choose $E_r$ to be around the energy of the first excited state in the $\bZ_2$ even sector. An estimate for this energy was obtained by extrapolating the earlier obtained results for lower values of the radius or the coupling, or by performing a quick computation with a smaller $\Delta_{\max}$. 

We also discussed in section \ref{sec:RG-improv} the subleading dependence on $(\Delta_i + \Delta_j)/R$ and on $\bar E - E_r$. This dependence is taken into account by adding to the correction Hamiltonian the additional non-local terms given in \eqref{eq:nonlocal}. Their coefficients, which we denote as $g_i^{(\HCFT)}$ and $g_i^{(H)}$, are determined by solving separate RG flow equations. For example, for $i=0$ we have
\beq
\begin{split}
\frac{\delta g_0^{(\HCFT)} (\Lambda)}{\delta \Lambda} &= \frac {(d-h_{220})\tilde f_{220}\, g_2^2 (\Lambda)}{\Lambda^{d-h_{220} + 1}(\Lambda - E_r)} + 
\frac {(d-h_{440})\tilde f_{440}\, g_4^2 (\Lambda) }{\Lambda^{d-h_{440}+1}(\Lambda - E_r)}\,,\\
\frac{\delta g_0^{(H)} (\Lambda)}{\delta \Lambda} &= \frac {\tilde f_{220}\, g_2^2 (\Lambda)}{\Lambda^{d-h_{220}}(\Lambda - E_r)^2} + 
\frac {\tilde f_{440}\, g_4^2 (\Lambda)}{\Lambda^{d-h_{440}}(\Lambda - E_r)^2}\,.
\end{split}
\eeq
The equations for other $i$ are determined by modifying the corresponding equation in \eqref{eq:rgflowlocalphi4} in a similar manner. We integrate these equations with boundary conditions zero at infinity. Notice that we ignore the backreaction of these terms in the sense that they do not appear on the right-hand sides of the flow equations.\footnote{For $g_0^{(\HCFT)}$ and $g_0^{(H)}$, which multiply local terms in the Hamiltonian, it would be possible to incorporate such a backreaction rather easily. At every step of RG one should factor out the modified coefficient of $\HCFT$, which leads to an overall rescaling of the remaining couplings. The product of all rescalings $Z$ should be stored separately to undo the rescaling at the end of the computation. This procedure resembles wavefunction renormalization in perturbative RG. We implemented it, but found the numerical effect of this improvement to be very small.}

\begin{figure}[htbp]
\begin{center}
\includegraphics[scale=0.7]{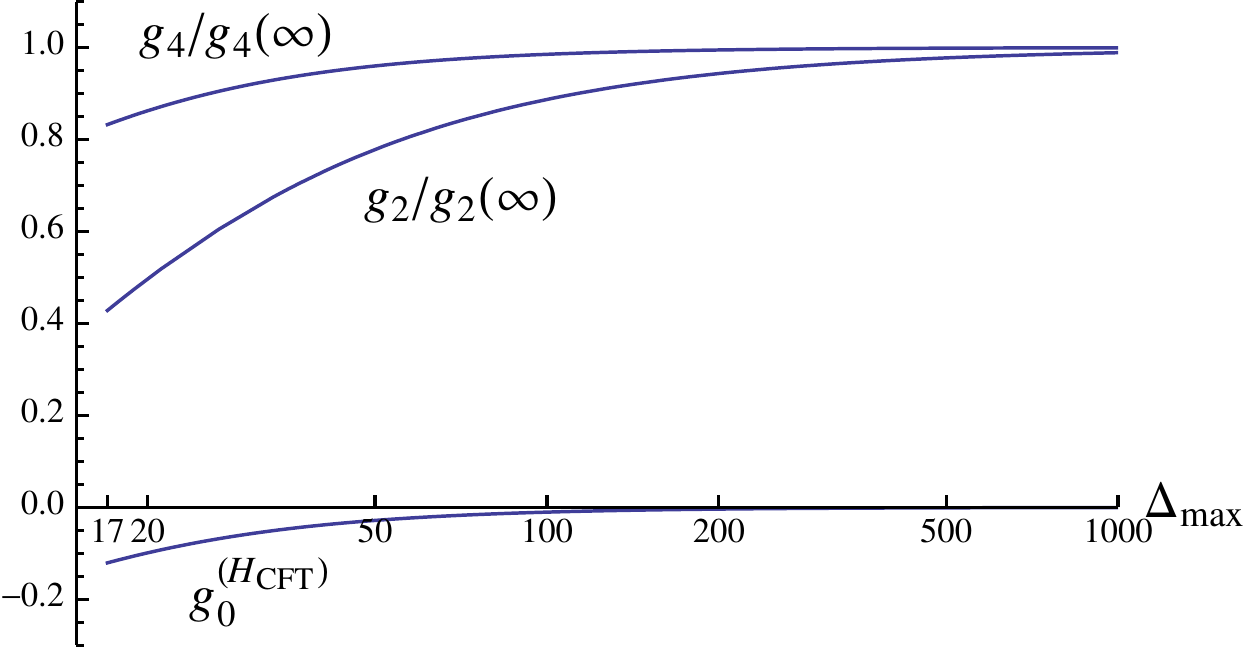}\\[3pt]
\caption{The relative change in a few of the running couplings as a function of the cutoff. In this example we set $R = 3$, and $\lambda = 0.7$ and $m^2 = 1$ in the UV. We used $\bar E_r = -6$.}
\label{fig:runningcouplings}
\end{center}
\end{figure}

In figure \ref{fig:runningcouplings} we present an example of the flow of several couplings. We see that $g_2$ receives substantial corrections, demonstrating the need for the RG improvement. On the other hand, we see that the relative change in $g_4$ is small, and that $g_0^{(\HCFT)}$ remains small compared to 1 (coefficient of $\HCFT$ in the bare Hamiltonian) throughout the flow.

With the correction terms described up to now, the results we obtained already looked reasonable. We were however able to take into account one further correction, which turned out to give a noticeable improvement mainly for small values of $\lambda$. Namely, we constructed a new nonlocal counterterm which completely takes into account the dependence on $(\Delta_i + \Delta_j)/R$ and $(\bar E - E_r)$, beyond expanding to first order as above. The coefficient of this correction is computed by running the RG flow again (once more ignoring the backreaction of this nonlocal term) but \emph{separately} for each value of $\kappa=\half(\Delta_i + \Delta_j)/R$ and $\bar E$. In equations, this means that we update this extra nonlocal correction in each RG step as follows,
\begin{multline}
\frac{\delta \Delta H^{\rm nl}(\kappa,\bar E){}_{ij}}{\delta \Lambda} = \sum_{a,b,c} g_a(\Lambda) g_b(\Lambda)\tilde f_{abc} \left(\int _{S^{d-1}}\calV_c\right)_{ij}\nn\\
\times 
\left\{\frac{1}{[\Lambda-\kappa]^{d-h_{abc}}(\Lambda-\bar E)}
-\frac{1}{\Lambda^{d-h_{abc}}(\Lambda-E_r)}
-\frac{\bar E-E_r}{\Lambda^{d-h_{abc}}(\Lambda-E_r)^2}
-\frac{(d-h_{abc})\kappa}{\Lambda^{d-h_{abc}+1}(\Lambda-E_r)}\right\} \,.
\end{multline}
Inside curly brackets, we subtract the zeroth- and first-order terms in $\kappa$ and $\bar E-E_r$, since these terms were already taken into account above.

Let us recap. We integrate all the above RG equations from $\Lambda=\infty$ to $\Lambda = \LUV$ and obtain the correction terms. We divide them into four groups: 
\begin{itemize}
\item 
$\Delta H_{\rm loc}$ which reflects the change in all local couplings;
\item 
$\Delta H_{1}$ and $\Delta H_2$ which include the nonlocal terms proportional to $\HCFT.V_c+V_c.\HCFT$
and $(H-E_r).V_c+V_c.(H-E_r)$, respectively;
\item 
$\Delta H^{\rm nl}$.
\end{itemize}
It is now time for numerical diagonalization. We add $\Delta H_{\rm loc}$ and $\Delta H_1$ directly to the bare TCSA Hamiltonian, and diagonalize. Let's call the resulting eigenvalues and eigenvectors $E_{n}$ and $c_{n}$. In principle, we would also have liked to add $\Delta H_2$ before the diagonalization. Unfortunately, we found that doing this destabilizes the numerics. This instability must have its origin in the fact that the factor $(H-E_r)$ is not small for states of high energy, and even for states of low energy it's not manifestly small, being a difference of two separately large quantities. We therefore chose to add the effect of $\Delta H_2$ only after the numerical diagonalization. We found it necessary, and sufficient, to do this to the second order in $\Delta H_2$. The correction is computed by the usual Hamiltonian perturbation formula:
\beq
\Delta E_{n} = c_{n}.\Delta H_2. c_n + \sum_{m \neq n} \frac{(c_{n}.\Delta H_2.c_{m})(c_{m}.\Delta H_2.c_{n})}{E_n - E_m}\,.
\eeq
The sum over $m$ in the second-order term is rapidly convergent, and it's enough to sum over the first few eigenstates. Notice that one has to appropriately insert the right and left eigenvectors. This is not reflected in the notation but explained in detail in section \ref{sec:RG-phi2}.

Finally, we compute one last correction due to $\Delta H^{\rm nl}$, which turns out to be very small, so doing it to first order is sufficient:
\beq
(\Delta E_n)^{\rm nl} = c_{n}.\Delta H^{\rm nl}(E_n). c_n= (c_{n})_i (\Delta H^{\rm nl}(E_n))^i{}_j (c_n)^j\,.
\eeq
When evaluating this correction, we are supposed to set $\bar E$ in the definition of $\Delta H^{\rm nl}$ to the energy of the state we are correcting. Also recall that $\Delta H^{\rm nl}$ depends on $\kappa=\half(\Delta_i + \Delta_j)/R$, and this dependence comes into play when evaluating the scalar product. 

This completes the description of the renormalization procedure used to produce the plots in section \ref{sec:phi4-num}. As the above discussion shows, an efficient implementation of the leading-order truncation effects given in \eqref{eq:rgmaster} is subject to various subtleties, mostly due to the non-negligible presence of $(\Delta_i+\Delta_j)/R$ and $\bar E$ in the correction terms. In this exploratory paper we have not aimed to present a complete analysis of these effects. Instead, we discussed various recipes for dealing with them at a practical level. The details we provided should be sufficient to reproduce our results. In the future it would certainly be interesting to perform a more systematic study of all the subtleties. 

\subsection{Non-Unitarity and Complex Energy Levels}
\label{sec:phi4-complex}
\begin{figure}[htbp]
\begin{center}
\includegraphics[scale=1.0]{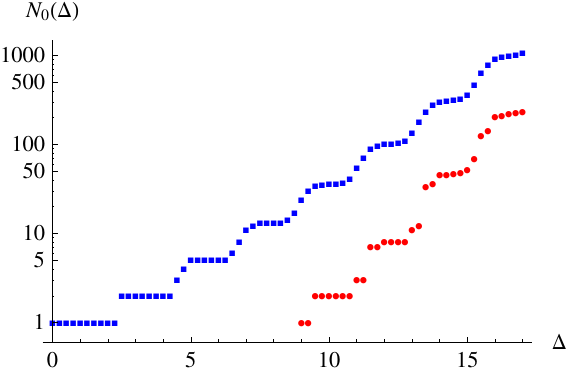}
\end{center}
\caption{The number of scalar $P$-even states in the Hilbert space of free massless scalar theory in $d=2.5$ on the cylinder.
Blue squares: all states. Red dots: negative-norm states.}
\label{fig:states25d}
\end{figure}

As we observed in section \ref{sec:nonun}, the free massless scalar theory in fractional $d$ is not unitary---its Hilbert space contains negative norm states. In $d=2.5$ the lowest negative-norm state occurs for $\Delta=9$. In figure \ref{fig:states25d} we show the total number of scalar states and the number of negative norm states as a function of $\Delta$.

What are the consequences of having these negative norm states? One expected consequence is that once we perturb the theory, we will get complex eigenvalues. The purely massive perturbation $\half m^2\phi^2$ is an exception, since in this case we expect that the spectrum agrees with the canonical quantization spectrum from section \ref{sec:can}, and thus is real.\footnote{In principle, numerical spectrum for a finite $\LUV$ could contain eigenvalues with small imaginary parts, approaching zero as $\LUV\to\infty$. However, in our numerical studies at $d=2.5$ we observed that even the truncated spectrum was real for the purely massive perturbation.}
What if we turn on $\lambda \phi^4$? As we saw in the previous sections, numerics indicate that the low-energy spectrum is still real. This may not be so surprising, since the negative norm states are secluded at high energies. So to see complex eigenvalues, we may expect to have to go to high energies. We will now present several computations which show that complex eigenvalues do occur.

Let us first of all examine the case of very small $R$. In this limit we can treat $m$ and $\lambda$ as perturbations, with dimensional couplings $m^2 R^2$ and $\lambda R^{d-4 \nu}$. The second coupling decreases less slowly as $R\to0$, and will dominate at very small $R$. The effects of the perturbation is to split the degenerate energy levels of the CFT Hamiltonian. The splittings are proportional to the eigenvalues of the perturbation diagonalized within each degenerate subspace. In high energy subspaces, which contain negative norm states, some of the eigenvalues may and do turn out to be complex. We find that this happens for the first time at $\Delta=11.5$, which is an 88 dimensional subspace with 7 negative norm states. We find that the matrix of the $\phi^4$ perturbation within this subspace has one pair of complex conjugate eigenvalues $\approx 1.85\pm 0.04 i$. This implies that for very small $R$ the energy levels will be complex.  

As a side remark, we note that it would be interesting to carry out a similar computation in $4-\eps$ dimensions. Free massless scalar theory in $4-\eps$ dimensions perturbed by $\lambda\phi^4$ flows to a weakly coupled Wilson-Fisher conformal fixed point. This is a short flow, and the energy levels at the IR fixed point, which are in one-to-one correspondence with the IR operator dimensions, are computable in perturbation theory. Once again, we expect that some of the IR operator dimensions will be complex. Likely this will happen already to first order in $\eps$, and it would be interesting to identify the first operator dimension for which this happens. The first null state in $d=4$ has dimension 15. To get a complex anomalous dimension to first order in $\eps$ we have to go to even higher $\Delta$. This computation is a bit more difficult to perform than the above computation in $d=2.5$, because in $d=4-\eps$ there are $O(\eps)$ splittings already in the unperturbed spectrum (like between $(\del\phi)^2$ and $\phi^4$). So one should use the near-degenerate rather than degenerate perturbation theory. We will come back to this question in future work \cite{MSB2}.

\begin{figure}[htbp]
\begin{center}
\includegraphics[scale=0.14]{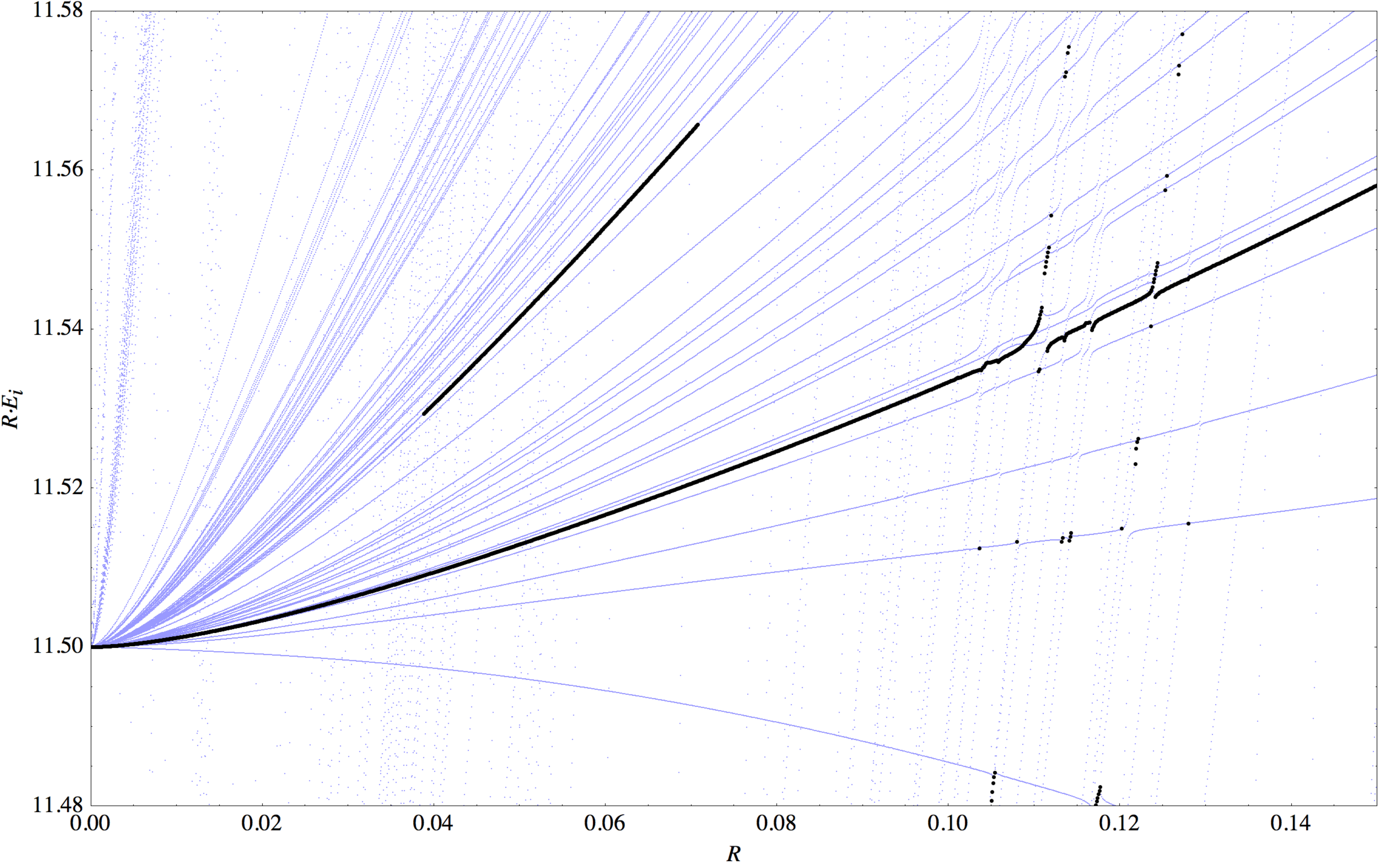}
\end{center}
\caption{The spectrum around $\Delta=11.5$ for $m^2=1$, $\lambda=0.55$, and $0<R<0.15$ with a step of $10^{-4}$. We are plotting energy levels multiplied by $R$. Light blue: real eigenvalues. Black: real part of eigenvalues with nonzero imaginary part. These are raw TCSA data with $\Delta_{\rm max}=12$.
}
\label{fig:imaginary1}
\end{figure}

Going back to $d=2.5$, the above argument is confirmed numerically in figure \ref{fig:imaginary1}, where we show the spectrum around $\Delta=11.5$ for $m^2=1$, $\lambda=0.55$, and $0<R<0.15$. We see precisely one pair of complex conjugate eigenvalues emerging out of the $\Delta=11.5$ group for small $R$. For larger $R$, the spectrum shows intricate structure. We see many beautifully resolved level crossing avoidances in the real part of the spectrum. We also see a second pair of complex conjugate eigenvalues appearing at $R\approx 0.04$ and then disappearing at $R\approx 0.07$. Zooming in on this line of complex eigenvalues, one notices that it joins collision points for pairs of real eigenvalues. 

This last observation may seem to create a minor paradox. Didn't we say that the Landau-Ginzburg flow is not integrable, and that in non-integrable flows energy levels don't cross? The resolution is that this last statement requires a qualification in presence of negative norm states. If two energy levels which head for a collision are both positive-norm (or both negative-norm), they will generically repel. However, in a subspace with non-sign-definite Gram matrix, no-level-crossing rule does not apply. To see this, consider a toy-model $2\times 2$ symmetric generalized eigenvalue problem
\beq
H.c=E\, G.c\,,\qquad H=\left(\begin{array}{cc}
h_{11} & h_{12} \\
h_{12} & h_{22}
\end{array}\right)\!,\quad 
G=\left(\begin{array}{cc}
1 & 0 \\
0 & \sigma
\end{array}\right),
\eeq
where $\sigma=\pm 1$ depending on whether we are dealing with a subspace of positive or non-sign-definite norm. We are assuming that the Hamiltonian matrix is symmetric and real. The distance between the two eigenvalues is controlled by the discriminant:
\beq
D=(h_{11}-\sigma h_{22})^2+\sigma h_{12}^2\,.
\eeq
For $\sigma=1$ the discriminant is a sum of two squares, and level crossing cannot generically happen. On the other hand, for $\sigma=-1$ the discriminant is not positive definite, and can readily change sign if the off-diagonal matrix element increases beyond a critical value. When this happens, we go from having two real eigenvalues to a complex conjugate pair. 

\begin{figure}[htbp]
\begin{center}
\includegraphics[scale=0.14]{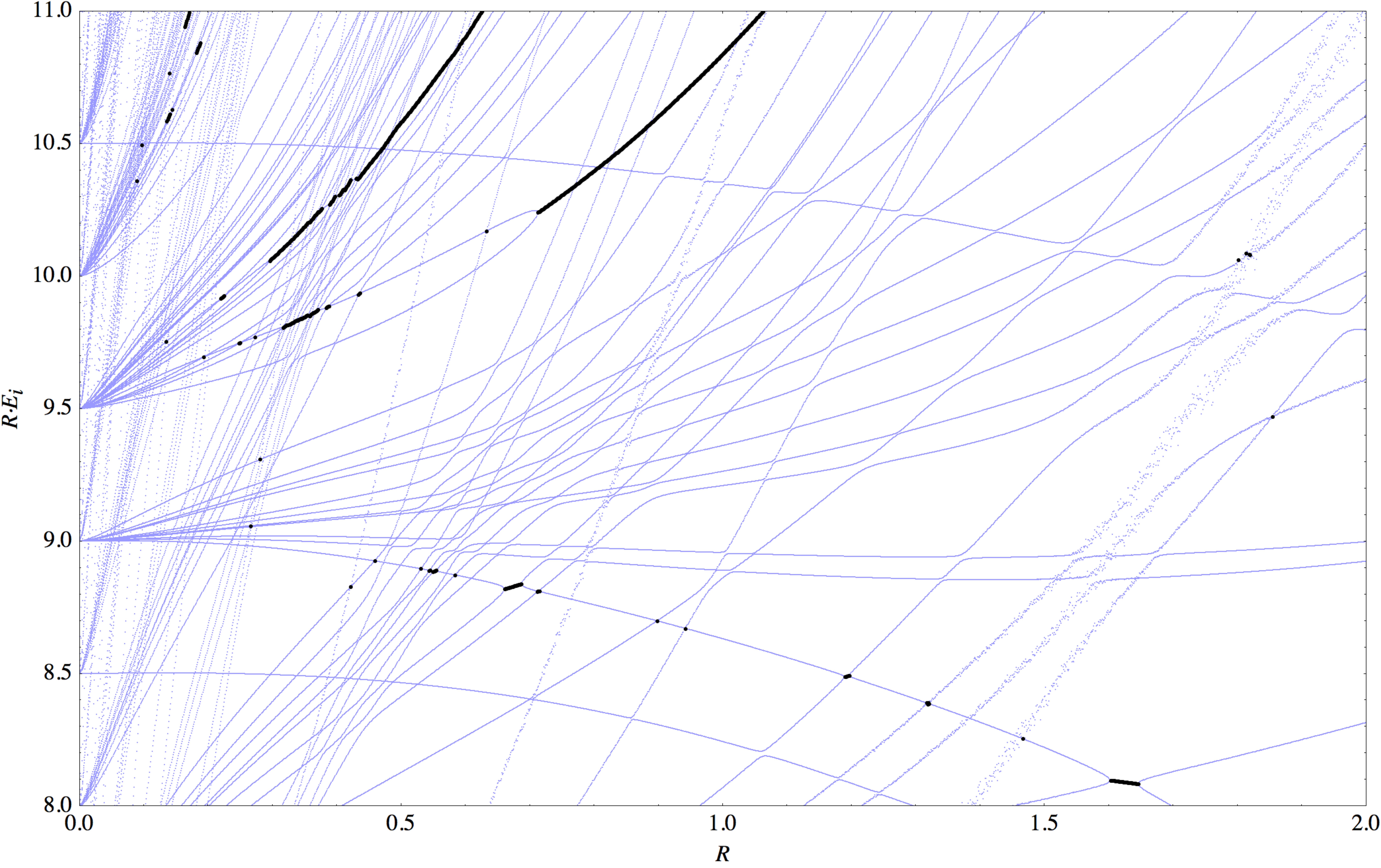}
\end{center}
\caption{The spectrum at $R\, E_i\in[8,11]$ for $m^2=1$, $\lambda=0.55$, and $0<R<2$ with a step of $10^{-3}$. We are plotting energy levels multiplied by $R$. Light blue: real eigenvalues. Black: real part of eigenvalues with nonzero imaginary part. These are raw TCSA data with $\Delta_{\rm max}=11$. The jittering spread noticeable in some of the eigenvalue curves at $R\gtrsim 1.5$ is due to numerical instabilities in the {\tt Mathematica} diagonalization routine.  }
\label{fig:imaginary2}
\end{figure}

As another illustration, in figure \ref{fig:imaginary2} we show the spectrum with $R\, E_i\in[8,11]$ for the same couplings as above but in a wider range $0<R<2$. We clearly see several points where real eigenvalues collide and form a complex conjugate pair, which sometimes reemerges as a pair of real eigenvalues for a slightly larger value of $R$. The most prominent such collision happens at $R\approx 0.7$.

To resolve the multitude of eigenvalue curves in figures \ref{fig:imaginary1}, \ref{fig:imaginary2}, we had to compute the spectrum with a very small $R$ step. For reasons of speed and numerical stability, we have performed these bulky computations with a relatively small $\Delta_{\rm max}$. Since the complex eigenvalues observed in these plots lie relatively close to the cutoff, their energies are likely to shift considerably when the cutoff is increased. However, we don't expect the complex eigenvalues to disappear. In fact, we performed checks for a few selected values of $R$, computing the spectrum with a higher cutoff and, for higher numerical stability, with a higher number of digits rather than at machine precision. The complex eigenvalues were always present.

Notice that the eigenstates corresponding to the complex eigenvalues necessarily have zero norm (computed with respect to the Gram matrix). In a \emph{unitary} theory a state of zero norm has zero overlap with any other state. Such a state is unphysical; it can be kept in the Hilbert state or thrown out without physical consequence. This was the situation with the scalar theory in $d=3$ in section \ref{sec:phi2}, whose extended Hilbert state included some null states, but only as a matter of convenience. In a non-unitary theory, as the one we are discussing now, states of zero norm do not in general have zero overlap with other states. They cannot be removed from the theory without modifying it.

To summarize, the Landau-Ginzburg theory in $d=2.5$ dimensions is a non-unitary interacting quantum field theory. Its spectrum on a sphere of finite radius contains negative-norm states with real energies, as well as zero-norm (but physical) states with complex energies. The negative-norm and zero-norm states belong to the high-energy part of the spectrum, and so their effect on the low-energy physics may not be huge, but the mere presence of these states is a proof that the theory is not unitary. We expect complex eigenvalues to be present also in the limit $R\to\infty$. In particular, the critical point of the theory should have operators with complex scaling dimensions. The same should be true for theories in any fractional $d$. As mentioned above, we expect that these facts are not difficult to check in $4-\eps$ dimensions.

\section{Discussion}
\label{sec:disc}
We have presented an initial study of the feasibility of the TCSA for theories in more than two spacetime dimensions. Both for the massive flow and for the Landau-Ginzburg flow we have obtained promising results. It appears that the TCSA does work, and that its region applicability extends far beyond the realm of perturbation theory.

The most important challenge to the TCSA is the exponential growth in the number of states, see figures \ref{fig:states3d} and \ref{fig:states25d} for examples. This exponential growth cannot realistically be overcome by simply employing more computational resources. Consequently, it appears difficult to raise the cutoff much beyond the values used in this paper. Improving the accuracy of the TCSA method will therefore hinge on our ability to invent techniques that circumvent this problem. This was realized already in investigations of the TCSA in $d=2$, see appendix \ref{sec:2d} for a summary. In this paper we have approached the issue by following standard renormalization principles, and found that adding the right, analytically computed, counterterms can drastically improve the numerical results. Our methods worked well in the given examples, but a more systematic investigation is certainly called for. 

The number of possible problems which can be attacked with the TCSA is huge. Where to start? In this paper we considered the Landau-Ginzburg theory in a sufficiently low number of dimensions ($d=2.5$) where it is UV finite. This was reflected in the fact that the counterterms that we had to add were proportional to \emph{negative} powers of the UV cutoff.
The next natural step would be to carry out the same study in $d=3$ dimensions, where the perturbing operator $\phi^4$ will have dimension $2$. Since this is above $d/2$, the ground state energy will require infinite renormalization (it will be linearly divergent). In addition, the mass term will be logarithmically divergent. The presence of these divergent leading counterterms implies that also the first \emph{subleading} counterterms, suppressed with respect to the leading ones by only one power of $\LUV$, will be numerically more important than in $d=2.5$. It would be interesting to see if good numerical accuracy can be achieved in spite of these complications. This will likely require further improvement of the renormalization procedure.

After $\phi^4$ in $d=3$, the next step might be to consider Yukawa interactions. In principle, dealing with fermions (including chiral fermions) in TCSA is completely straightforward. Because in $d<4$ the Yukawa interaction is less relevant $\phi^4$ (in $d=3$ it has dimension 2.5), the cutoff dependence for Yukawa theories is expected to be more severe than for the Landau-Ginzburg flow, and achieving good accuracy will probably be more challenging.

After the Yukawa theory, one could try to use TCSA to study the 3d gauge theories. The easiest problem in this class might be three-dimensional QED---the $U(1)$ gauge theory with fermionic matter. The interaction term in the Lagrangian, $A_{\mu}\bar \psi\psi$, has dimension 2.5, and so the cutoff dependence will be as severe as for the Yukawa interactions. In addition, it may not be entirely straightforward to treat the gauge interaction term in TCSA. Such issues obviously need to be resolved before one can start attacking nonabelian gauge theories in $d=4$.

Another problem for the future is as follows. In this paper, we were focusing on extracting the mass spectrum of strongly interacting QFTs. Once the accuracy of the method is improved, and the mass spectrum is under total control, it will make sense to turn to the problem of recovering the scattering matrix, from L\"uscher-type corrections to the masses on a sphere of finite radius. As we mentioned in section \ref{sec:phi4-fixedlambda}, the theory of these corrections is not yet available, and it would be interesting to work it out.

Let us conclude on a philosophical note. Few methods are available to study non-supersymmetric strongly coupled quantum field theories. Some of these, like the gap equation, are analytical but merely qualitative, and hardly more reliable than dimensional analysis. On the other hand there are lattice measurements, which are performed from first principles but require vast computer resources.
In this paper we pointed out that there exist alternative algorithms, which are computationally much cheaper but nevertheless defined with a level of mathematical rigor equal to that of the lattice. Here we highlighted the promise of the TCSA. However, the TCSA is just one representative of a family of Hamiltonian truncation methods in quantum field theory. In Appendix \ref{sec:other} we review the existing work on related methods, some of which also look promising and are currently under active development. It seems worthwhile to explore these methods, perhaps not necessarily in order to \emph{compete} with the lattice but rather with the hope that they can significantly reduce the time and resources required to answer interesting nonperturbative QFT questions.
We must break free from the view that some questions can only be answered by the lattice, roll up our sleeves and start constructing alternative tools.

\section*{Acknowledgements}

We thank Paolo Lodone and Xinyi Chen for the collaboration at the initial stages of this project.
We thank John Cardy, Ami Katz, Martin L\"uscher, Giuseppe Mussardo, Volker Schomerus, G\'abor Tak\'acs and Gerard Watts for useful discussions and comments. We thank Brendan McKay for communications concerning graph isomorphisms. We would also like to acknowledge the KITP in Santa Barbara for its hospitality. The results of this paper were first presented at the workshop ``Back to the Bootstrap IV" at Porto University, June 30-July 11, 2014; we thank the organizers and the participants for the stimulating atmosphere. MH thanks the ITFA in Amsterdam for its hospitality.

\appendix

\section{TCSA in $d=2$}
\label{sec:2d}

In this appendix we review the existing TCSA literature in $d=2$. The focus will be on highlighting ideas as they were introduced. Many of these ideas, directly or indirectly, found application in our work. We will also point out the differences between the way the computations are done in $d=2$ and in $d>2$.

TCSA was introduced by Yurov and Al.~Zamolodchikov in \cite{Yurov:1989yu}, where they used the method to study the Lee-Yang CFT $\calM_{2,5}$ perturbed by the only relevant scalar of the theory, of scaling dimension\footnote{In this section $\Delta=h+\bar{h}$, where $h$ and $\bar h$ are the holomorphic and antiholomorphic conformal dimensions. For scalar fields $h=\bar h$.} $\Delta =-2/5$. 
Even though this theory is non-unitary, it was a perfect example on which to demonstrate the method, because: 1) the UV CFT spectrum is very sparse; 2) the perturbing operator is very relevant, so that excellent convergence is achieved for very modest cutoffs; 3) the flow is integrable, which allowed comparing the predictions to an exact solution in the IR.

The next important paper was \cite{Lassig:1990xy} which studied flows originating at the tricritical Ising model $\calM_{4,5}$. This model has several interesting relevant perturbations. Using TCSA, they managed to chart out the intricate phase structure in the IR largely in agreement with expectations. They were the first to observe level repulsion for non-integrable flows, and spontaneous global symmetry breaking (via exponential degeneracy of ground states in finite volume). They also pointed out that for a given UV cutoff the method fails above certain $R$ and gave a practical criterion to determine this $R$, by looking for a change of exponent in the eigenvalue dependence on $R$.  

One case where Ref.~\cite{Lassig:1990xy} encountered a difficulty was the perturbation by the subleading energy operator $\eps'$ of dimension $\Delta=6/5$, which is known to flow in the IR to the Ising model CFT $\calM_{3,4}$. They observed a large UV cutoff dependence of the results which was obscuring the IR behavior. The basic reason for this was soon explained in Ref.~\cite{Klassen:1991ze}: only for $\Delta<d/2 = 1$ is the perturbation UV finite, and we can expect naive TCSA to converge as the cutoff is taken to infinity. For $\Delta\ge d/2$ UV divergences appear, and the TCSA Hamiltonian needs to be renormalized. For $\Delta$ just above $d/2$, the only UV divergence is in the vacuum energy density. Since this affects every energy level in the same way, a quick fix is to consider energy level \emph{differences}. As $\Delta$ is raised further, more nontrivial divergences are expected to appear; it was not discussed at the time how to deal with them.

After a several year's pause, a series of interesting papers appeared where TCSA was applied to perturbations of the free scalar boson theory (up to that moment only minimal model perturbations had been studied). In \cite{Feverati:1998va,Feverati:1998dt} the sine-Gordon perturbation was studied, and an agreement with the exact spectrum known from integrability was observed.
In \cite{Bajnok:2000ar}, the two-frequency sine-Gordon model was studied. This model is non-integrable and has a phase transition in the 2d Ising universality class, which the authors were able to locate and study using TCSA. An integrable SUSY sine-Gordon model was studied using TCSA methods in \cite{Bajnok:2003dk}.

We would also like to mention a nice paper \cite{Lassig:1990wc} (see also \cite{Yurov:1991my}), which first discussed how one can use L\"uscher corrections to extract scattering phases from TCSA data.

How does one do TCSA computations in $d=2$? The space of local scalar operators of the UV CFT is obtained by acting with raising operators on the Virasoro primaries:
\beq
\ket{i}= L_{-n_1}\ldots \bar L_{-\bar{n}_1}\ldots \ket{h,\bar h}\,.
\eeq
In minimal models, some of the states constructed in this way will be null; they are usually separated away. In early works \cite{Yurov:1989yu,Lassig:1990xy}, one also separated the states into quasi-primaries and higher Virasoro descendants. This is not strictly necessary, and the required computation effort may easily outweigh the subsequent speedup in the evaluation of the Gram and the Hamiltonian matrices. When we built a code to check some of the $d=2$ results, we avoided doing this step. In the free scalar theory, one can construct states using the $U(1)$ Kac-Moody algebra rather than the Virasoro algebra; this is simpler since the $U(1)$ Kac-Moody primaries are just the exponentials $\exp(i\alpha \phi)$ while there are many more Virasoro primaries. 

One computes the Gram matrix for these states $G_{ij}=\bra{i}j\rangle$
using the Virasoro (or Kac-Moody) algebra to commute all lowering operators to the right until they hit the primary. Finally, one computes the matrix elements of the perturbing operator, $\bra{i} \phi(1) \ket{j}$.
Here one uses the commutation relations with a Virasoro primary
\beq
[L_n,\phi(z,\bar z)]=z^n[(n+1) h+z\del_z]\phi(z,\bar z)\,,
\eeq
or an analogous Kac-Moody relation. Clearly the conformal algebra plays a big role in the $d=2$ TCSA computations, while it did not feature prominently in our discussion of TCSA in $d>2$. 

The earliest TCSA computations were done in Hilbert spaces of a very modest size: 17 in \cite{Yurov:1989yu}, about 200 in \cite{Lassig:1990xy}. The sine-Gordon papers cited above used much larger Hilbert spaces of a few thousand states. Trying to increase the number of states further by brute force is a game of rapidly diminishing returns, since the number of states grows exponentially with the cutoff, while the error goes down only as a power law. 

How can one tame the growth of the number of states? It is natural to employ the ideas of renormalization in this context, and the current literature already contains several proposals. One approach \cite{Konik:2007cb} (described in more detail in \cite{brandino2010}) is Numerical RG, inspired by the namesake method used in Wilson's famous solution of the Kondo problem. The idea is to add states to the Hilbert space in manageable batches, and after each addition rediagonalize the Hamiltonian and throw out the least important states so that the total number of \emph{retained} states never grows more than a few thousand, while the total number of \emph{explored} states may be several orders of magnitude larger. This procedure is not exact: some systematic error is accumulated because the subsequent batches of states are not allowed to talk to each other directly, but only through the retained states. Empirically, this error seems to be small, since numerical RG procedure is in good agreement with the exact results when available. It would be nice to better understand theoretically why this happens, and to get a quantitative estimate on the error.

An alternative opinion, which is also ours, is that one should invest more effort into purely analytic approaches to renormalization. The point is that, if the theory is weakly coupled at the UV cutoff, one should be able to integrate the energy states above the cutoff \emph{analytically}, and explicitly construct the correction terms needed to improve the accuracy of TCSA. This idea was first studied in \cite{Feverati:2006ni, Watts:2011cr} (see also \cite{Toth:2006tj}). This was done in the context of boundary RG flows, which can be studied via a variant of TCSA. For bulk flows, analytic renormalization was discussed in \cite{Giokas:2011ix} and more recently in \cite{Lencses:2014tba} (see also \cite{Beria:2013hz}, which in particular includes a study of marginally relevant perturbations). Our discussion of renormalization in section \ref{sec:RG} was inspired by \cite{Giokas:2011ix}, although, as we discussed in section \ref{sec:comments}, our method is different in several aspects from \cite{Giokas:2011ix,Lencses:2014tba}. 

\section{Other Hamiltonian Truncation Techniques}
\label{sec:other}

We would like to list here works which applied other Hamiltonian truncation methods in quantum field theory. These methods are conceptually close to TCSA even though their implementations may be quite different, for example because the UV theory is massive instead of conformal. The earliest such work known to us is \cite{Brooks:1983sb}, which studied a 2d Yukawa model---a massive scalar and a massive fermion with a Yukawa interaction $y\phi \bar \psi\psi$---in the Hilbert space of the free \emph{massive} theory on a circle of length $L$. 

Then, in a series of interesting papers \cite{Lee:2000ac,Lee:2000xna} (see also \cite{Salwen:2002dx}) this approach was applied to the massive $\phi^4$ theory in $d=2$ and $d=3$ dimensions. The reader should compare the spectra in figure 8 of \cite{Lee:2000ac} ($d=2$) and figure 1 of \cite{Lee:2000xna} ($d=3$) with our figure \ref{fig:phi4R2.5} in $d=2.5$. We believe that it is worth revisiting the approach of \cite{Lee:2000ac,Lee:2000xna} and possibly improve on some of the implementation details.\footnote{For example, their ``quasi-sparse eigenvector method" does not seem to take into account the high density of states present at high energy, which may compensate the smallness of individual contributions of each one of these states. Their ``stochastic error correction" procedure \cite{Lee:2000xna} computes the sums of squares of high-energy matrix elements of the type encountered in Eq.~\reef{eq:cc1} via a Monte-Carlo procedure. As we observed in section \ref{sec:RG-count} one should be able to compute such corrections analytically.} As we mentioned in section \ref{sec:can} we are planning to do this at least in $d=2$ where TCSA is not directly applicable \cite{Lorenzo}.

Then there is a series of papers by Fonseca and A.~Zamolodchikov \cite{Fonseca:2001dc,Fonseca:2006au,Zamolodchikov:2013ama} on the ``Ising field theory", which is the 2d Ising model perturbed by both relevant operators $m \eps +h\sigma$. In principle, this model can be studied using TCSA. However, these papers find it more efficient to take the $\eps$ perturbation into account exactly from the start, using the fact that it corresponds to turning on mass in the fermionic description of the 2d Ising model.

Finally, we would like to mention a large body of work which apply Hamiltonian truncation methods to field theories quantized on the light front. A time-honored technique is DLCQ, which compactifies $x^-$ to get a discrete spectrum for the Hamiltonian evolution in the $x^+$ direction \cite{Pauli:1985pv,Pauli:1985ps}. While this idea has been around for a while (see \cite{Brodsky:1997de} for a review), it has not yet become a viable alternative to the lattice above $d=2$ dimensions. In 2d it did lead to several interesting analyses, for example of the $\phi^4$ theory \cite{Harindranath:1987db}, or of the large $N$ QCD with matter in the adjoint \cite{Bhanot:1993xp}.

Recently, another approach to Hamiltonian truncation on the light front was proposed in \cite{Katz:2013qua,Katz:2014uoa} in the context of 2d QCD (large or finite $N$) coupled to massless matter. Instead of compactifying a light-cone direction, they use a discrete basis of multiparton wavefunctions, which are in one-to-one correspondence with the left-moving quasiprimary operators of the matter CFT. This appearance of conformal operators creates at least a superficial similarity with TCSA, although it seems that the light-front physics is rather different from that in a finite volume. At any rate, it turns out that the conformal basis of \cite{Katz:2013qua,Katz:2014uoa} leads to a dramatic improvement in the convergence rate of the method. Empirically, convergence seems to be exponentially fast. In comparison, DLCQ is always affected by $1/\text{(lightcone volume)}$ corrections. As we saw in this work convergence rate of TCSA is also power-like in the cutoff. It would be extremely interesting to understand if the successes of the light-front conformal basis can be extended to 2d theories where the right- and left-moving sectors talk to each other, and to theories in $d>2$.

\section{Asymptotics of $C(\tau)$}
\label{sec:tausq}

In this appendix we will give some details concerning the derivation of Eq.~\reef{eq:Bh}. We start from the definition of $C(\tau)$ in Eq.~\reef{eq:ctau}. We do the Weyl transformation which maps the correlation function on the cylinder into one in flat space. In flat space the operation insertions look like:
\beq
w\int_{S_r^{d-1}}dx\, \calV_a(x) \int_{S_{1/r}^{d-1}}dy\, \calV_b(y)\,,\qquad r=e^{\tau/2}\,,
\eeq
where 
\beq
w=r^{\Delta_a}(1/r)^{\Delta_b}=1+(\Delta_a-\Delta_b)\tau/2+O(\tau^2)\,.
\eeq
 is the product of factors picked up by the operators under the Weyl transformation. To the shown order in $\tau$, which is the one we need, the effect of $w$ will average to zero when summed over $a\leftrightarrow b$. We next use the flat space OPE:
\beq
\calV_a(x)  \calV_b(y)\to |x-y|^{-h}[V_c(x)-\half (x-y)^\mu \del_\mu\calV_c(x)+\ldots]\,.
\eeq
Instead of inserting the RHS operator at the middle-point as in \reef{eq:leadOPE}, we put it at one of the endpoints, since this facilitates the subsequent integration. However, the needed accuracy then requires the inclusion of the shown first subleading term.
 
We now have to do the integral over $y$ running over the sphere of radius $1/r=e^{-\tau/2}$. By rotation invariance we can take $x=(\vec{0},r)$. The nonanalytic behavior of the integral as $\tau\to0$ will come from the region of the $y$ sphere closest to $x$, i.e.~its northern cap, which we parameterize as
\beq
y=(\vec{\rho}, \sqrt{e^{-\tau}-\rho^2})\,,\quad |x-y|^2\approx \rho^2(1+\tau)+\tau^2\,,
\eeq
where we kept the approximation needed to get $C(\tau)$ to $O(\tau^2)$. We are led to evaluate the integral
\beq
\int_0^\infty d\rho \frac{{\rm S}_{d-1}\rho^{d-2}}{[\rho^2(1+\tau)+\tau^2]^{h/2}}[V_c(x)-(\tau/2) \del_z V_c(x)]\,.
\eeq
Rescaling $\rho\to \rho/\sqrt{1+\tau}$ and doing the integral over $\rho$ we obtain
\beq
(1+\tau)^{-(d-1)/2} \tau^{d-1-h} {\rm S}_{d-1} \frac{\Gamma((d-1)/2)\Gamma((h-d+1)/2)}{2\Gamma(h/2)}[V_c(x)-(\tau/2) \del_z V_c(x)]\,.
\label{eq:interm0}
\eeq
Next we replace $V_c(x)-(\tau/2) \del_z V_c(x)\approx V_c(\bar x)$, $\bar x=e^{-\tau/2}x$. The factor $(1+\tau)^{-(d-1)/2}$ is absorbed when we transform the remaining integral in $x\in S_r^{d-1}$ into an integral in $\bar x$ over the unit sphere. The remaining coefficient in \reef{eq:interm0} equals $B(h)\Gamma(h-d+1)$ in \reef{eq:Bh}.
\bibliographystyle{utphys}

\section{$M_n$ sequence for $\phi^2\times\phi^2$}
\label{sec:phi2tail}
The purpose of this appendix is to derive Eq.~\reef{eq:Mnij}, which gives the exact asymptotics for the $M_n$ sequence in the case of the $\phi^2$ perturbation. We thus consider the matrix $V^i{}_j$ defined as
\beq
\label{eq:Vmat} \int_{|x|=1} \no{\phi^2(x)} \ket{\Oo_j} \equiv V^i{}_j \ket{\Oo_i}.
\eeq

We will describe two methods to get the answer. The first one is direct: we will study the matrix elements $V^k{}_j$ and $V^i{}_k$ and identify the states of energy $\Delta_k\gg \Delta_i,\Delta_j$ which contribute to the sum defining $M_n$.
To compute $V^k{}_j$, we consider the OPE $\phi^2(x) \times \Oo_j(0)$. By Wick's theorem, we can write:
\begin{align} 
\no{\phi^2(x)} \Oo_j(0) \, &= \, \no{ \phi^2(x) \Oo_j(0)} \nn\\
&+ \sum \expec{\phi(x) \pd_{(\alpha)}\phi(0)} \no{\phi(x) \hat \Oo_j(0)} \nn \\
&+ \sum \expec{\phi(x) \pd_{(\a)}\phi(0)} \expec{\phi(x) \pd_{(\beta)}\phi(0)} \doublehat{\Oo}_j(0)\,.  
\end{align}
Here in the second and third line we put terms where one or two $\phi$'s out of $\phi^2(x)$ are contracted with the $\phi$'s making up $\Oo_j$, which can possibly carry several derivatives denoted collectively as $(\alpha)$, $(\beta)$. The operators $\hat \Oo_j$ and $\doublehat{\Oo}_j$ are $\calO_j$ minus the contracted parts. 

The operators in the third line have all dimension $<\Delta_j$, so they are not relevant for the asymptotics. The operators coming from the second line will appear by expanding $\phi(x)$ under the normal ordering sign into the Taylor series and picking up terms which will not vanish upon integration over the unit sphere, with $\expec{\phi(x) \pd_{(\alpha)}\phi(0)}$ as a weight. A moment's thought shows that the only surviving operators will be $\no{\del_{(\alpha)}\phi(x) \hat \Oo_j(0)}$, i.e.~where $\phi(x)$ is expanded at order $\alpha$. These operators have dimension $\Delta_j$ and are also irrelevant for the asymptotics.

Thus, all the operators with asymptotically large dimensions come from the first line. Expanding around $x=0$ and integrating, we get:
\begin{align} 
\int_{|x|=1} \no{ \phi^2(x) \Oo_j(0)} \ket{0} = \sum_{p = 0}^\infty 
\frac{\Sd}{p! \,4^p (d/2)_p}
 \ket{ \Box^{p}(\phi^2)\Oo_j} \,.
\label{eq:int21} \end{align}
Here $\Box\equiv \del^2$, and $(a)_n = \Gamma(a+n)/\Gamma(a)$ is the Pochhammer symbol. We used
\beq \int_{|x|=1} x_{\mu_1} \dotsm x_{\mu_{2n}}  =  \frac{\Sd}{2^n (d/2)_n} \, [g_{\mu_1 \mu_2}\dotsm g_{\mu_{2n-1} \mu_{2n}} + \text{ permutations}], \eeq
the total number of permutations on the right hand side being $(2n-1)!!$. 

We conclude that the large dimension states appearing in the OPE are the states $\ket{ \Box^{p} (\phi^2) \Oo_j}$, whose dimension is $\DD_j + \Delta(\phi^2) + 2p$.
To complete the calculation, we need to compute $V^i{}_k$ for $k$ being one of these states and $\calO_i$ an operator of low scaling dimension. We have
\begin{align} \label{eq:OPE2} \int_{|x|=1} \no{\phi^2(x)} \no{\Box^{p} (\phi^2) \Oo_j(0)} \, &= \, \int_{|x|=1} 2^{p+1} \expec{\phi(x) \pd_{\mu_1}\dotsm \pd_{\mu_p} \phi(0)}^2 \no{\Oo_j(0)} \nn \\
& \qquad + \text{operators of high dimension}. 
\end{align}
This equation expresses the fact that the only way to lower the dimension drastically is to contract both $\phi$'s out of $\phi^2$ with $\phi$'s under the $\Box^{p}$ sign. The first line evaluates to
\beq
\Sd \Nd^2 \, 2^{2p+1} (2\nu)_p (\nu)_p \ket{\Oo_j}\,.  \label{eq:int22}
\eeq
Multiplying the factors in \reef{eq:int21} and \reef{eq:int22}, we obtain exactly Eq.~\reef{eq:Mnij}.

The second method of deriving Eq.~\reef{eq:Mnij} proceeds by analyzing the correlation function $C(\tau)$, defined in Eq.~\reef{eq:ctau}. Let us focus on one particular contribution to this correlation function, when we contract the $\phi$'s in the $\phi^2$ insertions among themselves, and the $\phi$'s in $\calO_i$ and $\calO_j$ among each other. Going to flat space, the integral over $\bn$ and $\bn'$ in \reef{eq:ctau} can be performed exactly, giving a hypergeometric function ${}_2F_1$. It turns out that expanding this ${}_2F_1$ 
in a series in $r$ and reading off the coefficients gives exactly the sequence \reef{eq:Mnij}. This means that all other ways of contracting $\phi$'s give contributions to the correlation function whose expansion in $r$ does not give rise to terms with arbitrarily high powers. These other options would involve only one or zero contractions between the two $\phi^2$ operators. One can show that for these terms the ${}_2 F_1$ truncates to a polynomial, and therefore they do not correspond to an infinite sequence of scalar operators.

The second way of computing the asymptotic behavior of $M_n$ looks much more economical than the direct method given above. It must also be easier to generalize to other perturbing operators. This would allow one to construct a renormalization procedure for the Landau-Ginzburg flow which takes into account the discreteness of the sequence $M_n$, and is thus more accurate than the one described in section \ref{sec:RG-phi4}.

\footnotesize

\bibliography{tcsa-biblio}

\end{document}